\documentclass[a4paper,11pt]{article}
\usepackage{pos}
 \usepackage{mathtools}
 \usepackage{multirow}
 \newcommand{\nn}{{\nonumber}}

\newcommand{\beq}{\begin{equation}}
	\newcommand{\eeq}{\end{equation}}
\newcommand{\bea}{\begin{eqnarray}}
	\newcommand{\eea}{\end{eqnarray}}

\newcommand{\gsim}{\lower.7ex\hbox{$\;\stackrel{\textstyle>}{\sim}\;$}}
\newcommand{\lsim}{\lower.7ex\hbox{$\;\stackrel{\textstyle<}{\sim}\;$}}
\newcommand{\be}{\begin{equation}}
	\newcommand{\ee}{\end{equation}}
\newcommand{\ba}{\begin{eqnarray}}
	\newcommand{\ea}{\end{eqnarray}}

\newcommand{\ov}{\overline}

\def\vo{\mathcal{V}}

\title{Seeking de-Sitter Vacua in the String Landscape} \ShortTitle{String vacua}

\author*[a]{George K. Leontaris}
\author[b,c]{Pramod Shukla}

\affiliation[a]{Physics Department, University of Ioannina\\
 University Campus, Ioannina 45110, Greece}
 
\affiliation[b]{Department of Physics, University of Allahabad,\\ 
University Road, Old Katra, Prayagraj 211002, India}

\affiliation[c]{Department of Physics, Unified Academic Campus, Bose Institute,\\ 
EN 80, Sector V, Bidhannagar, Kolkata 700091, India}

\emailAdd{leonta@uoi.gr}
\emailAdd{pshukla@jcbose.ac.in}

\abstract{      
In this report, we present a concise review on the various moduli stabilisation schemes proposed in the context of type IIB superstring compactifications using Calabi-Yau orientifolds. We discuss the details of the known schemes by classifying them into two categories; the first one includes non-perturbative superpotential contributions leading to the well-known class of models like KKLT, Racetrack and LVS, while the second one includes only perturbative corrections arising from the series of $\alpha^\prime$- and string-loop effects, leading to a new scheme - known as perturbative LVS. In addition, we motivate and briefly discuss about the global embedding requirements for all these moduli stabilisation schemes, and present some details for the perturbative LVS scheme by focusing on a concrete CY orientifold setting. In this context, the resulting 4D effective supergravity model indeed does not receive any non-pertubative superpotential effects due to the lack of necessary divisor topologies, and the K\"ahler moduli stabilisation is performed entirely by perturbative effects, leading to AdS minimum with exponentially large VEV for the overall volume of the internal CY manifold. It is further shown how this class of AdS solutions can be uplifted to de-Sitter solutions by using a couple of prescriptions such as $D$-term uplifting and the $T$-brane uplifting.
}

\FullConference{%
  Corfu Summer Institute 2022 "School and Workshops on Elementary Particle Physics and Gravity",\\
  28 August - 1 October, 2022\\
  Corfu, Greece}

\begin{document}
\maketitle

\section{Preface}

Realising de-Sitter (dS) solutions in string (inspired) models has been in the centre of attraction since more than two decades, and the tireless efforts made in this program have resulted in a huge amount of literature. In fact, this task of searching for stable dS solutions has been found to be notoriously difficult, especially due to the presence of several no-go results arising from the theoretical constraints while working with the minimal set of ingredients at hand; e.g., the Maldacena-Nu\~nez no-go theorem \cite{Maldacena:2000mw}. Likewise, a series of no-go theorems obstructing the dS realisation has been observed/proposed in the context of type IIA superstring compactifications using Calabi Yau orientifolds \cite{Hertzberg:2007wc, Hertzberg:2007ke, Haque:2008jz, Flauger:2008ad, Caviezel:2008tf,  Covi:2008ea, deCarlos:2009fq, Caviezel:2009tu, Danielsson:2009ff, Danielsson:2010bc, Wrase:2010ew,  Shiu:2011zt, McOrist:2012yc, Dasgupta:2014pma, Gautason:2015tig, Junghans:2016uvg, Andriot:2016xvq, Andriot:2017jhf, Danielsson:2018ztv,Shukla:2019dqd,Shukla:2019akv,Marchesano:2020uqz,Shukla:2022srx}. However, it has been reported at plenty of occasions that considering the possible loopholes in the minimal dS no-go scenarios, one can construct (relatively less simple) models with more ingredients which may realise (stable) dS vacua \cite{Kachru:2003aw, Burgess:2003ic, Achucarro:2006zf, Westphal:2006tn, Silverstein:2007ac, Rummel:2011cd, Cicoli:2012fh, Louis:2012nb, Cicoli:2013cha, Cicoli:2015ylx, Cicoli:2017shd, Akrami:2018ylq, Antoniadis:2018hqy,Antoniadis:2018ngr, Antoniadis:2019rkh, Basiouris:2020jgp, Antoniadis:2020stf, Cicoli:2018kdo, Crino:2020qwk, Cicoli:2021dhg, Heckman:2019dsj, Heckman:2018mxl,Andriot:2022way}. Nevertheless, the simple type IIA no-go results with minimal ingredients have been always useful as a guiding tool to narrow down the suitable region in the flux landscape. Moreover, these no-go scenarios have also played a central role in the recent revival of the swampland conjectures  \cite{Ooguri:2006in, Obied:2018sgi} and the proposal of the Quintessence alternative~\footnote{For a detailed review on the subject, see \cite{Cicoli:2018kdo,Cicoli:2023opf}.} \cite{Garg:2018reu, Agrawal:2018own, Andriot:2018wzk, Andriot:2018ept, Denef:2018etk, Conlon:2018eyr, Roupec:2018mbn, Murayama:2018lie, Choi:2018rze, Hamaguchi:2018vtv, Olguin-Tejo:2018pfq, Blanco-Pillado:2018xyn,Lin:2018kjm, Han:2018yrk, Raveri:2018ddi, Dasgupta:2018rtp, Danielsson:2018qpa, Andriolo:2018yrz, Dasgupta:2019gcd, Andriot:2019wrs,Palti:2019pca,Cicoli:2012tz,Junghans:2022exo}. 

This review will mainly focus on the physics of de-Sitter vacua and the stabilisation of the moduli fields in effective field theory models of string origin. To briefly explain the importance of such a project, in the beginning we should like to make a small digression and present a few  of many  possible underlying motivations.

In cosmology, observational  data already known more than two decades ago~\cite{SupernovaSearchTeam:1998fmf}, have undoubtly confirmed that there is a continuous accelerated expansion  of the universe that began before its reheating. At first glance, this phenomenon contradicts our intuition about what could happen long after the big bang when the attractive force of gravity prevails. According to the widely accepted interpretation, the observed accelerated expansion indicates that the universe is entering an era dominated by dark energy that permeates its entire space.
In the context of Einstein's general relativity, the density of dark energy can be attributed to a positive value of the cosmological constant,  $\Lambda>0$, with a current value of $\Lambda \sim 10^{-122}M_P^4$, where $M_P$ is the four-dimensional Planck mass (for latest update see~\cite{Dimopoulos:2022wzo,Cunillera:2022hdb} and relevant refereneces therein). It is perhaps worth noting that the order of magnitude of $\Lambda$  coincides with that of the mass square differences of the light neutrinos $\Delta m_{\nu_{ij}}^2\lesssim 10^{-30}M_P$ which might indicate some connection between Planck scale and EW physics~\cite{Gonzalo:2021zsp}.

Dark energy has an important role in the theory of General Relativity and Einstein's equations in particular. To understand its effect,  let's recall that in the action of a scalar field within a simple theory context,  only potential differences matter. In General Relativity, however, adding a cosmological constant in the Einstein action, a $\sqrt{-g}\Lambda$ term is induced in the Lagrangian,  i.e.,  proportional to $\sqrt{-g} $ which describes the geometry of space-time. Hence, any change in the potential energy modifies the vacuum energy.

In the context of effective field theories, the simplest scenario describing the basic features of cosmological observations consists of a scalar field acquiring a potential $V(\phi)$ where there are two viable possibilities: (i).~the potential is a slow varying function of $\phi$ whose present  value 
is equal to a slowly varying cosmological constant $V(\phi_{today})=\Lambda$, and (ii).~the potential has a local  minimum with a positive vacuum energy equal to  $\Lambda$. In the first case the potential has a gentle (almost zero) slope, whilst in the second one the scalar potential  has a stable or metastatic dS-type vacuum. Provided that some additional requirements are met in the second case, the $V(\phi)$ potential could be suitable for a successful slow-roll inflationary scenario with the scalar field  $\phi$ playing the role of the inflaton field. Recall  that cosmic  inflation is the theory of the exponential expansion of space in the early universe and explains the origin of the large-scale structure of Cosmos (see for example~\cite{Dimopoulos:2022wzo} and for string inflation~\cite{Baumann:2014nda,Cicoli:2023opf}). 

Based on the preceding observations and remarks one can draw the logical inference  that a variety of fundamental open questions involving vast range of scales are intertwined. Thence, it would  be desirable to contemplate an effective theory with ultra-violet (UV) completion where Planck-scale Physics are naturally integrated.  Currently, the most successful and robust candidate towards a UV completion is string theory which  unifies all fundamental forces including gravity, however, it predicts six extra dimensions which must be compactified to a minuscule size. Thus the ten-dimensional space is written as 
\[ {\cal R}^{3,1}\times {M}_6~, \]
where ${\cal R}^{3,1}$ the four-dimensional Minkowski
space  and $ {M_6} $ is associated with the extra compactified dimensions which escape our detection. In the context constructing four-dimensional (semi)realistic models using superstring compactifications, the so-called Calabi-Yau (CY) threefolds have been found to be directly useful, and studying the properties of such geometries has been among the most important tasks in string model building. In fact, ``Which CY threefold is better or  more suitable for constructing realistic models ?" can be considered to be one of the longstanding questions one still seeks an answer for. Nevertheless, one can always make a classification of the known CY threefolds to single out ``some" classes of examples which can be more useful for certain requirements as compared to the other ones. In this context, apart from studying the global properties of the CY threefolds, the study of co-dimension one topologies also plays a crucial role due to a variety of reasons, e.g. the various scalar potential contributions which are used for moduli stabilisation purpose are governed by such divisor topologies, and a pheno-motivated classification has been recently presented in \cite{Shukla:2022dhz}. 

While  effective models derived from string theory naturally satisfy the criterion of UV completion, a number of problems  must be remedied
before they can be considered as successful low energy theory candidates. Here we focus on  two of the most important issues, the moduli fields 
and the vacuum of the theory which must be of de-Sitter type to ensure a positive $\Lambda$.  
A generic problem in  string derived models is the appearance of several scalar (moduli) fields  in the massless spectrum. In compactifications on a Calabi-Yau manifold, for example, deformations respecting the Ricci flatness conditions give rise to moduli fields that do not acquire a potential at the classical tree-level and do not affect the four-dimensional action. Massless fields in effective theory, however, might have unwanted consequences. 
If they  couple to the gravitational field, they could mediate long-range forces, yet not observed. Furthermore, specific  classes of moduli fields  control the volume and the shape of  the extra dimensions. At the level of the four-dimensional effective field  theory such moduli fields control the magnitude of the  gauge and Yukawa couplings and other parameters  such as masses, the hierarchy of scales,  etc. 
Therefore, in order to construct a realistic model in the framework of superstring-derived field theory, it is important to create a potential with dS-type vacuum energy, and to ensure a positive mass squared $m^2>0$ for the various moduli fields. This has been called the moduli stabilisation problem.  Seeking a solution to these issues one faces a number of additional challenges. As already pointed out, compactification happens on appropriate Calabi-Yau (CY)  manifolds  and some choice of quantum fluxes such as the electromagnetic ones $F_3, H_3$ in type IIB string theory. However, 
there exists a large number of choices of CY manifolds
and fluxes which eventually  determine the properties of the effective quantum field theory and, in particular, the scalar potential thereof. As a result, string theory  is characterised by an enormous number of vacuum ground states  which constitute the so-called ``string landscape''.   
However, the last few years there is considerable activity on whether all possible consistent quantum field theories do belong to the string landscape (see for example~\cite{Danielsson:2018ztv} and references therein). 
According to recent conjectures  it is quite possible that not all  consistent quantum theories emerge as low-energy limits of string compactifications   and a vast majority of them  lies outside of the {\it landscape}, the so called  {\it swampland}~\cite{Vafa:2005ui,Vafa:2019evj,Palti:2019pca,vanBeest:2021lhn}.
Whatever the theory might be, according  to the cosmological data  accepted by the majority of the  scientific community today, the  universe is of  dS type. 
For all these reasons, 
currently, the fundamental problem of moduli stabilisation and the quest for dS vacua  are  a subject of intensive research  activity  in string theory. 
Despite the continuous efforts over the last couple of decades, both issues remain open  to this day.
Possible solutions, if they exist at all~\footnote{The recent literature is vast. For comprehensive analysis and related work on these issues see the reviews~\cite{Danielsson:2018ztv,Palti:2019pca}.}, are sought beyond the classical level and are based on quantum corrections  which  modify   the K\"ahler potential
and the superpotential. During the last few decades, 
a broad spectrum  of perturbative and non-perturbative contributions have been implemented to confront  these issues.
However, it is reasonable to anticipate that an elegant and viable solution should be achieved only with a minimum number of well defined ingredients.   From this perspective, it would be  of particular interest to investigate whether  a successful  outcome can emerge by incorporating only perturbative quantum corrections. 

Building on these lines, the central task for having a genuine dS solution may be divided into three steps; (i).~ensuring the existence by evading any possible no-go restrictions which could come on the way in a given setting, (ii).~ensuring the stability, e.g. via checking the absence of tachyons (and possible flat directions which may have the potential of decompacifying the effective supergravity model), and (iii).~arguing for the viability of the dS solution along with the underlying effective supergravity construction against any possible (un-)known sub-leading corrections or any other new obstacle which  the underlying model may face in course of an improved understanding of the current state-of-the-art. These aspects become more relevant especially for the cases in which very little is known about the series of infinitely many corrections inducing the scalar potential terms responsible for performing moduli stabilisation, and subsequently this presents a challenge to ensure whether the dS vacua which pass the tests in the first two steps are really genuine. For instance, one may take the issues like scale separation and field excursions in moduli space \cite{Blumenhagen:2017cxt, Blumenhagen:2018nts, Blumenhagen:2018hsh, Palti:2017elp, Conlon:2016aea, Hebecker:2017lxm, Klaewer:2016kiy, Baume:2016psm, Landete:2018kqf, Cicoli:2018tcq, Font:2019cxq, Grimm:2018cpv, Hebecker:2018fln, Banlaki:2018ayh, Junghans:2018gdb,Junghans:2020acz,Apers:2022zjx}, tadpole conjecture \cite{Bena:2020xrh,Plauschinn:2021hkp,Plauschinn:2020ram,Marchesano:2021gyv} etc.

The structure of present review is as follows: we start with collecting the necessary ingredients for type IIB model building using CY orientifolds, and discuss the stabilisation of the complex-structure moduli and axio-dilaton using the background fluxes in section \ref{sec_basic}. In section \ref{sec_no-scale-breaking} we present a brief summary of the various (non-)perturbative corrections known in the literature, which can generically be useful for breaking the so-called no-scale-structure in order to facilitate the K\"ahler moduli stabilisation. We present a detailed and classified revisit of the known moduli stabilisation schemes such as KKLT, Racetrack, LVS and perturbative LVS in section \ref{sec_schemes}. Followed by the same, in section \ref{sec_global} we motivate the need of global embedding of these schemes and present a concrete CY orientifolds which fulfils the requirements of realising perturbative LVS, and subsequently uplifting the AdS solution into dS one via including additional corrections from the $D$-term effects and the so-called $T$-brane uplifting. Finally we summarise the report with conclusions and future directions in section \ref{sec_conclusions}.

%%%%%%%%%%SSS%%%%%%%%%%%%%SSS%%%%%%%%%%%%%%SSS%%%%%%%%%%%%%%
%%%%%%%%%%SSS%%%%%%%%%%%%%SSS%%%%%%%%%%%%%%SSS%%%%%%%%%%%%%%

\section{Ingredients for type IIB superstring model building}
\label{sec_basic}

To begin with, in this section, the basic geometric setup and the moduli field content is briefly reviewed within the type IIB  
framework with six internal dimensions compactified on a Calabi-Yau  threefold, denoted hereafter with ${X}$.
The moduli fields associated with deformations of ${X}$ relevant to this work are the dilaton field $\phi$, 
the K\"ahler moduli $T_i$, and the complex structure (CS) moduli, $U^\alpha$.
We start with a brief discussion on the  K\"ahler- and complex-structure moduli and then proceed with the rest of the bosonic spectrum of type IIB. 

\subsection{CY compactifications and moduli fields}

In this section, first we will give a brief review about the origin of moduli fields where we mainly follow the analysis of~\cite{Candelas:1990pi} (see also~\cite{Greene:1996cy}).
First we recall that a K\"ahler manifold $M$ is associated with a (closed) K\"ahler form 
$$J =  g_{i\bar j} d z^i\wedge d \bar z^{\bar j}~,\;d J=0~.$$
A Calabi-Yau (CY) manifold is a compact K\"ahler manifold of
vanishing first Chern class, and in order to dimensionally reduce a 10-dimensional superstring theory to 4-dimensions one needs CY threefolds. For F-theory compactifications to 4-dimensions, the compactifying CY manifold is of complex dimension four, while having the three K\"ahler manifold as the base of the fibration which is not necessarily a CY.
Once the manifold  $M$ and a K\"ahler form are determined 
it can be proven that
there is a unique Ricci-flat K\"ahler metric $g$ on $M$ (i.e. $R_{i{\ov j}}(g) = 0$) such that
the associated K\"ahler
form $J$ is in the same cohomology class as the given one.
In view of this we may
consider the parameter space of Calabi-Yau manifolds to be the parameter space
of Ricci-flat K\"ahler metrics~\footnote{i.e., deformations of
	the background metric
	$g+\delta g$ and the NS-NS two-form $B_2$ variations that do not
	change the Calabi-Yau condition  $R(g)=R(g+\delta g)=0$.}. Thus,
requiring $g_{ij}$ and $g_{ij}+\delta g_{ij}$ to be Ricci flat metrics, so $\nabla^i\delta g_{ij}=0$ and this implies that $g_{ij}$ must satisfy a particular type of differential equation (the Lichnerowicz equation)~\cite{Lichnerowicz1963,Hitchin},
\[   \nabla^k\nabla_k \delta g_{ij} +2R_{i\;\,j}^{\,\,m\,n}\delta g_{mn}=0~.\]
Due to the properties of the K\"ahler manifold, the mixed 
$\delta g_{\bar i j}$ and pure $\delta g_{ i j}, \delta g_{\bar i \bar j}$ type
zero modes satisfy independently the Lichnerowicz equation. Now, variation of the K\"ahler structure $ \delta { g}_{i\bar j}$ (mixed type),
gives ${h^{1,1}} $  parameters~\footnote{The hodge numbers 
	${ h^{r,s}} $ determine the dimension  of Dolbeault cohomology $ H^{r,s}=\frac{\{\omega^{r,s}|\bar\partial \omega^{r,s} =0\}}{\{\alpha^{r,s}|\alpha^{r,s}=\bar\partial \beta^{r,s-1}\}}~\cdot$}, which counts the number of K\"ahler moduli in the 4D effective supergravity theory. On the other hand, 
pure type metric variations ${ { g}_{ij},\; {g}_{\bar i\bar j}}$ give rise to complex structure ({CS}) parameters
counted by ${h^{2,1}} $ of the CY threefold, implying  a nowhere vanishing holomorphic harmonic
3-form ${\Omega}_3$ induced via:
\[   {\Omega_{ijk}} { g}^{k\bar l} {\delta}{ g}_{\bar l\bar m}\,{ dz^i\wedge dz^j\wedge d\bar z^{\bar m}},\]
which is a basic constituent of the Calabi-Yau space.
For the case of a $CY_3$ manifold with $SU(3)$ holonomy, $h^{3,0}=h^{0,3}=1$, hence $\Omega$ is unique. Since more than three decades, a huge amount of effort has been continuously made for constructing and classifying the CY geometries  \cite{Green:1986ck, Candelas:1987kf, Green:1987cr, Candelas:1993dm, Batyrev:1993oya,Candelas:1994hw,Hosono:1994ax,Kreuzer:2000xy,Gray:2013mja} resulting in two broadly defined classes: 
\begin{itemize}
\item
First, the class of CY threefolds realised as multi-hypersurfaces in the product of projective spaces are popularly known as CICYs (Complete Intersection CY threefolds), and this class of CYs involves a total of 7890 geometries \cite{Green:1986ck,Candelas:1987kf,Green:1987cr}. These CICYs have been used not only for constructing local MSMS-like models \cite{Anderson:2011ns, Anderson:2012yf, Anderson:2013xka} but also for addressing other phenomenological issues such as moduli stabilisation, inflation etc. \cite{Anderson:2010mh,Anderson:2011cza,Bobkov:2010rf,Carta:2021sms,Carta:2021uwv, Carta:2022web,Carta:2022oex,Carta:2020ohw}.

\item
The second class of CY threefolds corresponds to those which are realised as hypersurfaces in toric varieties, and are known as THCYs (Toric Hypersurface CY threefolds) \cite{Batyrev:1993oya}. Explicit examples of these CYs can be easily constructed using the four-dimensional reflexive polytopes of the Kreuzer-Skarke (KS) database \cite{Kreuzer:2000xy}. Interestingly, these CY geometries have been studied in great detail since more than a decade ago~\cite{Candelas:1993dm, Candelas:1994hw,Hosono:1994ax} and have been used for explicit ``global embedding'' of many of the earlier proposed models useful for moduli stabilisation, dS realisation, inflationary embedding etc.~see e.g.~\cite{Blumenhagen:2008zz,Collinucci:2008sq,Cicoli:2011qg,Cicoli:2012vw,Blumenhagen:2012kz,Blumenhagen:2012ue,Cicoli:2013mpa,Cicoli:2013cha,Gao:2013rra,Cicoli:2016xae,Cicoli:2017axo,Cicoli:2017shd,AbdusSalam:2020ywo,Crino:2020qwk,Cicoli:2021dhg,Leontaris:2022rzj, Braun:2017nhi,Demirtas:2018akl,Demirtas:2020dbm}. This has led to the development of several efficient/powerful tools/packages such as ``Package for analyzing lattice polytopes" (PALP) \cite{Kreuzer:2002uu,Braun:2011ik}, \texttt{cohomCalg} \cite{Blumenhagen:2010pv,Blumenhagen:2011xn} and \href{https://cytools.liammcallistergroup.com/about/}{\texttt{CYTools}}.

\end{itemize}

\noindent
Let us finally mention that, there exists a very huge number of CY threefolds that can be possible candidates for compactifying the internal six-dimensions of the superstring theory based models, and each of these compactifications can generically result in models with a large number of K\"ahler- and complex-structure moduli, whose stabilisation is typically a very tricky task. In this regard, an estimated upper bound for the number of topologically inequivalent Calabi-Yau hypersurfaces in toric varieties, arising from the Kreuzer-Skarke (KS) database \cite{Kreuzer:2000xy}, is found to be $N_{CY}^{max} \simeq 10^{428}$ \cite{Demirtas:2020dbm}. We will get back to the discussion on constructing explicit global models with some concrete moduli stabilisation and dS realisation in one of the upcoming sections.

\subsection{IIB bosonic spectrum and the flux superpotential}

\noindent
The type IIB superstring compactification using CY threefolds results in ${\cal N} = 2$ supergravities in four-dimensions, and in order to further reduce the amount of supersymmetry one instead uses the orientifolds of the CY threefolds. It turns out that the massless states in the type IIB spectrum are in one-to-one correspondence with the harmonic forms which can be either even or odd under the holomorphic involution $\sigma$ acting on the internal CY threefold. Subsequently, the dimensions of the equivariant cohomology groups $H^{p,q}_\pm (X)$ counts the number of fields in the low energy spectrum. We fix our notations by denoting the bases of even/odd 2-forms as $(\mu_i, \, \nu_a)$ while the dual 4-forms are denoted as $(\tilde{\mu}^i, \, \tilde{\nu}^a)$ where $i=1,..., h^{1,1}_+(X), \, a=1,..., h^{1,1}_-(X)$. Configurations with $h^{1,1}_-(X) \neq 0$ have been found to be quite interesting, e.g. see \cite{Lust:2006zg,Lust:2006zh,Blumenhagen:2008zz,Cicoli:2012vw,Gao:2013rra,Gao:2013pra, Carta:2020ohw, Altman:2021pyc, Crino:2022zjk}. Also, we denote the bases for the even and odd cohomologies of 3-forms $H^3_\pm(X)$ respectively as the symplectic pairs $(a_K, b^J)$ and $({\cal A}_\Lambda, {\cal B}^\Delta)$. Subsequently we fix the normalisation in the various cohomology bases by the following relations:
\bea
&& \int_X \, \mu_i \wedge \tilde{\mu}^j= \delta_i^{\, \, \,j} , \quad \int_X \, \nu_a \wedge \tilde{\nu}^b = {\delta}_a^{\, \, \,b}, \quad \int_X \, \mu_i \wedge \mu_j \wedge \mu_k = k_{ijk},  \nn \\
&& \int_X \, \mu_i \wedge \nu_a \wedge \nu_b = \hat{k}_{i a b}, \qquad \int_X a_K \wedge b^J = \delta_K{}^J, \qquad \int_X {\cal A}_\Lambda \wedge {\cal B}^\Delta = \delta_\Lambda^\Delta. 
\label{eq:intersection}
\eea
Note that, there are two types of orientifold choices; one involving the O3/O7-planes in the fixed point set of the involution corresponding to $K=1, ..., h^{2,1}_+$ and $\Lambda=0, ..., h^{2,1}_-$, while the other one with O5/O9-planes such that $K=0, ..., h^{2,1}_+$ and $\Lambda= 1, ..., h^{2,1}_-$. 

Given that various fields/fluxes can be expanded in suitable bases of the equivariant cohomologies, the involutively-odd holomorphic 3-form $\Omega$, which generically depends on the complex-structure moduli $U^\alpha$ counted by $h^{2,1}_-(X)$, can be written in terms of the period vectors in the following form: 
\be
\label{eq:Omega3}
\Omega\, \equiv  {\cal X}^\Lambda \, {\cal A}_\Lambda - \, {\cal F}_{\Lambda} \, {\cal B}^\Lambda \,,
\ee
where ${\cal F} = ({\cal X}^0)^2 \, \, f({U^\alpha})$ is a generic pre-potential, with $U^\alpha =\frac{\delta^\alpha_\Lambda \, {\cal X}^\Lambda}{{\cal X}^0}$ and with $f({U^\alpha})$ being some function dependent on the complex-structure moduli \cite{Hosono:1994av}. Similarly, one can consider the K\"ahler form $J$ expanded as $J = t^i\, \mu_i$, where $t^i$ denotes 2-cycle volume moduli. In addition to the complex-structure and K\"ahler moduli, there exist other fields in the closed string spectrum of effective 4D supergravity theory rising from the type IIB superstring compactifications using CY threefolds. The latter is obtained by combining left and right moving open strings with Neveu-Schwarz (NS) and Ramond (R) boundary conditions. The following four combinations are possible: 
\[(NS_+, NS_+), \; (R_-,R_-),\;  (NS_+,R_-),\; (R_-,NS_+).\]
From the $(NS_+, NS_+)$ sector we obtain the graviton $g_{\mu\nu}$,  the dilaton $\phi$ and the Kalb-Ramond field $B_2$ which is a rank-2 antisymmetric tensor.
The $(R_-,R_-)$ sector 
gives rise to the $p$-form potentials 
$ C_p,\, p=0, 2, 4$. The $C_0$ potential  and the dilaton field, in particular, define the usual axio-dilaton combination:
\be 
S= C_0+ i\,e^{-\phi}\equiv {{ C_0}+\frac{i}{ g_s}}~,
\label{axidil}
\ee 
where $g_s=e^{\phi}$ corresponds to the string coupling after dilaton $\phi$ receives a VEV. The other NS-NS and RR $p$-form potentials can be expanded as:
\bea
B_2= b^a\, \nu_a\,, \qquad C_2 =c^a\, \nu_a\,,  \qquad C_4 = \rho_i \, \tilde\mu^i + \dots, %+ D_2^i \wedge \mu_i + V^K \wedge a_K + U_K\wedge b^K \,,
\label{eq:fieldExpansions}
\eea
while $b^a$, $c^a$ and $\rho_i$ are various axions, and $\dots$ denotes pieces from quantities (such as a dual-pair of spacetime 1-forms, and the 2-form dual to the scalar field $\rho_i$) which are not directly relevant for our purpose. 

The dynamics of the ${\cal N} = 1$ type IIB 4D effective supergravity theory can be described by using the following chiral variables ($U^\alpha, S, G^a, T_i$) defined as \cite{Benmachiche:2006df}:
\bea
U^\alpha &=& v^\alpha -i\, u^\alpha\,, \qquad S = C_0 + \, {\rm i} \, e^{-\phi} = C_0 + {\rm i}\, s\, , \qquad G^a= c^a + S \, b^a \,, \nn \\
T_i &=& \left(\rho_i +\, \hat{{k}}_{i a b} c^a b^b + \frac12 \, C_0 \, \hat{{k}}_{i a b} b^a \, b^b \right) -{\rm i}\, \left(\tau_i - \frac{s}{2} \, \hat{{k}}_{i a b} \, b^a \, b^b \right)\,, 
\label{eq:N=1_coords}
\eea 
where $\tau_i = \frac12 \, {k}_{ijk} t^j t^k$ is an Einstein frame 4-cycle volume. Notice also that Im$({T}_i)=-\tau_i$ as we follow the conventions of \cite{AbdusSalam:2020ywo}. Let us also mention that throughout this review article, we will limit our discussion to the compactifying CY threefolds with holomorphic involutions such that $H^{1,1}_-(X)$ is trivial, which means that all the fluxes/moduli/axions counted by $a \in h^{1,1}_-$ indices are absent. 

The tree-level K\"ahler potential ${\cal K}_{0}$ can be expressed in the following manner which depends logarithmically on the various moduli fields,
\begin{eqnarray} 
	{\cal K}_{0} &=&-\log\left(-i\int_X{\Omega}\wedge {\bar{\Omega}}\right) - \log(-i(S-\bar S))-2\log{(\cal {\cal V})},
	\label{Kahler0}
\end{eqnarray}
where the internal volume ${\cal V} $ of the CY threefold can be given in terms of the two-cycle volume moduli as follows:
\begin{equation}
	{\cal V}= \frac{1}{6} k_{ijk} t^it^jt^k ~,
	\label{Vol}
\end{equation}
where $k_{ijk} $ denotes the classical triple intersection numbers on ${X}$.

From the $p$-form potentials and the RR-field one gets the field strengths $ F_p= d\,{ C_{p-1}}$, and ${ H_3}= d\,{ B_2}$.
A particular combination   defined as $ G_3=F_3+ S\,  H_3$
appears in type IIB action  and the flux-induced superpotential of the moduli  fields.  The latter 
is a  holomorphic  function which depends on the axio-dilaton modulus $S$, and the CS moduli $U^\alpha$  given by the Gukov-Vafa-Witten (GVW)  formula~\cite{Gukov:1999ya}:
\begin{eqnarray} 
	W_0=  \int_X\, G_3\wedge { \Omega}({U^\alpha}).
	\label{SupW} \label{W0}
\end{eqnarray} 

\noindent
Let us also mention here the fact that the $T$-dual completion of the  GVW superpotential suggests the presence of several other fluxes in a generalised flux framework \cite{Shelton:2005cf, Ihl:2007ah, Robbins:2007yv, Aldazabal:2008zza,Grana:2006hr, Benmachiche:2006df,Aldazabal:2006up, deCarlos:2009fq, deCarlos:2009qm, Aldazabal:2010ef, Lombardo:2016swq, Lombardo:2017yme,Shukla:2019wfo,Shukla:2019dqd} which also includes (non-) geometric fluxes, in addition to the standard S-dual pair of $(F_3, H_3)$ fluxes. In fact, most of the well-known dS no-go scenarios which have been initially proposed in the (geometric) type IIA setting, have been $T$-dualised to some non-geometric type IIB settings in \cite{Shukla:2019dqd}, which may be helpful in narrowing down the vast flux landscape while hunting for dS vacua. However, let us also admit that the naive thought of enriching the model via incorporation of various (non-) geometric fluxes for creating better model building possibilities may not be always helpful, as one does not clearly know how many and which types of fluxes can be consistently turned-on, in the lights of satisfying the the full set of constraints, e.g. those induced from the various Bianchi identities and the tadpole conditions \cite{Ihl:2007ah,Robbins:2007yv, Shukla:2016xdy,Gao:2018ayp}. 

From the above general picture we may infer that due to the existence of many CY-manifolds leading to many moduli, axions and fluxes in the model, it is possible to have a large number of possibly realistic models arising from CY orientifold compactifications. Moreover,  there are also many choices of  fluxes that in principle could satisfy the required restrictions (such as tadpole cancellation etc). 
These possibilities entail an enormous number of  ``String Vacua"  which constitute the so called ``String Landscape''. A long standing question therefore is whether there are any stable dS solutions in the vast String Landscape. We certainly know, however, that  even if the answer is  `yes',  dS vacua are certainly scarce.  If this is the case indeed, it would be interesting to search for unique observational or experimental signatures associated with dS vacua. 

Given the above, a  reasonable sequence of tasks would be 
to ensure all moduli/axionic fields getting non-tachyonic masses in a dynamical fashion, resulting in a  dS vacuum  in  some string (inspired) setting. If possible this should be: (i).~based only on perturbative corrections, (ii).~can be argued to be controllable, and (ii).~can be generically present in effective underlying models. If these  requirements are fulfilled one may also examine the cosmological implications of the effective model, such as inflation.

The effective potential is computed from the K\"ahler- and super-potential expressions using the standard supergravity formula,
\begin{eqnarray} 
	{ V}_{\rm eff }&=&e^{\cal K}\left(\sum_{\alpha,\beta} ({\cal D}_{\alpha}{W}){\cal K}^{\alpha\bar \beta}({\cal D}_{\bar \beta}\overline{W}) - 3 |{W}|^2\right)\,, \label{KahlerV}
\end{eqnarray}
where ${\cal D}_{\alpha}=\partial_{\alpha}+{\cal K}_{\alpha}$ denotes the covariant derivatives with the index 
${\alpha}$ running over all moduli fields.  Now using (\ref{Kahler0}) and (\ref{W0}) through the standard two-step procedure, one initially fixes the axio-dilaton $S$ and the CS moduli $U^\alpha$
via the supersymmetric (flatness) conditions imposed to preserve the supersymmetry,
\be 
{\cal D}_{S}{W}_0=0,\; \qquad {\cal D}_{U^\alpha}{W}_0=0~.
\label{susycond}
\ee 
However, the K\"ahler moduli  do not appear in the perturbative GVW superpotential. Furthermore, at the classical level, i.e., when  non-perturbative (in $W$)- or any perturbative (in ${\cal K}$)- corrections are not present,   the potential (\ref{KahlerV}) vanishes identically $V_{\rm eff }\equiv 0$, (by virtue of the no-scale structure), hence, the K\"ahler moduli   cannot be stabilised   at this level. In a second step, one considers the sub-leading (quantum) corrections which are used to stabilise the K\"ahler moduli.  At this stage we tacitly assume that the masses of the dilaton $S$ the CS moduli $U^\alpha$ have been 
fixed at much larger values than those of the K\"ahler moduli $T_k$ to be fixed by sub-leading effects, so that this two-step procedure makes sense.  In the next section, we are going to consider  various possible corrections suitable for breaking the no-scale structure, and hence for fixing all the moduli.

%%%%%%%%%%SSS%%%%%%%%%%%%%SSS%%%%%%%%%%%%%%SSS%%%%%%%%%%%%%%
%%%%%%%%%%SSS%%%%%%%%%%%%%SSS%%%%%%%%%%%%%%SSS%%%%%%%%%%%%%%

\section{Breaking the no-scale structure via various corrections} 
\label{sec_no-scale-breaking}

The GVW superpotential (\ref{W0}) induced through the background fluxes $F_3$ and $H_3$ can generically help in stabilising the complex-structure moduli along with the axio-dilaton, however all the K\"ahler moduli remain unfixed due to the so-called no-scale structure in the underlying 4D effective supergravity theory. In order to dynamically stabilise these moduli the subleading (quantum) corrections play an important role through the generation of additional pieces in the scalar potential.  Such corrections can be either perturbative or non-perturbative in nature, and can be further classified in terms of two (possibly intertwined) series of terms: ($i$)  via an expansion with respect to the string tension (where the leading order perturbative corrections have been found to be proportional to $ {\alpha'}^3$) and ($ii$) an expansion where the string coupling $g_s$ acts like expansion parameter. Before proceeding to the discussion on their role in moduli stabilisation, let us elaborate a bit more on the form of these corrections which can be leading order in regards to the breaking the no-scale structure.

\subsection{Non-perturbative corrections}
One class of non-perturbative corrections can arise from the E3-brane instaton effects or gaugino condensation through D7-brane, wrapping some suitable 4-cycles inside the compactifying CY threefold. Such a superpotential term can be expressed as \cite{Witten:1996bn} (see \cite{Blumenhagen:2010ja} also for zero-mode analysis):
\bea
& & W_{\rm np} = \sum_k{A}_k e^{- i \, a_k T_k},
\label{WNP} 
\eea
where the coefficients ${A}_k$ may generically depend on complex-structure moduli and the axio-dilaton (or any other moduli, e.g. open string deformations), but in most cases they are considered constants under the assumption that these moduli (unlike the K\"ahler moduli) do not have no-scale structure and subsequently can be stabilised by the leading order flux superpotential. Further, the coefficients $a_k$ on the exponents are equal to $a_k=2\pi$ for E3-instanotn contributions and $a_k=2\pi/N$ for gaugino codensation on stacks of $D7$-branes, $N$ being the rank of the corresponding gauge group. Such non-perturbative superpotential contributions have been proven to be the central ingredients for various popular K\"ahler moduli stabilisation schemes, e.g. KKLT \cite{Kachru:2003aw}, racetrack \cite{Denef:2004dm, BlancoPillado:2006he} and LARGE volume scenarios (LVS) \cite{Balasubramanian:2005zx}.

In addition to these, there are non-perturbative contributions appearing as exponential corrections on top of the usual $E3$-instanton or gaugino-condensation effects leading to the following schematic form of the superpotential \cite{Blumenhagen:2012kz},
\bea
& & W_{\rm np}^{\rm poly} = \sum_k{A}_k e^{- i \, a_k \left(T_k + \sum_w{A}_w e^{- i \, a_w T_w}\right)},
\label{WNP1} 
\eea
where ${A}_w$'s are again some parameters similar to ${A}_k$'s while $T_w$ corresponds to the complexified four-cycle volume of the so-called ``Wilson divisor" which are essential for generating this effect \cite{Blumenhagen:2012kz}. Such effects are known as poly-instanton corrections to the superpotential which can generate sub-leading contributions for K\"ahler moduli stabilisation and also help in driving inflation \cite{Blumenhagen:2012ue,Gao:2013hn,Cicoli:2011ct,Lust:2013kt,Gao:2014fva}.

Moreover, there are non-perturbative effects to the K\"ahler potential which can also generate contribution to the scalar potential, and have the potential of stabilising (many of) the moduli, e.g. the so-called worldsheet- and D1-instanton corrections, e.g. see \cite{RoblesLlana:2006is,Grimm:2007xm,Carta:2021sms}. Such corrections have received a lot of attention recently, in the context of realising the so-called perturbatively flat flux vacua \cite{Demirtas:2019sip,Demirtas:2021nlu,Broeckel:2021uty,Carta:2021kpk,Carta:2022oex}.

After enumerating the possible non-perturative corrections which can generically induce scalar potential terms useful for moduli stabilisation, let us now review the possible leading order perturbative effects. 

\subsection{Perturbative ${\alpha'}^3$-corrections}

\subsubsection{BBHL corrections}

Let us start by recalling that the tree level K\"ahler potential~(\ref{Kahler0}) -as far as the K\"ahler moduli are concerned- exhibits a no-scale structure~\footnote{The no-scale is ensured since  ${\cal K}_i {\cal K}^{i\bar j} {\cal K}_{\bar j}=3$ for ${\cal K}= {\cal K}_0$, and the K\"ahler moduli part does not mix with the pieces involving the moduli $U^\alpha$ and $S$. The no-scale  structure of the ${\cal N}=1$ model is established if also the superpotential  does not depend on $T_i$, hence, when non-perturbative terms are ignored, i.e., ${W}={W}_0$.}.
It is well known, however, that the prepotential for the K\"ahler deformations receives corrections on the worldsheet which 
have been computed long time ago~\cite{Candelas:1990rm}. In the K\"ahler potential these $\alpha'$-corrections are captured 
by a shift in the volume ${\cal V}$ of equation~(\ref{Kahler0}),  given in reference~\cite{Becker:2002nn},  where it was shown  that they arise from a term proportional to ${\alpha'}^3$ in the ten-dimensional
action. In fact, the equation of motion for the dilaton to order ${\alpha'}^3$ is derived from the action~\cite{Becker:2002nn}
\[ S= \int d^{10} x \sqrt{-g^{10}}\, e^{2\phi} (R+4(\partial\phi)^2 +{\alpha'}^3J_0)~,\]
with solution $\phi=\phi_0+\zeta(3)Q/16$,
and $Q$ defined through the 6-d Euler  integrand $\chi= \int_{CY_3} d^6x \sqrt{g} Q$.
Upon compactification, ${\alpha'}^3$-corrections are incorporated in the redefinition of 4D dilaton:
\ba
e^{-2{ \phi_4}}&=&  %e^{-2{\blue\phi_{10}}}({\blue\cal V} +{\red\xi}/2) \nn\\
%	&=& 
e^{-\frac 12{\phi_{10}}}\,({{\cal V}} +\hat{\xi}/2), %\;\;\;{(\rm Einstein\,frame)}
\nn
\ea 
where ${\cal V}$  is the CY volume in the Einstein frame and the coefficient $\hat\xi$ appearing as a shift in the overall volume via a redefinition of the dilaton expressed in terms of the Euler characteristic of the manifold given as below,
\[{ \cal V}=\frac{1}{3!}\kappa_{ijk}{t^i t^j t^k}, ~ \qquad \hat\xi = -\frac{\zeta(3) \, \chi(X)}{2(2\pi)^3\, g_s^{3/2}}. \]
Note that the Einstein-frame volumes are related with their respective string-frame quantities as ${\cal V} = g_s^{-3/2} \, {\cal V}_s$ and  $t^i = g_s^{-1/2} \, t_s^i$, where ${\cal V}_s$ denotes the string-frame CY volume while $t_s^i$ denotes the string-frame 2-cycle volume moduli.  Subsequently, the $\alpha'$-correction is
incorporated into the { K\"ahler} potential through the shift~\footnote{${\xi}$ appears into the { prepotential}~ ${\cal F}=\frac{i}{3!}k_{abc}\frac{{\cal X}^a {\cal X}^b {\cal X}^c}{{\cal X}^0}+{\xi} ({\cal X}^0)^2$~\cite{Candelas:1990pi}.}:
$${\cal V}\to  {\cal Y}_0 = {\cal V} +\frac{\hat{\xi}}{2}\equiv {\cal V} +\frac{{\xi}}{2} \left(\frac{{ S}-\bar{ S}}{2i}\right)^{3/2}\equiv {\cal V} +\frac{{\xi}}{2 g_s^{3/2}}~.$$
As a consequence,  the $\alpha'$-corrected  K\"ahler potential  acquires the following form:
\begin{eqnarray} 
	{\cal K}_{\alpha'} &=&-\log(-i\int_X{\Omega}\wedge {\bar{\Omega}})-\log(-i(S-\bar S))-2\log( {\cal Y}_0)\,.
	\label{KahlerP}
\end{eqnarray}

Provided that the supersymmetric conditions~(\ref{susycond}) are implemented,   when $\alpha'$  corrections are taken into account  break the no-scale invariance and generate a non-vanishing scalar potential, $V_{\rm eff }\ne 0$. In the large volume limit scenario (LVS)~\cite{Conlon:2005ki,Balasubramanian:2005zx,Cicoli:2007xp,Burgess:2020qsc,AbdusSalam:2020ywo}, for example, the leading term of the scalar  potential 
is proportional to $V_{\rm eff}^{\rm LVS}\propto \frac{\hat{\xi}}{{\cal V}^3}|{W}_0|^2$.
Furthermore with the inclusion of non-perturbative corrections in the superpotential as in~(\ref{WNP1}) and an uplifting term 
to the scalar potential emerging from appropriate contributions either from $\overline{D3}$ branes  or D-terms, it is possible to stabilise the K\"ahler moduli and uplift the vacuum to a de-Sitter one. We will get back to this aspect in some detail later on.

\subsubsection{Higher derivative \texorpdfstring{$F^4$}{F4}-corrections}
Apart from the BBHL correction which appears at the two-derivative level through the K\"ahler potential, there are additional higher derivative $\mathcal{O}(F^4)$ effects \cite{Ciupke:2015msa}. These corrections are also induced at $({\alpha^\prime})^3$ order like BBHL, and can be argued to be generic for a given CY orientifold model\footnote{In the context of CY threefolds having divisors of specific topologies, such corrections may be (partially) absent; for example see \cite{Shukla:2022dhz, Cicoli:2023abc}.}. The scalar potential can be expressed in the following simple form: 
\beq
\label{eq:VF4}
	 V_{F^4} = - \frac{g_s^2}{4} \frac{\lambda\,|W_0|^4}{g_s^{3/2} {\cal V}^4} \Pi_i \, t^i,
\eeq
where the $t^i$'s are the volumes of the $2$-cycles for the generic CY manifold $(X)$, $\lambda$ is an unknown combinatorial factor whose value is expected to be around $10^{-3}$ \cite{Grimm:2017okk}, and the $\Pi_i$'s are topological numbers, also called \textit{second Chern numbers}, defined as:
\beq
	\Pi_i =\int_{D_i}c_2(X)\:.
\eeq
In recent years, these corrections have received a significant amount of attention in the context of LVS moduli stabilisation and inflation, e.g. see \cite{Cicoli:2017axo,Leontaris:2022rzj,AbdusSalam:2022krp}.

\subsection{Perturbative string-loop corrections}
The known string-loop corrections can be classified into two parts; first, those which are of the so-called ``KK-type" and ``winding-type" as proposed in \cite{Berg:2004ek, Berg:2005ja, Berg:2005yu, Berg:2007wt}, while the second class consists of those which are of ``logarithmic-type" as proposed in \cite{Antoniadis:2018hqy,Antoniadis:2019rkh}. Let us present the relevant pieces of information about those classes one-by-one.

\subsubsection{KK-type and winding-type corrections}
The initial computations of the string-loop corrections to the K\"ahler potential have been done for the toroidal models, e.g. see \cite{Berg:2004ek, Berg:2005ja, Berg:2005yu, Berg:2007wt}. Subsequently, using some guided route, those have been eventually conjectured for generic CY orientifolds as well \cite{Cicoli:2007xp}. These so-called KK- and winding-type string-loop corrections have been conjectured to take the following form in the Einstein-frame \cite{Cicoli:2007xp}: 
\beq
\label{eq:KgsE}
	\delta {\cal K}_{g_s}^{\rm KK} = g_s \sum_i \frac{C_i^{\rm KK} \, t^i_\perp}{\cal V} \,, \qquad \delta {\cal K}_{g_s}^{\rm W} =  \sum_i \frac{C_i^W}{{\cal V}\, t^i_\cap}\,,
\eeq
where the quantities $C_{i}^{KK}$ and $C_{i}^W$ are some generic functions depending on the complex-structure moduli (and the other possible moduli as well, e.g. the open string moduli). The 2-cycle volume moduli $t^{i}_\perp$ denote the transverse distance among stacks of non-intersecting $D7$-branes and $O7$-planes while $t^i_\cap$ denotes the size of the non-contractible 1-cycles lying on the intersection loci of the two $D7/O7$ stacks. A couple of concrete global realisations of these Ans\"{a}tze have been presented in explicit CY orientifold models in~\cite{Cicoli:2016xae, Cicoli:2017axo}. The scalar potential terms arising from the corrections in \eqref{eq:KgsE} can be given as~\cite{Cicoli:2007xp}:
\bea
\label{eq:VgsKK-W}
& & \delta V_{g_s}^{\rm KK} =  \frac{g_s^3}{2} \frac{|W_0|^2}{{\cal V}^2} \sum_{ij} C_i^{\rm KK} C_j^{\rm KK} \, {\cal K}_{ij}^0 \,, \\
& &\delta V_{g_s}^{\rm W} = -g_s\, \frac{|W_0|^2}{{\cal V}^2} \, \delta {\cal K}_{g_s}^{\rm W} = -g_s\, \frac{|W_0|^2}{{\cal V}^3} \, \sum_i \frac{C_i^W}{t^i_{\cap}}\,. \nonumber
\eea
Here, ${\cal K}_{ij}^0$ is the tree-level K\"ahler metric:
\bea
\label{eq:Kij-tree}
& & {\cal K}_{ij}^0 = \frac{1}{16\, {\cal V}^2}\left(2\,t^i\, t^j - 4 {\cal V}\, k^{ij} \right)\:,
\eea
with $k^{ij}=(k_{ij})^{-1}=(k_{ijk}t^k)^{-1}$. It has been found that the scalar potential is protected against the leading-order pieces of the KK-type corrections due to the so-called ``extended no-scale structure"  \cite{vonGersdorff:2005bf,Cicoli:2007xp}, and subsequently the leading order string-loop pieces in the scalar potential are subleading as compared to those of the BBHL's ${\alpha^\prime}^3$ effects despite the later being leading order in the K\"ahler potential.

The initial arguments of \cite{Berg:2004ek, Berg:2005ja, Berg:2005yu, Berg:2007wt} have been such that the so-called {winding} loop corrections can arise whenever two divisors wrapped by $O$-planes or $D$-branes intersect each other, admitting a non-contractible one-cycle at the intersection locus while the KK-corrections can arise instead from the exchange of KK modes between non-intersecting $D$-branes/$O$-planes. However, building on the field theoretic approach of \cite{vonGersdorff:2005bf}, in a recent revisit \cite{Gao:2022uop}, it has been found that the winding-type effects can appear more generically as to what is expected from \cite{Berg:2004ek, Berg:2005ja, Berg:2005yu, Berg:2007wt}.

\subsubsection{Logarithmic corrections}

\begin{figure}[h!]
	\centering
	\includegraphics[width=0.4\columnwidth]{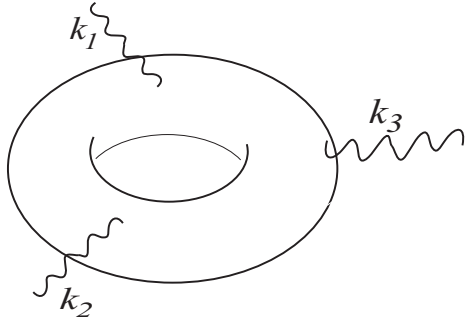}
	\caption{
		\footnotesize
		{Non-zero contribution from 1-loop; 3-graviton scattering amplitude of 2 massless and 1 KK.}
	}
	\label{xx0}
\end{figure}

Apart from the KK-type and winding-type loop corrections discussed so far, there are additional one-loop effects which can appear as logarithmic corrections in the K\"ahler potential \cite{Antoniadis:2018hqy, Antoniadis:2019rkh}. These effects can serve as an alternative way to stabilise the K\"ahler moduli using the perturbative effects only~\cite{Antoniadis:2018hqy,Antoniadis:2019rkh}~\footnote{In this subsection, for completeness, a few computational details  are included. The reader not  interested in these details can skip this section.}.  In the context of the  suggested geometric  configuration of intersecting $D7$-brane stacks, such corrections emerge when 
higher-order curvature terms  in the ten-dimensional effective action are also taken into account. 
The leading correction term of this type 
is  proportional to the fourth power of curvature $  R^{4}$. 
Including this term, the type IIB action takes the following form:
\begin{eqnarray} 
	{\cal S }&{ \supset }&\frac{1}{(2\pi)^7 \alpha'^4}  \int\limits_{ M_{10}} e^{-2\phi_{10}} { R}_{(10)}  - \frac{6}{(2\pi)^7 \alpha'} \int\limits_{M_{10}} \left(-2\zeta(3) e^{-2\phi_{10}} - 4\zeta(2)\right)   { R}_{(10)}^4 \wedge e^2\,,
\end{eqnarray}
where ${R}_{(D)}$ is the $D$-dimensional Ricci scalar, $\phi_{(D)}$ the D-dimensional dilaton,  and ${ R}_{(10)}^4 \wedge e^2$ is a short hand 
expression  in differential forms~(see~\cite{Antoniadis:2019rkh}, as well as~\cite{Kiritsis:1997em} and~\cite{Antoniadis:2002tr}).
Compactifying six dimensions on a CY threefold $({X})$  while  taking into account the tree-level and one-loop generated EH terms,  the ten-dimensional action reduces to
\begin{eqnarray} 
	{\cal S }_{\rm grav}&=& \frac{1}{(2\pi)^7 \alpha'^4} \int\limits_{M_{4} \times {{X}}} e^{-2\phi_{10}} { R}_{(10)} + \frac{\chi}{(2\pi)^4 \alpha'} \int\limits_{M_{4}} \left(2\zeta(3) e^{-2\phi_{10}}  + 4\zeta(2) \right) { R}_{(4)}\,,  
	\label{IIBAction} 
\end{eqnarray} 
Here, $M_{10}={ M}_4\times {X}$, and $\chi$ is 
the Euler characteristic which defined as
\[	
\chi= 	\frac{3}{4\pi^3}\int\limits_{{X}} R\wedge R\wedge R~\cdot 
\]
One  readily observes that $\chi$ contains three powers of the Ricci scalar  $R$, thus it follows that the effective EH term ${R}_{(4)}$ emerging  from the corresponding one ${R}_{(10)}^4$, is possible only in four spacetime dimensions. Besides this, ${R}_{(10)}^4$  can be condidered as a vertex localised at the  specific points in the  bulk space where $\chi\ne 0$.  As such, these vertices can emit gravitons and Kaluza-Klein (KK) excitations within the compactified six-dimensional space.
 The one-loop corrections  are extracted from  the graviton scattering amplitude involving one Kaluza-Klein (KK)-excitation in the $(-1,-1)$-ghost picure as well as 
two massless gravitons in the $(0,0)$ picture as it is  
depicted in Figure~\ref{xx0}.    Performing the computations 
in the orbifold limit ${\mathbb T}^6/Z_N$~\cite{Antoniadis:2019rkh,Antoniadis:2002tr} we can express the final result  as follows: 
\begin{eqnarray} 
	\langle V^2_{(0,0)}V_{(-1,-1)}\rangle = -{ C_R} \frac{1}{N^2}
	\sum_{f,
		k}e^{i \gamma^kq \cdot  {x}_f} \int_{\cal F}
	\frac{d^2\tau}{\tau_2^2}\,
	\int \prod_{i=1,2,3}\frac{d^2z_i}{\tau_2} {\sum_{(h,g)}}' e^{\alpha' q^2 F_{(h,g)}(\tau, z_i)}~. \nonumber  
\end{eqnarray} 
On the right-hand side of the above equation 
${C_R}$ is a constant related to the tensor structure and the  $N$ is the integer related to the orbifold lattice $Z_N$, while the symbol $q$ appearing in the two exponents refers to the KK-momentum. The first sum over $(f,k)$ is associated with the   the orbifold's fixed points $x_f$ and  the representation $\gamma^k$  of  the orbifold's group action. The second sum is over the pairs $(h,g)$ of the orbifold sectors and the prime  means that we exclude the untwisted sector $(0,0)$.  In addition, 
the function $F_{(h,g)} $  appearing in the exponent, indicates  the contribution of the twisted sector $(f,g)=(l,m)\frac{v}{N}$.  Notice that the integration over  the world-sheet torus modulus  $\tau=\tau_1+i\tau_2$ should be restricted  in the fundamental domain of the modular group.
In order to express the localisation property associated with the aforementioned vertex it is convenient to indroduce the definition of the localisation width, denoted hereafter with $w$~\cite{Antoniadis:2019rkh,Antoniadis:2002tr},   
\begin{eqnarray}  w^2 \approx \alpha' F_{(h,g)}(\tau,z_a)\mid_{\rm min} \sim \frac{l_s^2}{N}~.
	\label{Width}
\end{eqnarray} 
Here, as usual,  $l_s= 2 \pi\sqrt{\alpha'}$  stands for the fundamental string length. Then,  computing the amplitude 
 it is found that the coefficient of the induced EH action ${\cal R}_{(4)}$ 
exhibits a Gaussian profile
\begin{eqnarray}
	C_R\frac{ N}{w^6} \,e^{-w^2q_{\bot}^2/2 }~,
	\label{1lcont}
\end{eqnarray}
where $q_{\bot}$ is the internal momentum along the directions transverse  to the D7-brane stack. Then, adopting the intepretation of $N$ as the Euler  characteristic  $ N\rightarrow \chi$, the one loop correction   (expressed in terms of the width $w$ and ${\chi}$)  obtains the following form
\begin{eqnarray}
	\frac{4\zeta(2)}{(2\pi)^7\alpha'}\chi \int_{M_4\times {\cal X}}\frac{e^{-y^2}/(2w^2)}{w^6} \, {\cal R}_{(4)}~,
	\label{Asformula}
\end{eqnarray}

\begin{figure}[h!]
	\centering
	\includegraphics[width=0.65\columnwidth]{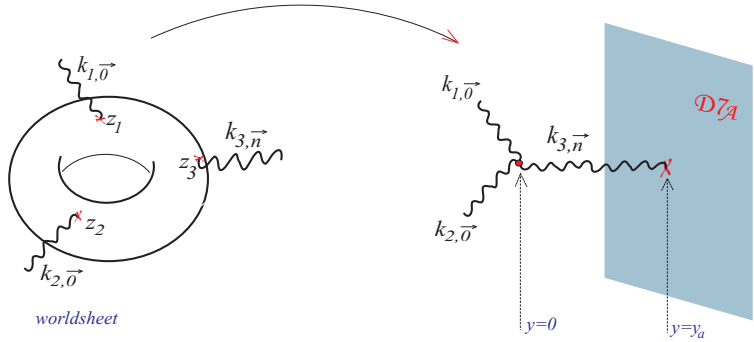}
	\caption{
		\footnotesize
		{Non-zero contribution from 1-loop; 3-graviton scattering amplitude of 2 massless and 1 KK
			%Graviton scattering $\langle V_{(0,0)}^2 V_{(-1,-1)}\rangle$ KK-
			propagating in 2-dimensions  towards a  D7 brane.}
	}
	\label{xx01}
\end{figure}

\noindent 
Within the type IIB  framework, the geometric configuration may also include $D7$ branes and orientifold $O7$ planes. In this context  there can be an exchange of KK-modes between the localised gravity vertex  and the various stacks of $D7$-branes.  Since  the latter reside in four internal dimensions,  KK-excitations transmitted towards each one of them propagate  in a  2-dimensional bulk transverse to  $D7$. For a visual representation this is depicted in Figure~\ref{xx01}. In this way,   logarithmic corrections  are generated.  The magnitude  of each one of them depends on the size of the two dimensional transverse space. Using  (\ref{IIBAction}) and (\ref{Asformula}) while replacing $N\to \chi $, we can compute   the 
corresponding contribution  which takes the form (see the relevant works for the details~\cite{Antoniadis:2019rkh,Antoniadis:2002tr})
\begin{eqnarray}
	{\cal A_S }
	&=&- { C_R} g_s^2 {\cal T}\frac{2\pi}{\sin{\frac{2\pi}{N}}}  \left\{-\frac{\gamma}2 + \ln{\left(\frac{R_\perp \sqrt{2}}{w}\right)} + {\cal O}\left(\frac{w^2}{R^2_\perp}\right) \right\}\,,
	\label{ASint}
\end{eqnarray}
where ${\cal T}$ represents the $D7$-tension and $R_{\bot}$  stands for the size of the transverse space. We observe that the amplitude has a logarithmic dependence on $R_{\bot}$.  Summing up the two contributions~(\ref{1lcont}) and (\ref{ASint})  one finds: 
\begin{eqnarray}
	\frac{4\zeta(2)}{(2\pi)^3}\chi \int_{M_4} \left(1-\sum_k e^{2\phi} {\cal T}_k \ln(R^k_{\bot}/w)\right)\,R_{(4)}\,.
	\label{allcor}
\end{eqnarray}

For the implementation of the prescribed 
plan in the subsequent analysis, a geometric configuration of three intersecting $D7 $ brane stacks 
with quantised 3-form fluxes proposed in~\cite{Antoniadis:2018hqy} can be considered. 
Each  $D7$ brane-stack spans four compact dimensions while it is localised at the remaining two. Table~\ref{Tx}  shows how the setup of the D7 stacks is arranged  in the internal space.  
\begin{table}
	\begin{center}
		\begin{tabular}{l|cccc|cccccc|}
			%	\hline
			\multirow{2}{*}{D7s} &
			\multicolumn{4}{c}{4d  Minkowski}&
			\multicolumn{6}{c}  6 Compact Dimensions
			\\
			&{ 0} &{  1} & { 2} &{ 3} &{ 4 }&{ 5}&{ 6}&{7}&{8} &{ 9} \\
			\hline
			$D7_1$ & & $*$  & $*$  & $*$& ${ *}$  & ${ *}$  & ${*}$& ${*}$  &.  &.\\
			$D7_2$ & & $*$  & $*$  & $*$& ${*}$  & ${*}$   & . & .& ${*}$& ${*}$ \\
			$D7_3$ & & $*$  & $*$   & $*$ & .& .& ${*}$  & ${*}$  & ${*}$& ${*}$\\
			\hline
		\end{tabular}
		\caption{D7- brane configuration of the model under consideration. $D7_1$ stack  resides in `4'`5'`6'`7'  internal dimensions and intersects with $D7_3$ along `6'`7'. } 	\label{Tx}
	\end{center}
\end{table}
Secondly, 
a novel Einstein-Hilbert (EH) term  stemming from $R^4$ terms in the 10D action, can be taken into account which acts as a graviton vertex in the 6D bulk. In fact the graviton excitations transmitted towards the $D7_i$-brane stacks can induce loop-corrections which generate the minima of the F-term scalar potential. Subsequently it turns out that the logarithmic one-loop corrections~\cite{Antoniadis:2019rkh}  discussed in the this section take the general form~$C_{\rm log}=\sigma+\sum_{k} \eta_k \log(\tau_k)$. Here, $\tau_k$ are the  K\"ahler (volume) moduli defined in the previous section, $\sigma\propto\chi$ and the coefficients $\eta_k $ depend on the specific geometric setup and the tension of the branes, e.g. the  D7-brane tension ${\cal T}_i={ g_s} T_0$  and $R_\perp^i$ describing the D7-transverse 2-dimensions. We will detail these aspects more in the upcoming moduli stabilisation section.

%%%%%%%%%%SSS%%%%%%%%%%%%%SSS%%%%%%%%%%%%%%SSS%%%%%%%%%%%%%%
%%%%%%%%%%SSS%%%%%%%%%%%%%SSS%%%%%%%%%%%%%%SSS%%%%%%%%%%%%%%

\section{Moduli stabilisation schemes in type IIB models}
\label{sec_schemes}

\subsection{Models with non-perturbative effects}

As we discussed, the effective theory model is described by ${\cal N}=1$ supergravity with 
the tree-level superpotential given by~(\ref{W0}), which is a function of
the axio-dilaton $S$ and the complex-structure moduli $U^\alpha$ but does not depend on the K\"ahler moduli $T_i$. 
The K\"ahler moduli can appear through various (non-)perturbative corrections. In this case the superpotential
takes the form (e.g. see \cite{Blumenhagen:2009qh}): 
\be 
W = W_0 + \sum_k{A}_k e^{- i \, a_k T_k}.
\label{WNP2} 
\ee 
with $ {W}_0 $  given by~(\ref{W0}), and the coefficients 
${A}_k$ may be considered as some constant parameters while $a_k=2\pi$ for E3-instanton contributions and $a_k=2\pi/N$ for gaugino condensation 
on $D7$ branes. 

Using the generic superpotential (\ref{WNP2}) along with the BBHL corrected K\"ahler potential (\ref{KahlerP}), a master formula for the scalar potential has been presented in \cite{AbdusSalam:2020ywo}. This takes the following form,
\be
V = V_{\mathcal{O}(\alpha'^3)} + V_{\rm np1} + V_{\rm np2} \,,
\label{eq:Vgen-nGen}
\ee
where considering $W_0=|W_0|\, e^{{\rm i} \, \theta_0}$ and $A_i = |A_i|\, e^{{\rm i}\, \phi_i}$ we get:
\bea
\label{MasterF}
V_{\mathcal{O}(\alpha'^3)} &=&  e^{{\cal K}} \, \frac{3 \, \hat\xi ({\cal V}^2 + 7\,{\cal V}\, \hat\xi +\hat\xi^2)}{({\cal V}-\hat{\xi }) (2{\cal V} + \hat{\xi })^2}\, \,|W_0|^2\,, \\
V_{\rm np1} &=& e^{{\cal K}}\, \sum_{i =1}^n \, 2 \, |W_0| \, |A_i|\, e^{- a_i \tau_i}\, \cos(a_i\, \rho_i + \theta_0 - \phi_i) \nn \\
&& \times ~\biggl[\frac{(4 {\cal V}^2 + {\cal V} \, \hat\xi+ 4\, \hat\xi^2)}{ ({\cal V} - \hat\xi) (2{\cal V} + \hat\xi)}\, (a_i\, \tau_i) +\frac{3 \, \hat\xi ({\cal V}^2 + 7\,{\cal V}\, \hat\xi + \hat\xi^2 )}{({\cal V}-\hat\xi) (2{\cal V} + \hat\xi)^2}\biggr]\,, \nn \\
V_{\rm np2} &=& e^{{\cal K}}\, \sum_{i=1}^n\,  \sum_{j=1}^n \, |A_i|\, |A_j| \, e^{-\, (a_i \tau_i + a_j \tau_j)} \, \cos(a_i\, \rho_i - a_j \, \rho_j -\phi_i + \phi_i) \,\nn \\
&& \times \biggl[ -4 \left({\cal V}+\frac{\hat\xi}{2}\right) \, (k_{ijk}\,t^k) \, a_i\, a_j\, + \frac{4{\cal V} - \hat{\xi}}{({\cal V} - \hat\xi)} \left(a_i\, \tau_i) \, (a_j\,\tau_j \right) \nn \\
&& + \,\frac{(4{\cal V}^2  + {\cal V} \, \hat{\xi} + 4\, \hat{\xi}^2)}{({\cal V} - \hat{\xi}) (2{\cal V} + \hat{\xi})}\, (a_i\, \tau_i +a_j\, \tau_j) +\frac{3 \, \hat{\xi} ({\cal V}^2 + 7\,{\cal V}\, \hat{\xi}  +\hat\xi^2)}{({\cal V}-\hat{\xi }) (2{\cal V} + \hat{\xi })^2}\biggr]~. \nn
\eea
Here we assume that the complex-structure moduli and the axio-dilaton are supersymmetrically stabilised by the leading order effects. Notice that the BBHL term \cite{Becker:2002nn} vanishes for $\hat\xi = 0$, reproducing the standard no-scale structure, and for very large volume ${\cal V} \gg \hat\xi$, this term takes the standard form which plays a crucial r\^ole in LVS models \cite{Balasubramanian:2005zx}:
\be
V_{\mathcal{O}(\alpha'^3)} \simeq \frac{e^{K_{\rm cs}}}{2\,s\, {\cal V}^2}\, \times \frac{3\,\hat\xi\, |W_0|^2}{4\, {\cal V}}\,.
\ee
Moreover the master formula (\ref{MasterF}) for can be directly used for moduli stabilisation purposes, and can be considered to be having the following advantages:
\begin{itemize}
\item{Moduli stabilisation is more naturally performed in terms of the 2-cycle moduli $t^i$ which helps in bypassing the ``$t^i$-to-$\tau_i$" conversion needed for defining the chiral coordinates to be used for the K\"ahler metric computations.}

\item{Given that the master formula (\ref{MasterF}) completely determines the scalar potential by merely specifying some topological quantities (such as $k_{ijk}$ and $\chi$ of the CY), and the number $n$ regarding the non-perturbative contributions to $W$, it can elegantly reproduce known moduli stabilisation schemes by choosing parameters like $h^{1,1}_+$, $n$ and $\hat\xi$ as can be seen from Table \ref{tab_known-models}. 

\begin{center}
\begin{tabular}{ |c||c|c|c| } 
\hline 
Model \qquad &  $h^{1,1}_+$ \qquad & $n$ \qquad & $\hat\xi$ \qquad \\ 
\hline
1-modulus KKLT \cite{Kachru:2003aw} \quad &  $1$ \qquad & $1$  & \quad $\hat\xi =0$ \qquad \\
1-modulus $\alpha'$-uplift \cite{Westphal:2006tn, Rummel:2011cd, Ben-Dayan:2013fva} \qquad &  $1$ \quad & $1$ & \quad $\hat\xi>0$ \quad \\
2-moduli KKLT  \cite{Denef:2004dm, BlancoPillado:2006he} \qquad &  $2$ \quad & $2$ & \quad $\hat\xi=0$ \quad \\
2-moduli $\alpha'$-uplift \cite{Louis:2012nb} \quad &  $2$ \quad & $2$ & \quad $\hat\xi>0$ \quad \\
2-moduli Swiss cheese LVS \cite{Balasubramanian:2005zx,Cicoli:2012vw, Cicoli:2013mpa, Cicoli:2013cha} \quad &  $ 2$ \quad & $1$ & \quad $\hat{\xi} > 0$ \quad \\
3-moduli Swiss cheese LVS \cite{Conlon:2005jm, Cicoli:2017shd} \qquad &  $3$ \quad &  $2$ & \quad $\hat\xi>0$ \quad \\
3-moduli fibred LVS \cite{Cicoli:2008va} \qquad &  $3$ \quad &  $2$ & \quad $\hat\xi>0$ \quad \\
\hline
\end{tabular}
\captionof{table}{A set of models for which the scalar potentials can be read-off from the master formula \cite{AbdusSalam:2020ywo}.}
\label{tab_known-models}
\end{center}
}

\item{A general analysis for multi-axion models can be performed using the master formula.}

\end{itemize}

\subsubsection{KKLT-like scenarios}
In order to get the so-called KKLT potential \cite{Kachru:2003aw}, one simply needs to consider $h^{1,1}_+ = n = 1$ and $\hat\xi = 0$ for which case, the 3 contributions to the general potential given in (\ref{MasterF}) become:
\bea
V_{\mathcal{O}(\alpha'^3)} &=& 0, \quad V_{\rm np1} =  4 \, e^{{\cal K}}\, |W_0| \, |A_1|\,a_1\, \tau_1\,e^{- a_1\, \tau_1}\, \cos(a_1 \rho_1 + \theta_0 - \phi_1),  \nn \\
V_{\rm np2} &=& 4\, e^{{\cal K}}\, |A_1|^2 \, e^{- 2 a_1 \tau_1} \left( - {\cal V}\, a_1^2\,k_{111}\,t_1 +  (a_1\, \tau_1)^2 +  a_1 \, \tau_1\right),
\eea
where:
\be
{\cal V} = \frac16\,k_{111}\, t_1^3\,, \qquad \tau_1 = \frac12\,k_{111}\, t_1^2\,, \qquad e^{{\cal K}} = \frac{e^{K_{\rm cs}}}{2\, s \, {\cal V}^2}\,.
\ee
This subsequently leads to the standard KKLT scalar potential,
\be
V_{\rm KKLT} = \frac{9\, e^{K_{\rm cs}}\, a_1\, k_{111}\, |A_1|}{s\, \tau_1^2}\, e^{-a_1\, \tau_1} \left[ |W_0| \, \cos(a_1 \rho_1 + \theta_0 - \phi_1) + \frac{|A_1|}{3} \, e^{-a_1\, \tau_1} \left(a_1\, \tau_1 +3 \right)\right],
\ee
which admits a supersymmetric AdS minimum.

\subsubsection{Racetrack scenarios}

Another moduli stabilisation scheme, known as ``racetrack" \cite{Denef:2004dm, BlancoPillado:2006he} can be considered using the master formula for the choice of parameters $h^{1,1}_+ = n = 2$ and $\hat\xi = 0$ resulting in:
\bea
V &=& \frac{e^{K_{\rm cs}}}{2 s {\cal V}^2}\left[ \sum_{i=1}^2 4 a_i \tau_i |W_0| |A_i|\, \cos(a_i \rho_i + \theta_0 - \phi_i)\, e^{- a_i \tau_i}\right. \nn \\
&+& \sum_{\substack{i=1 \\ i\neq j}}^2 \frac43\,|A_i|^2 a_i^2 \tau_i^2 \, e^{-2 a_i \tau_i} \left(1 + 2 \sqrt{\frac{k_{iii}}{k_{jjj}}}\,\left(\frac{\tau_j}{\tau_i}\right)^{3/2} + \frac{3}{a_i \tau_i} \right) \nn \\
&+& \left. 8 a_1 a_2 |A_1| |A_2| \, e^{-(a_1 \tau_1 + a_2 \tau_2)}\cos(a_1 \rho_1 - a_2 \rho_2 - \phi_1 +\phi_2) \left(\tau_1 \tau_2 + \frac{a_1\tau_1+a_2\tau_2}{2 a_1 a_2} \right) \right]. \nn
\eea
where, like the swiss-cheese CY realised as degree-18 hypersurface in $\mathbb{C}{\rm P}^4[{1,1,1,6,9}]$, we assumed that only $k_{111}\neq 0$ and $k_{222}\neq 0$ leading to the overall volume being given as,
\bea
{\cal V} &=& \frac16\left(k_{111}\,t_1^3 + k_{222}\,t_2^3\right)  =  \frac{\sqrt{2}}{3\sqrt{k_{222}}}\left( \tau_2^{3/2} - \sqrt{\frac{k_{222}}{k_{111}}}\,\tau_1^{3/2}\right),
\eea
where the minus sign in the last piece is dictated by the K\"ahler cone condition $t_1 < 0$, for the choice of basis having the exceptional divisor with the corresponding volume $\tau_1$.

\subsubsection{LARGE volume scenarios (LVS)}

In order to arrive at the standard  LVS scalar potential \cite{Balasubramanian:2005zx,Cicoli:2012vw, Cicoli:2013mpa, Cicoli:2013cha} using the master formula, one needs to consider the parameters as: $h^{1,1}_+=2$, $n=1$ and $\hat\xi>0$ which results in the following pieces,
\bea
V_{\mathcal{O}(\alpha'^3)} &=&  e^{{\cal K}} \, \frac{3 \, \hat\xi \, ({\cal V}^2 + 7\, \hat\xi \,{\cal V} +\hat\xi^2)}{({\cal V}-\hat\xi) (2{\cal V} +\hat\xi)^2}\, \,|W_0|^2,\\
V_{\rm np1} &=& 2\, e^{{\cal K}} \, |W_0| \, |A_1|\, e^{- a_1 \tau_1}\, \cos(a_1 \rho_1 + \theta_0 - \phi_1) \nn \\
& & \times \biggl[\frac{(4\,{\cal V}^2  + {\cal V} \, \hat\xi +4\, \hat\xi^2)}{({\cal V} - \hat\xi) (2\, {\cal V} + \hat\xi)}\, (a_1 \tau_1) +\frac{3 \, \hat{\xi } \, ({\cal V}^2 + 7\, \hat\xi \,{\cal V} +\hat\xi^2)}{({\cal V}-\hat\xi)\,(2{\cal V} +\hat\xi)^2}\biggr],  \nn \\
V_{\rm np2} &=& 4\,e^{{\cal K}}\, |A_1|^2 \, e^{- 2 a_1 \tau_1} \biggl[ - \left({\cal V}+\frac{\hat\xi}{2}\right) \,a_1^2\, k_{111}\,t_1  + \frac{4{\cal V} - \hat\xi}{4({\cal V} - \hat{\xi})} (a_1 \tau_1)^2 \nn \\
& & + \frac{(4\,{\cal V}^2  + {\cal V} \, \hat\xi + 4\, \hat\xi^2)}{2 ({\cal V} - \hat\xi) (2\, {\cal V}+ \hat\xi)}\, (a_1 \tau_1) +\frac{3 \, \hat\xi\, ({\cal V}^2 + 7\, \hat{\xi} \,{\cal V} + \hat\xi^2)}{4 ({\cal V} -\hat\xi) (2 \,{\cal V} + \hat\xi)^2}\biggr]. \nn
\eea
In the large volume limit, the leading order pieces can be given as:
\bea
& & V \simeq \frac{e^{K_{\rm cs}}}{2 s} \biggl[ \frac{3 \hat\xi |W_0|^2}{4 {\cal V}^3} - \frac{4 a_1^2 |A_1|^2 k_{111} t_1}{{\cal V}}\, e^{-2 a_1 \tau_1} \\
& & \hskip2cm + \frac{4 a_1 \tau_1 |W_0| |A_1|}{{\cal V}^2}\, e^{- a_1 \tau_1} \cos\left(a_1 \rho_1 + \theta_0 - \phi_1\right) \biggr]. \nonumber
\label{Vlvs}
\eea
Again considering the swiss-cheese type CY one gets:
\be
V \simeq \frac{e^{K_{\rm cs}}}{2 s} \left( \frac{\beta_{\alpha'}}{{\cal V}^3} + \beta_{\rm np1}\,\frac{\tau_1}{{\cal V}^2}\, e^{- a_1 \tau_1} \cos\left(a_1 \rho_1 + \theta_0 - \phi_1\right) \\
+ \beta_{\rm np2}\,\frac{ \sqrt{\tau_1}}{{\cal V}}\, e^{-2 a_1 \tau_1} \right), 
\label{VlvsSimpl}
\ee
with:
\be
\beta_{\alpha'} = \frac{3 \hat\xi |W_0|^2}{4}\,, \qquad \beta_{\rm np1} = 4 a_1 |W_0| |A_1|\,, \qquad \beta_{\rm np2} = 4 a_1^2 |A_1|^2 \sqrt{2 k_{111}}\,. 
\ee
Notice that (\ref{VlvsSimpl}) matches the form of the potential of standard Swiss cheese LVS models with 2 K\"ahler moduli \cite{Balasubramanian:2005zx} which results in stable non-supersymmetric AdS solutions. 

\subsection{Models without non-perturbative effects}
Although it is hard to argue about the superiority of using non-perturbative effects over the perturbative effects and vice-versa, however let us make the following points:
\begin{itemize}
\item{Non-perturbative effects are quite model specific as we will discuss in the motivation of the global embedding and construction of explicit CY orientifolds in one of the upcoming sections. On contrary, one may argue that perturbative effects may be relatively more generically present. One may take the example of BBHL which is non-vanishing for non-zero $\chi(X)$ while for E3-instanton, gaugino condensation or poly-instanton effects one needs `special' types of 4-cycle in the compactifying CY threefold.}
\item{It maybe fair to argue that one has better understanding of perturbative effects, e.g. in the sense of their 10D origin, as compared to the non-perturbative effects.}
\item{In the sense of lacking iterative behaviour in terms of some expansion parameter, the non-perturbative effects may lead to the issue of the overall control of the model, unless, either they are argued to be absent by specific local construction or if present, are suitably engineered in a useful manner, e.g. in LVS.} 
\end{itemize}

\noindent
Based on these reasons it is always good to seek for alternatives to the standard type IIB moduli stabilisation schemes like KKLT, racetrack and LVS where non-perturbative effects are used for stabilising the K\"ahler moduli. On these lines, in the remaining part of this section, we will discuss a couple of alternative schemes of purely perturbative nature.

\subsubsection{Non-geometric scenarios}

In the meantime, the so-called non-geometric flux compactification scenarios have also received some significant amount of attention for string-inspired model building \cite{Derendinger:2004jn,Grana:2012rr,Dibitetto:2012rk, Danielsson:2012by, Blaback:2013ht, Damian:2013dq, Damian:2013dwa, Hassler:2014mla, Ihl:2007ah, deCarlos:2009qm, Danielsson:2009ff, Blaback:2015zra, Dibitetto:2011qs}. The origin of such (non-geometric) fluxes is argued via a successive application of T-duality on the NS-NS  3-form $H_3$-flux which appears as \cite{Shelton:2005cf},
\bea
\label{eq:Tdual}
& & H_{ijk} \longrightarrow \omega_{ij}{}^k  \longrightarrow Q^{jk}{}_i  \longrightarrow R^{ijk} \, .
\eea
Here $\omega$ denotes the so-called geometric (or metric) flux while $Q/R$ denote non-geometric fluxes. In the presence of such fluxes, the GVW flux superpotential can receive terms in which the K\"ahler moduli can generically appear, e.g. taking the following form \cite{Aldazabal:2006up,Aldazabal:2008zza, Guarino:2008ik,Blumenhagen:2015kja,Shukla:2015hpa,Shukla:2016hyy},
\begin{eqnarray} 
	W_{\rm ng} =  \int_X\, \left[F + S H + \omega_a\, G^a + \,{\hat Q}^{i}\, \,{T}_i \right]_3 \wedge { \Omega}({U^\alpha}),
	\label{Wng1}
\end{eqnarray} 
where $[\cdots]_3$ denotes a combination of 3-forms involving the $F_3, H_3$ and the non-geometric $Q$-flux which can be expressed in a more suitable symplectic/cohomology notation as compared to using the real six-dimensional indices. The appearance of various such fluxes in the generalised superpotential (\ref{Wng1}) introduces a new set of parameters in the effective scalar potential which can subsequently lead to facilitating the moduli stabilisation process by fixing all moduli at tree level. This can be considered as one of the most attractive features of including non-geometric fluxes, which sometimes also creates the possibility of realising dS vacua in non-geometric setting \cite{Blaback:2013ht,Damian:2013dq, Damian:2013dwa, Hassler:2014mla, Blaback:2015zra, deCarlos:2009fq, Blumenhagen:2015xpa}. 

In addition to the set of fluxes $\{H, \omega, Q, R\}$, the modular completion arguments for the 4D effective type IIB supergravity theory demand to introduce a new type of non-geometric $P$-flux which forms a S-dual pair with the non-geometric $Q$-flux, similar to the S-dual pair of ($F_3, H_3$) fluxes \cite{Aldazabal:2006up, Aldazabal:2008zza,Font:2008vd,Guarino:2008ik, Hull:2004in, Kumar:1996zx, Hull:2003kr}. Subsequently, a modular completed version of the so-called generalised flux superpotential (\ref{Wng1}) takes the following form, e.g. see \cite{Aldazabal:2006up,Aldazabal:2008zza, Guarino:2008ik,Blumenhagen:2015kja,Shukla:2015rua},

\begin{eqnarray} 
	W_{\rm ng} =  \int_X\, \left[\left(F + S H \right) + \omega_a\, G^a + \,\left({\hat Q}^{i}\, + S\, {\hat P}^{i} \right)\,{T}_i - \frac{1}{2} {\hat P}^{i}\, \hat{k}_{iab} \, G^a\, G^b\right]_3 \wedge { \Omega}({U^\alpha}),
	\label{Wng2}
\end{eqnarray} 
where as we defined earlier $\left[\cdots\right]_3$ denotes a combination of 3-forms involving all the fluxes expressed in a symplectic/cohomology notation \cite{Ihl:2007ah}. Most of the non-geometric flux models studied in the literature have been initiated with some toroidal setting \cite{Aldazabal:2008zza,Font:2008vd,Guarino:2008ik,deCarlos:2009fq, Danielsson:2012by,Damian:2013dq, Damian:2013dwa}, e.g. using orientifolds of ${\mathbb T}^6/({\mathbb Z}_2 \times {\mathbb Z}_2)$ and ${\mathbb T}^6/({\mathbb Z}_4)$ etc. However, the 10D origin of the 4D effective type IIB scalar potentials have been explored in a series of works \cite{Blumenhagen:2013hva, Gao:2015nra, Shukla:2015rua, Shukla:2015bca, Shukla:2015hpa,Shukla:2016hyy, Andriot:2013xca, Andriot:2011uh, Blumenhagen:2015lta,Villadoro:2005cu}. In addition, applications towards moduli stabilisation, searching dS vacua as well as  building inflationary models using these fluxes have been initiated in \cite{Hassler:2014mla, Blumenhagen:2015qda, Blumenhagen:2015kja,Blumenhagen:2015jva, Blumenhagen:2015xpa,  Li:2015taa,Shukla:2022srx}.

\subsubsection{Perturbative LVS}

In the context of superstring compactifications consisting of some specific intersecting $D7$-brane stacks, a new localised Einstein-Hilbert term can be generated from the dimensional reduction of the $R^4$ terms in the effective ten-dimensional action.   As it has been shown in~\cite{Antoniadis:2019rkh} within this configuration logarithmic corrections appear due to local tadpoles induced by the localised gravity kinetic terms. In the presence of such logarithmic one-loop corrections proposed in ~\cite{Antoniadis:2019rkh}, the corrected K\"ahler potential can be expressed in the following form,
\be 
{\cal K} = -\log\left(-i\int_X \Omega\wedge\bar{\Omega}\right) -\log\left(-i(S-\bar S)\right)-2 \log{\cal Y}~,
\label{Kahlerfin}
\ee  
where both the $\alpha'^{3}$  and the logarithmic corrections are incorporated in the new function ${\cal Y}$ such that
\ba 
\mathcal{Y}&\equiv & \mathcal{Y}_0 + \mathcal{Y}_1, %= \left(\mathcal{V}+\frac{\hat\xi}2\right)+\hat\eta \log \mathcal{V}~.
\ea
where, 
\bea
& & {\cal Y}_0 = \vo +  \frac{\xi}{2} \, e^{-\frac{3}{2} \phi} = \vo + \frac{\xi}{2}\, \left(\frac{S-\ov{S}}{2\,{\rm i}}\right)^{3/2} \,, \nonumber\\
& & {\cal Y}_1 = e^{\frac{1}{2} \phi}\, f({\cal V}) = \left(\frac{S-\ov{S}}{2\,{\rm i}}\right)^{-1/2} f({\cal V})\,.
\label{eq:defY}
\eea
This leads to the following form of the K\"ahaler metric \cite{Leontaris:2022rzj},
\bea
\label{eq:simpKij-1}
& & \hskip-1cm {\cal K} _{S \ov{S}} = {\cal P}_1, \qquad {\cal K} _{T_i\, \ov{S}} = t^i\, {\cal P}_2 = {\cal K} _{S\,\ov{T}_i}, \qquad  {\cal K} _{T_i \, \ov{T}_j} = (t^i \, t^j)\, {\cal P}_3 \, - k^{ij}\, {\cal P}_4~,
\eea
and its inverse
\bea
\label{eq:simpinvKij-1}
& & \hskip-1cm  {\cal K} ^{S \ov{S}} = \tilde{\cal P}_1, \qquad {\cal K} ^{T_i\, \ov{S}} = k_i\, \tilde{\cal P}_2 = {\cal K} ^{S\,\ov{T}_i}, \qquad  {\cal K} ^{T_i \, \ov{T}_j} = (k_i \, k_j)\, \tilde{\cal P}_3 \, - k_{ij}\, \tilde{\cal P}_4~,
\eea
where the functions ${\cal P}_1, {\cal P}_2, {\cal P}_3$ and ${\cal P}_4$ are collected as below,
\bea
\label{eq:Pis}
& & {\cal P}_1 = \frac{1}{8\,s^2\, {\cal Y}^2}\, \left({\cal V} ({\cal Y} + {\cal V}) - 4 \hat\xi ({\cal Y} - {\cal V}) + 4 \hat\xi^2\right), \\
& & {\cal P}_2 = -\frac{1}{8\, s\, {\cal Y}^2} \left(\frac{3}{2}\, \hat\xi - s^{-\frac{1}{2}}\, f + \, s^{-\frac{1}{2}} \, ({\cal V} + 2\hat\xi) \frac{\partial f}{\partial {\cal V}} \right), \nonumber\\
& & {\cal P}_3 = \frac{1}{8\, {\cal Y}^2} \left(1 + s^{-\frac{1}{2}} \, \frac{\partial f}{\partial {\cal V}} -  {\cal Y} \,  s^{-\frac{1}{2}} \, \frac{\partial^2 f}{\partial {\cal V}^2} \right), \nonumber\\
& & {\cal P}_4 = \frac{1}{4\, {\cal Y}} \left(1 + s^{-\frac{1}{2}} \, \frac{\partial f}{\partial {\cal V}} \right), \nonumber
\eea
while the functions $\tilde{\cal P}_1, \tilde{\cal P}_2, \tilde{\cal P}_3$ and $\tilde{\cal P}_4$ are given as,
\bea
\label{eq:tildePis}
& & \tilde{\cal P}_1 = \frac{{\cal P}_4-6 {\cal P}_3 {\cal V}}{{\cal P}_1 {\cal P}_4 + 6 {\cal P}_2^2 {\cal V}-6 {\cal P}_1 {\cal P}_3 {\cal V}}\,, \\
& & \tilde{\cal P}_2 = \frac{{\cal P}_2}{{\cal P}_1 {\cal P}_4 + 6 {\cal P}_2^2 {\cal V}-6 {\cal P}_1 {\cal P}_3 {\cal V}}\,, \nonumber\\
& & \tilde{\cal P}_3 = \frac{{\cal P}_2^2-{\cal P}_1 {\cal P}_3}{{\cal P}_4\left({\cal P}_1 {\cal P}_4 + 6 {\cal P}_2^2 {\cal V}-6 {\cal P}_1 {\cal P}_3 {\cal V}\right)}\,, \nonumber\\
& & \tilde{\cal P}_4 = ({\cal P}_4)^{-1}. \nonumber
\eea
Now assuming that the complex-structure moduli and the axio-dilaton are fixed by the leading order effects, the scalar potential for the K\"ahler moduli can be generically given by the following master formula,
\bea
\label{eq:masterV}
& & \hskip-0.4cm \boxed{ V_{\alpha^\prime + {\rm log} \, g_s} %=   e^{{\cal K}} \, |W|^2\, \left(K_{T_i} \, K^{{T_i} \ov{T}_{\ov j}} \, K_{\ov{T}_j} -3 \,\right) \\ &&
 = e^{{\cal K}}\, |W|^2 \biggl[\frac{3\, {\cal V} }{2\, {\cal Y}^2} \left(1 + \frac{\partial {\cal Y}_1}{\partial{\cal V}}\right)^2 \left(6 {\cal V} \tilde{\cal P}_3 - \tilde{\cal P}_4 \right) - 3 \biggr].} 
\eea

This master formula can result in quite complicated terms for generic function $f({\cal V})$. However, assuming that $\eta_1 = \eta_2 = \eta_3 = \eta$ exploiting the underlying exchange symmetries within the model, we take the following form for $f({{\cal V}})$ \cite{Antoniadis:2019rkh},
\be
f({{\cal V}})   = {\sigma }+{\eta }\log({{\cal V}}),
\ee 
where coefficients ${\eta }$ and ${\sigma}$ are expressed in terms of ${\xi }~\propto { \chi}$, and the following relation holds \cite{Leontaris:2022rzj},
\be 
{\sigma }=-{\eta }=\frac{\zeta(2)}{\zeta(3)}{\xi}.
\ee
Subsequently, one gets the following ratio which turns out to be of particular interest:
\be
\frac{{\hat{\xi}}}{{{\hat\eta}}}=
-\frac{\zeta(3)}{\zeta(2)}\frac{1}{ g_s^2}\,,
\ee
with 
\be
{ \hat{\xi}}={\xi}{g_s^{-3/2}}~;\;\;{{\hat\eta}}={ g_s^{1/2}}{\eta }.
\ee
\noindent
Now, the (F-term) potential in the large volume limit results in the following leading order pieces,
\ba
\label{eq:pheno-potV3}
& & V_{\alpha^\prime +{\rm log} \, g_s}^{(1)}\equiv V_F = \frac{3\, \kappa\, \hat\xi}{4\, {\cal V}^3}\, |W_0|^2 - \frac{3 \, \kappa\, (2\,\hat\eta -\hat\eta  \ln{\cal V})}{2{\cal V}^3}\,|W_0|^2\approx C\,\frac{{\hat\xi}+ 2 \hat{\eta} \log {\cal V}}{{\cal V}^3}~,\label{VFex}
\ea
where  $C= 3 g_s \,e^{K_{cs}}|W_0|^2/(32 \pi) > 0$ follows from $\kappa = g_s e^{K_{cs}}/(8 \pi)$, and the last approximation is due to the fact that in the LSV(-like) scenario $\log{\cal V} \gg 1$.

\begin{figure}[h!]
	\centering
	\includegraphics[width=0.65\columnwidth]{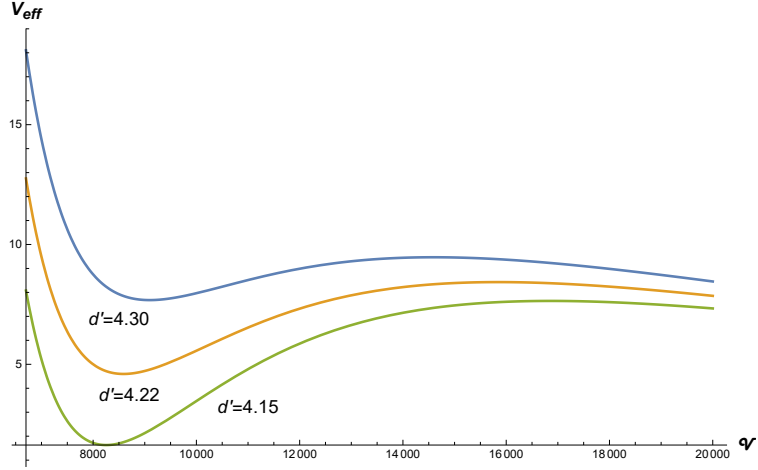}
	\caption{
		\footnotesize
		{The potential~(\ref{VFexd}) in arbitrary units vs ${\cal V}$ for $\xi=8,\eta=-1/2$ and three values of the parameter $ d'$ where $d'= 10^4 d$. The dS minima are restricted in the range $ d'\sim [4.15-4.42]$. For  $d'\lesssim 4.15$ the minimum of the potential is of AdS type. For values  $ d'\gsim 4.42$ the minimum is lost and the potential becomes a monotonic function of ${\cal V}$. }
	}
	\label{f1}
\end{figure}

A few properties of the potential (\ref{VFex}) are worth mentioning:  a minimum  can be ensured as long as  ${\hat\eta}<0$ (while $\hat\xi>0$). This confirms that the logarithmic correction plays a decisive role. Furthermore, stabilisation of the volume modulus occurs at {\bf large} values and in the  {\it  weak string-coupling  regime}.
Nevertheless, as long as we consider only the F-term potential, the minimum is of {\it AdS}-type.  Hence, new contributions are required to {\bf  uplift} the vacuum to {\bf dS}.
This can be uplifted either with 
a $\overline{D3}$ contribution \cite{Kachru:2003aw} or with a positive D-term  (see also~\cite{Burgess:2003ic}) which originates from the the universal $U(1)$ factors of the intersecting D7-branes~\cite{Antoniadis:2019rkh}. This term has a volume dependence of the form $V_D= \frac{Q}{{\cal V}^2}$  where $Q$ a positive constant, so the effective potential can be written
\be
V_{\rm eff}= C\left(\frac{\xi+ 2 \eta \log {\cal V}}{{\cal V}^3}+\frac{3 d}{{\cal V}^2}\right)\label{VFexd}
\ee
where    $d=Q/(3 C)>0$. This type of potential has been studied extensively in references~\cite{Antoniadis:2018hqy,Antoniadis:2019rkh}  and is outlined here in brief.  
Minimising  the effective  potential it is found that the volume ${\cal V}$ at the extrema is given by:
\be 
{\cal V}_{\rm extrema}=
\frac{\eta }{d} W_{0/-1}\left(\frac{d }{ \eta}e^{\frac{1}{3}-\frac{\xi}{2 \eta}}\right)
\equiv -\frac{|\eta| }{d} W_{0/-1}\left(-\frac{d }{ |n|}e^{\frac{1}{3}+\frac{\xi}{2 |\eta|}}\right)~,
\ee 
where $W_0, W_{-1}$ represent the two brances of the double-valued Lambert-W function.  The potential acquires a minimum as long as the coefficient 
$\eta$ takes  negative values, $\eta<0$. 
The volume at the minimum (denoted with ${\cal V_{\rm min}}$)  is associated with the $W_0$ branch, whlist there
is also a local maximum  located at ${\cal V_{\rm max}}$
associated with the $W_{-1}$ branch. 
Moreover, real  values for $W_{0/-1}(z)$ are ensured when  $z>-\frac{1}{e}$  which  implies the upper  bound on 
$d< |\eta| e^{-\frac{4}{3}-\frac{\xi}{2 |\eta|}}$.  The requirement for dS vacua, $V_{\rm eff}>0$, puts more stringent  constraints~\cite{Antoniadis:2018hqy} which can be solved numerically for each set of values of $\xi$ and $ \eta$. For typical values of $-\chi \sim [100$-$200]$ and $g_s\sim 0.1$ the values of $\xi\lesssim {\cal O}(10)$.
As an example, in Figure~\ref{f1} the  effective potential~(\ref{VFexd})  is depicted for $\xi=8,\eta=-1/2$ and several values of the parameter $d$. 

%%%%%%%%%%SSS%%%%%%%%%%%%%SSS%%%%%%%%%%%%%%SSS%%%%%%%%%%%%%%
%%%%%%%%%%SSS%%%%%%%%%%%%%SSS%%%%%%%%%%%%%%SSS%%%%%%%%%%%%%%

\section{Global embedding of perturbative LVS}
\label{sec_global}

Although the compactifying CY threefolds can be considered to part of the so-called hidden sector ingredients for the phenomenological model building, however the underlying geometries are omnipresent in the 4D effective models, and their imprints can be traced in physical observables such as masses and couplings which in the end depend on the moduli VEVs attained through an appropriate scheme such as KKLT and LVS. In addition, let us recall that while doing the pheno model building one (implicitly) assumes many things and it is always a challenging task to find concrete ``global" setting in which all those requirements are fulfilled. In this context, it has been found that apart from knowing the topological data for the compactifying CY threefold (such as hodge numbers which counts the field content in the spectrum), the study of divisor topologies is a very crucial step as it can help in understanding the various classes of scalar potential terms which can be induced at sub-leading order, and hence maybe useful or concerning for a given model\footnote{For example, see \cite{Shukla:2022dhz} regarding a recent classification of the divisor topologies of CY threefolds arising from the KS dataset, and their relevance for pheno-model building.}. In this section, we will briefly enumerate (some of) such requirements appearing in a couple of well known models and discuss the challenges one has to face in this regard.

\subsection{Why global embedding ?}
As a motivation for global model building, let us review some of the known corrections to the 4D type IIB scalar potential and their possible connections with the internal geometries of the compactifying CY threefold.

\begin{enumerate}

\item
In order to break the no-scale structure and facilitate the K\"ahler moduli stabilisation, the use of {non-perturbative effects} in the superpotential have been  witnessed, leading to interesting scenarios like KKLT \cite{Kachru:2003aw}, Racetrack \cite{Denef:2004dm,BlancoPillado:2006he} and LVS \cite{Balasubramanian:2005zx}. For facilitating these corrections one needs to find the some suitable class of divisors ($D$) satisfying the Witten's unit arithmetic genus condition \cite{Witten:1996bn}:
\bea
& & \chi_0 (D) = h^{0,0}(D) - h^{0,1}(D) + h^{2,0}(D) = 1.
\eea
For this reason, the so-called rigid divisors with topological data $\{h^{0,0}(D) =1, h^{0,1}(D) = 0, h^{2,0}(D) = 0\}$ are natural candidate to serve this purpose. In cases otherwise, there can be `rigidification' of non-rigid divisors as proposed in \cite{Bianchi:2011qh,Bianchi:2012pn,Louis:2012nb} in order to make the divisor suitable for wrapping E3-instantons or D7-branes inducing the gaugino condensation effects. In passing, let us also mention that the sub-leading poly-instanton effects to the superpotential are induced through a so-called `Wilson divisor' \cite{Blumenhagen:2012kz}, which are some complex two-surfaces realised as ${\mathbb P}^1$-fibrations over tori.

\item
The CY threefolds with the so-called Swiss-Cheese structure are found to be central ingredients for the LVS scheme of moduli stabilisation \cite{Balasubramanian:2005zx}. These divisors are known as ``diagonal'' del-Pezzo divisors and they can be shrinked to a point-like singularity by squeezing along a single direction \cite{Cicoli:2011it,Cicoli:2018tcq,Cicoli:2021dhg}. In fact, a del-Pezzo surface $dP_n$ is obtained by blowing up $\mathbb{P}^2$ at $n$ generic points, and a \textit{diagonal} del-Pezzo is a divisor on the threefold $X$ which is a $dP_n$ surface and can be arbitrarily contractible to a point. To be more specific, a del Pezzo divisor ($D_s$) satisfies the following topological conditions:
\bea
\label{eq:dP}
& & \int_{X} D_s^3 = k_{sss} > 0\, , \qquad \int_{X} D_s^2 \, D_i \leq 0 \qquad \forall \, i \neq s \,.
\eea
Here $k_{sss} = 9 - n \neq 0$ for a $\mathrm{dP}_n$ divisor is the degree of $\mathrm{dP}_n$, and the `diagonality' of $D_s$ is ensured by imposing the following useful condition \cite{Cicoli:2018tcq}
\bea
\label{eq:diagdP}
& & k_{sss} \, \, k_{s i j } = k_{ss i} \, \, k_{ss j} \, \qquad \qquad \forall \, \, \, i, j \:.
\eea
This condition can be understood as follows: whenever (\ref{eq:diagdP}) is satisfied, the volume of the diagonal dP (ddP) divisor $D_s$ turns out to be a complete-square quantity in terms of the 2-cycle volumes because of the following relation:
\bea
& & \tau_s = \frac{1}{2}\, k_{s ij} t^i \, t^j = \frac{1}{2 \, k_{sss}}\, k_{ssi} \, k_{s s j} t^i \, t^j = \frac{1}{2 \, k_{sss}}\, \left(k_{ss i} \,t^i \, \right)^2\,,
\eea
where we sum over $i,j$ but not over $s$. In the presence of ddP divisors in LVS models, the volume of the CY threefold can be expressed in the so-called \textit{Swiss-Cheese} form:
\beq
	{\cal V} =f_{3/2}(\tau_j)-\sum_{i=1}^{n_{\rm small}}\kappa_i\tau_i^{3/2}; \quad \text{with } j=1,...,N_{\rm large}\:.
\eeq
Here $n_{\rm small}$  denotes the K\"ahler moduli corresponding to the rigid ddP cycles, while the remaining $N_{\rm large}=h^{1,1}_+-n_{\rm small}$ K\"ahler moduli are clubbed in $f_{3/2}$ which is a homogeneous function of degree-3/2 in $\tau_i$ moduli; e.g. the CY realised as a degree-18 hypersurface in WCP$^4[1,1,1,6,9]$ has a ddP and $h^{1,1} = 2$ resulting in $f_{3/2} = \tau_b^{3/2}$. 

Using the topological data of the CY threefolds collected in the AGHJN-database \cite{Altman:2014bfa}, an initial classification for the diagonal dP divisors has been presented in \cite{Cicoli:2021dhg} as summarised in Table \ref{tab_ddPns-GstarM}, and have resulted in a  useful conjecture stated as below \cite{Cicoli:2021dhg}:

\noindent
``{\it The CY threefolds arising from the 4D reflexive polytopes listed in the KS database do not exhibit a `diagonal' $\mathrm{dP}_n$ divisor for $1 \leq n \leq 5$, in the sense of satisfying the eqn. (\ref{eq:diagdP}).}"

\noindent
\begin{table}[h!]
	%\centering
	\hskip1cm \begin{tabular}{|c||c|c||c|c|c|c|c|c||c|}
		\hline
		$h^{1,1}$ & Poly$^*$ &  Geom$^*$ & ${\mathbb P}^2$  & ${\mathbb P}^1\times{\mathbb P}^1$  & $\mathrm{ddP}_n$ & $\mathrm{ddP}_6$  & $\mathrm{ddP}_7$  & $\mathrm{ddP}_8$  & $n_{\rm LVS}$ with \\
		&  & ($n_{\rm CY}$) & ($\mathrm{ddP}_0$) &  & $1\leq n \leq 5$ &  &  &  & $\mathrm{ddP}_n\geq 1$ \\
		\hline
		1 & 5 & 5 & 0 & 0  & 0  & 0 & 0 & 0 & 0 \\
		2 & 36 & 39 & 9 & 2 & 0 & 2  & 4  & 5 & 22 \\
		3 & 243 & 305 & 55 & 16 & 0  & 16 & 37 & 34 & 132\\
		4 & 1185 & 2000 & 304 & 140 & 0 & 97 & 210 & 126 &  750 \\
		5 & 4897 & 13494 & 2107 & 901  & 0  & 486 & 731 & 374 & 4104 \\
		\hline
	\end{tabular}
	\caption{Number of CY geometries with a `diagonal' dP divisor suitable for LVS \cite{Cicoli:2021dhg}.}
	\label{tab_ddPns-GstarM}
\end{table}

\item
Another important class of CY threefolds consists of those which are $K3$-fibered and exhibit some peculiar properties leading to interesting phenomenological applications, including the well-known Fibre inflation models \cite{Cicoli:2008gp,Cicoli:2011it, Cicoli:2016xae, Cicoli:2017axo,Cicoli:2018tcq,Cicoli:2021dhg,Shukla:2022dhz,AbdusSalam:2022krp}. It is useful to mention a theorem of \cite{OGUISO:1993, Schulz:2004tt} which states that if the CY intersection polynomial is linear in the homology class $\hat{D}$ corresponding to a divisor $D$, then the CY threefold has the structure of a K3 or a ${\mathbb T}^4$ fibration over a ${\mathbb P}^1$ base, i.e. the self-cubic and all the self-quadrics vanish:
\bea
& & \int_{X_3} \hat{D} \wedge \hat{D} \wedge \hat{D} = 0, \qquad \int_{X_3} \hat{D} \wedge \hat{D} \wedge \hat{\cal{D}} = 0, \quad \qquad \quad \forall \, \hat{\cal D}.
\eea
These conditions are quite restrictive for the volume form, specially when one also has the requirement of a ddP for LVS. For example it leads to the following volume forms:
\bea
& h^{1,1} = 2, & I_3 = k_{122} \, D_1\, D_2^2, \quad \Longrightarrow \quad {\cal V} \propto \tau_2 \sqrt{\tau_1}, \\
& h^{1,1} = 3 \, \, {\rm with \, \, a \,\, ddP}, & I_3 = k_{122} \, D_1\, D_2^2 + k_{sss} \, D_s^3, \quad \Longrightarrow \quad {\cal V} \propto \tau_2 \sqrt{\tau_1}- \kappa_s \, \tau_s^{3/2}, \nonumber
\eea
where the divisor $D_1$ appearing linearly in the intersection polynomial $I_3$ is a K3 (and therefore its self-cubics and self-quadratic pieces are trivial), while $k_{122}$ denotes the corresponding triple intersection number. This structure in the volume form has been found to be important from the phenomenological point of view.

\item
The string-loop corrections obtained through toroidal computations \cite{Berg:2005ja,Berg:2007wt} have been used to extrapolate the main insights for the CY-orientifold-based models resulting in two classes of string-loop effects (known as KK-type and Winding-type) induced for some particular brane settings \cite{Berg:2007wt,Cicoli:2007xp}. For example, KK-type corrections need the presence of non-intersecting stacks of $D7/O7$ and $O3$ planes while Winding-type effects need intersecting stacks of $D7/O7$ configurations which intersect at some non-contractible two-cycles. However some of such requirements have been recently revisited in \cite{Gao:2022uop} arguing that Winding-type corrections can appear more generically than what is expected from the original proposal of \cite{Berg:2005ja,Berg:2007wt}, something which has been anticipated from a field theoretic argument earlier in \cite{vonGersdorff:2005bf}. 

For having a global model building of another class of string-loop effect, the so-called log-loop-corrections proposed in \cite{Antoniadis:2018hqy,Antoniadis:2019rkh}, one needs three stacks of $D7$-branes with non-trivial intersection loci as proposed in \cite{Leontaris:2022rzj}. 

\item
Let us recall that there are higher derivative $F^4$-corrections to the 4D scalar potential, which are beyond the two-derivative contributions induced via the K\"ahler- and the super-potentials \cite{Ciupke:2015msa}. Such corrections contribute directly to the scalar potential through a topological quantity defined in terms of the second Chern-class $c_2(X)$ of the CY threefold ($X$): $\Pi(D) = \int_{X} c_2(X) \wedge \hat{D}$ whre $\hat{D}$ denotes the dual $(1,1)$ class corresponding to the divisor $D$ of the CY threefold. A classification of such topologies can be found in \cite{Cicoli:2023abc}.

\item
Realising (an exponentially) small value of the so-called Gukov-Vafa-Witten flux superpotential $|W_0|$ in natural way has been always desired. This is because of the fact that low $|W_0|$ gets correlated with the VEV of the CY volume moduli, and hence becomes important to the issue of control over a series of perturbative and non-perturbative effects. One such recipe has been recently proposed in \cite{Demirtas:2019sip} leading to the so-called Perturbatively Flat Flux Vacua (PFFVs). Several follow ups like \cite{Carta:2021kpk,Carta:2022oex} have suggested that the models based on the K3-fibred CY threefolds have significantly large number of (physical) PFFVs as compared to those which are based on non-K3-fibred CY threefolds.

\item
Another issue which can be addressed only after taking the global model building aspects into account is related to satisfying the so-called tadpole cancellation conditions. In the lights of anti-D3-brane uplifting, the recent demand of constructing models with large tadpole charge \cite{Crino:2020qwk, AbdusSalam:2022krp} and the subsequent challenges \cite{Junghans:2022exo,Gao:2022fdi} have been interesting topics to be explored.

\end{enumerate}

\subsection{A concrete global model for perturbative LVS}
\label{sec_global-model}
In this section we present a concrete CY orienitifold model which realises (most of) the requirements (if not all) for stabilising K\"ahler moduli through log-loop effects  \cite{Antoniadis:2018hqy,Antoniadis:2018ngr,Antoniadis:2019doc,Antoniadis:2019rkh,Antoniadis:2020ryh,Antoniadis:2020stf, Antoniadis:2021lhi}. For this purpose, we consider the CY threefold corresponding to the polytope Id: 249 in the CY database of \cite{Altman:2014bfa} defined by the following Toric data:
\begin{center}
\begin{tabular}{|c|ccccccc|}
\hline
Hyp & $x_1$  & $x_2$  & $x_3$  & $x_4$  & $x_5$ & $x_6$  & $x_7$       \\
\hline
4 & 0  & 0 & 1 & 1 & 0 & 0  & 2   \\
4 & 0  & 1 & 0 & 0 & 1 & 0  & 2   \\
4 & 1  & 0 & 0 & 0 & 0 & 1  & 2   \\   
\hline
Div. topology& $K3$  & $K3$ & $K3$ &  $K3$ & $K3$ & $K3$  &   \\
\hline
\end{tabular}
\end{center}
The Hodge numbers are $(h^{2,1}, h^{1,1}) = (115, 3)$, the Euler number is $\chi=-224$ and the SR ideal is:
\be
{\rm SR} =  \{x_1 x_6, \, x_2 x_5, \, x_3 x_4 x_7 \} \,. \nn
\ee
Considering the basis of smooth divisors $\{D_1, D_2, D_3\}$ some relevant quantities such as intersection polynomial, the $c_2(X)$ and the K\"ahler cone are given as,
\bea
& & I_3 = 2\, D_1\, D_2\, D_3, \quad \Longrightarrow \quad {\cal V} = 2 \, t^1\, t^2\, t^3 = \frac{1}{\sqrt{2}}\,\sqrt{\tau_1 \, \tau_2\, \tau_3},  \\
& & c_2(CY) = 5 D_3^2+12 D_1 D_2 + 12 D_2 D_3+12 D_1 D_3, \nonumber\\
& & {\rm KC:} \quad \left\{ \, t^1 > 0\,, \quad t^2 > 0\,, \quad t^3 > 0 \, \right\}\,, \nonumber
\eea
where $\tau_1 = 2\, t^2 t^3,\, \tau_2 = 2\, t^1 t^3,\, \tau_3 = 2 \,t^1 t^2$. This shows that there is a toroidal-like exchange symmetry $1 \leftrightarrow 2 \leftrightarrow 3$ under which all the three $K3$ divisors which are part of the basis are exchanged, and the same happens with the other quantiities such as $I_3, {\cal V}, \tau_i$'s and KC as well. In addition, note that the volume ${\cal V}$ can also be expressed as:
\bea
{\cal V} = t^1 \, \tau_1 = t^2 \, \tau_2  = t^3 \, \tau_3,
\eea
which like the toroidal case translates into the fact that the transverse distance for the stacks of $D7$-branes wrapping the divisor $D_1$ is given by $t^1$ and similarly $t^2$ is the transverse distance for $D7$-branes wrapping the divisor $D_2$ and so on. Further, considering the involution $x_7 \to - x_7$ results in the only fixed point set being given as $\{O7 = D_7\}$ without any $O3$-planes. So one can satisfy the $D3/D7$ tadpole cancellation conditions by having the brane setting involving 3 stacks of $D7$-branes wrapping each of the three divisors $\{D_1, D_2, D_3\}$ present in the basis such that,
\bea
& {\rm D7\, \, tadpole:} & \quad 8\, [O_7] = 8 \left([D_1] + [D_1^\prime] \right) + 8 \left([D_2] + [D_2^\prime] \right) + 8 \left([D_3] + [D_3^\prime] \right)\,, \\
& {\rm D3\, \, tadpole:} & \quad N_{\rm D3} + \frac{N_{\rm flux}}{2} + N_{\rm gauge} = 0 + \frac{240}{12} + 8 + 8 + 8 = 44\,, \nonumber
\eea
which shows some flexibility with turning on background flux as well as the gauge flux. Moreover the divisor intersection analysis shows that all the three $D7$-brane stacks intersect at ${\mathbb T}^2$ while each of those intersect the $O7$-plane on a curve with $h^{0,0} = 1$ and $h^{1,0} = 9$. 

Noting the fact that there are no ddP or Wilson divisors present, this setup does not receive non-perturbative superpotential contributions from E3-instanton, gaugino condensation or poly-instanton effects- making this CY to be not suitable for the K\"ahler moduli stabilisation using the conventional schemes \footnote{However after including all the known/possible perturbative effects arising from the $\alpha^\prime$ as well as $g_s$ series, one can indeed fix all the volume moduli. While we will present some relevant features/insights of this model, we refer the reader to \cite{Leontaris:2022rzj} for more details.}. Further one finds that there are no non-intersecting $D7$-brane stacks and the $O7$-planes along with no $O3$-planes present, and so this model does not induce the KK-type string-loop corrections to the K\"ahler potential. However, as the $D7$-brane stacks intersect on non-shrinkable two-torus, there are string-loop effects of the winding-type along with the non-trivial higher derivative $F^4$-corrections. Subsequently, the leading order pieces in the scalar potential can be collected as below,
\bea
\label{eq:Vfinal-simp}
& & \hskip-1cm V \simeq {\cal C}_1 \left(\frac{\hat\xi - 4\,\hat\eta + 2\,\hat\eta \, \ln{\cal V}}{{\cal V}^3}\right) + 6 \, {\cal C}_1 \left(\frac{3 \hat{\eta } \hat{\xi }+4 \hat{\eta }^2+\hat{\xi }^2-2 \hat{\eta } \hat{\xi} \ln{\cal V} -2 \hat{\eta }^2 \, \ln{\cal V}}{{\cal V}^4}\right) \\
& & + \frac{{\cal C}_2}{{\cal V}^3} \left(\frac{1}{t^1} + \frac{1}{t^2}+\frac{1}{t^3} \right) +  \frac{{\cal C}_3\,\left(t^1 + t^2 + t^3\right)}{{\cal V}^4} + {\cal O}({\cal V}^{-5}) + \dots \,, \nonumber
\eea
where
\bea
\label{eq:calCis}
& & \hskip-1cm {\cal C}_1 = \left(\frac{g_s}{8\pi}\right)\frac{3\,|W_0|^2}{4}, \qquad {\cal C}_2 =  \left(\frac{g_s}{8 \pi}\right)|W_0|^2, \qquad {\cal C}_3 = - \left(\frac{g_s}{8 \pi}\right)^2 \frac{\lambda\,|W_0|^4}{g_s^{3/2}}~\cdot
\eea
Now, using $\chi(X) = -224$ and the following useful relations
\bea
& & \hskip-1cm \hat\xi %= - \frac{\chi(CY)\, \zeta[3]}{2(2\pi)^3\, g_s^{3/2}}
= \frac{14\, \zeta[3]}{\pi^3\, g_s^{3/2}}, \qquad \hat\eta %= \sqrt{g_s}\, \frac{\chi(CY)\, \zeta[2]}{2(2\pi)^3} 
= - \frac{14 \sqrt{g_s} \zeta[2]}{\pi^3}, \qquad \frac{\hat\xi}{\hat\eta} = -\frac{\zeta[3]}{\zeta[2]\, g_s^2}\,,
\eea
it follows that the leading order (first) piece of scalar potential (\ref{eq:Vfinal-simp}) gives a large volume AdS minimum described as:
\bea
\label{eq:pert-LVS}
& & \langle {\cal V} \rangle \simeq e^{a/g_s^2 + b}, \qquad a = \frac{\zeta[3]}{2 \zeta[2]} \simeq 0.365381, \quad b = \frac73~,
\eea
while the remaining moduli are fixed by the sub-leading winding and $F^4$-corrections. For having a numerical estimate, we present the following AdS minimum,
\bea
& & \chi(X) = -224, \quad e^{K_{cs}} =1, \quad W_0 = 1, \quad C_1^W = C_2^W = C_3^W = -1, \quad \lambda = -0.01, \\
& & g_s = 0.20,  \quad \hat\xi= 6.06818, \quad \hat\eta = -0.332156, \quad \langle {\cal V} \rangle_{\rm aprrox.} = 95594.5, \nonumber\\
& & \langle t^i \rangle = 38.553, \quad \langle {\cal V} \rangle = 114605.0, \quad \langle V \rangle = - 9.29 \cdot 10^{-19}, \nonumber
\eea
where $\langle {\cal V} \rangle_{\rm aprrox.}$ corresponds to the approximated VEV obtained from (\ref{eq:pert-LVS}) in the large volume limit while $\langle {\cal V} \rangle$ denotes the value obtained via  the 3-field numerical analysis.

\subsection{On de-Sitter uplifting}
Given that the CY orientifold at hand does not have any $O3$-planes, the anti-$D3$ uplifting proposal of \cite{Crino:2020qwk,AbdusSalam:2022krp} does not apply to our case. However, having three stacks of $D7$-branes intersecting at three ${\mathbb T}^2$'s, there can be various other ways of inducing uplifting term which can result in dS solution. 

For example, assuming that the matter fields receive vanishing VEVs along with each one of the $D7$-brane stack being appropriately magnetised by suitable gauge fluxes to generate a moduli-dependent Fayet-Iliopoulos term leads to the following $D$-term contributions,
\bea
& & V_D \propto \sum_{i =1}^3 \left[\frac{1}{\tau_i} \left(\sum_{j \neq i}\, q_{ij}\, \frac{\partial K}{\partial \tau_j}\right)^2 \right] \simeq \sum_{i =1}^3 \frac{d_i}{f_i^{(3)}},
\eea
where $q_{ij}$ are some model dependent charges, and $f_i^{(3)}$ is some homogeneous cubic polynomial in generic 4-cycle volume $\tau_i$. Using the underlying exchange symmetry $t^1 \leftrightarrow t^2 \leftrightarrow t^3$ of the $F$-term scalar potential (\ref{eq:Vfinal-simp}), one can take $d_\alpha$ as $d_1 = d_2 = d_3 \equiv d$ which self consistently leads to $\tau_1 = \tau_2 = \tau_3 \equiv \tau$, i.e. $f_\alpha^{(3)} \sim \tau^3$, and hence can facilitate an uplifting of the AdS vacua à la ~\cite{Antoniadis:2018hqy}. One such tachyon-free dS solution is given as below:
\bea
& & g_s = 0.2, \quad \hat\xi = 6.06818, \quad \hat\eta = -0.332156, \quad d = 1.24711\cdot 10^{-8}, \, \nonumber\\
& & \langle t^\alpha \rangle = 48.0191 \, \quad \forall \alpha, \quad \langle {\cal V} \rangle = 221447.96, \quad \langle V \rangle = 1.54074\cdot10^{-31}, \nonumber\\
& & {\rm Eigen}(V_{ij}) = \{5.04286\cdot 10^{-22},\, 5.04286\cdot 10^{-22}, \, 5.04286\cdot 10^{-22}\}.
\eea

Alternatively, one can use the so-called $T$-brane uplifting mechanism in the presence of non-zero gauge flux on the hidden sector D7-branes via inducing a non-vanishing Fayet-Iliopoulos term \cite{Cicoli:2015ylx,Cicoli:2017shd,Cicoli:2021dhg} such that $|\varphi| \simeq {\cal V}^{-1/3}$ leading to a positive semidefinite term that can uplift the AdS,
\bea
\label{eq:T-brane-pot}
& & \hskip-1cm V_{\rm up} \simeq \left(\frac{g_s}{8 \pi}\right)\,\frac{{\cal C}_{up}\, |W_0|^2}{{\cal V}^{8/3}} \geq 0,
\eea
where  ${\cal C}_{up}$ denotes a model dependent coefficient. The numerical estimates for one such stable dS solution is given below:
\bea
& & g_s = 0.3, \quad \hat\xi = 3.3031, \quad \hat\eta = -0.406806, \quad {\cal C}_{up} = 0.0814039, \, \nonumber\\
& & \langle t^\alpha \rangle = 19.5862 \, \quad \forall \alpha, \quad \langle {\cal V} \rangle = 15027.3, \quad \langle V \rangle = 3.74709\cdot10^{-30} \nonumber\\
& & {\rm Eigen}(V_{ij}) = \{6.81793\cdot 10^{-18},\, 4.68145\cdot 10^{-19}, \, 4.68145\cdot 10^{-19}\}.
\eea

%%%%%%%%%%SSS%%%%%%%%%%%%%SSS%%%%%%%%%%%%%%SSS%%%%%%%%%%%%%%
%%%%%%%%%%SSS%%%%%%%%%%%%%SSS%%%%%%%%%%%%%%SSS%%%%%%%%%%%%%%

\section{Summary and conclusions}
\label{sec_conclusions}

In this work, we have presented a concise review on the various (conventional) K\"ahler moduli stabilisation schemes proposed in the literature, along with the possibility of realising de-Sitter vacua in the context of type IIB superstring compactifications using the orientifolds of Calabi-Yau threefolds. However, given the vast nature of the subject it is understood that the presented review covers a subset of topics with some (limited) model building details, and is certainly not an exhaustive collection of all the interesting proposals known in the literature.

First we have presented some basic relevant details about the 4D type IIB effective supergravity arising from a generic CY orientifold compactification, and the interconnected origin of various possible (closed-string) moduli, such as complex-structure modulu, axio-dilaton and the K\"ahler moduli, which are counted by the topological invariant such as hodge numbers of the internal manifold. This motivates the tight correlation of string model building with the study of the algebraic geometry and topology related to the compactifying threefolds. Given that these moduli appear in the 4D masses and couplings it is necessary to stabilise those in some stable vacuum having VEVs consistent with the model building requirements such as overall control of the effective theory against the various (un-)known corrections. The various attempts made in this program of moduli stabilisation and subsequently finding dS solutions have been presented in the following four steps:

\begin{enumerate}

\item
In the first step, we discuss the stabilisation of the axio-dilaton and complex-structure moduli via introduction of the conventional $S$-dual pair of the NS-NS and RR fluxes, namely the $H_3$ and $F_3$ fluxes, which can induce superpotential terms depending on these moduli and hence generate a dynamical VEV for each of these moduli. However in 4D type IIB effective theory, the K\"ahler moduli are protected by the so-called no-scale structure, and turning-on these fluxes does not induce scalar potential contributions for the K\"ahler moduli.

\item
In the second step, we discuss the possible corrections which can break this no-scale structure via generating some K\"ahler moduli dependent scalar potential terms, which can subsequently facilitate fixing these moduli. We present such corrections in two classes: 

\begin{itemize}

\item
{\underline{Non-pertubative effects}:} In this class we summarise the possible non-perturbative effects arising from the E3-instantons and gaugino condensation effects generated via suitable (rigid) 4-cycles wrapping E3-brane and D7-brane stacks in appropriate setting. Another class of non-perturbative effects known as poly-instanton corrections is also discussed with the necessary superpotential ingredients.

\item
{\underline{Perturbative effects}:} In this class we summarise a number of interesting corrections such as the so-called BBHL's ${\alpha^\prime}^3$-correction which introduces a shift in the overall volume of the CY threefold appearing in the expression of the K\"ahler potential. In addition, we discuss the possible higher derivative $F^4$-corrections which can be induced to contribute in the scalar potential beyond the two-derivative effects arising from the K\"ahler- and the super-potentials.

Moreover, apart from the $\alpha^\prime$-corrections, we summarise three classes of perturbative string-loop effects: (i).~``KK-type", (ii).~``winding-type", and (iii).~``logarithmic-type". These corrections have received great amount of interest in recent years, especially towards model building with moduli stabilisation, dS realisation, and inflationary embeddings.

\end{itemize}

\item
In the third step, we discuss how the various possible corrections presented in step-2 can be utilised for K\"ahler moduli stabilisation, leading to several popular schemes which we further classify in two categories:

\begin{itemize}

\item
{\underline{Models with non-perturbative effects:}} This class of schemes utilises the non-perturbative superpotential contributions, such as KKLT, Racetrack and LVS. 

\item
{\underline{Models without non-perturbative effects:}} In this class we discuss two possible (K\"ahler) moduli schemes which do not utilise any non-perturbative effects:

(i).~The first one corresponds to the non-geometric models in which the superpotential couplings are introduced for all moduli, including the K\"ahler moduli, even at the tree level, which facilitates the possibility of fixing all moduli via tree-level effects. However such constructions have other unresolved issues such as overall control of the EFT, lack of knowledge about a complete set of Bianchi identities for the so-called cohomology formulation of the generalised fluxes etc. 

(ii).~The second scenarios corresponds to what we call as ``perturbative LVS". This is very similar to the standard LARGE volume scenario which realises an exponentially large VEV for the overall volume of the CY threefold via a competition between the BBHL's $\alpha^\prime$-correction in the K\"ahler potential and a non-perturbative correction in the superpotential. However this perturbative LVS uses string-loop effects of ``logarithmic-type" instead of using any non-perturbative correction in the superpotential. Nevertheless the overall volume still receives an exponentially large VEV given as:  ${\cal V} \simeq e^{a\, g_s^{-2} + b}$ where $a = 0.37$ and $b= 7/3$, and results in a stable AdS minimum.

\end{itemize}

\item

In the final step, we present the arguments about the need of global embedding for the pheno-inspired models which is a very crucial ingredient of string model building, given that one simply assumes many things while building the models based on phenomenological interests, and it is very important to ensure that those assumptions can be consistently realised in some concrete and explicit CY orientifold setting. We illustrate these arguments for several corrections which are guaranteed only when certain topological constraints are satisfied in a given model. In addition, we present a particular CY orientifold with specific ingredients such as a volume form similar to the toroidal case, brane setting with 3 stacks of D7-branes wrapping K3-surfaces, and each stacks intersecting on 2-tori etc. It turns out that these ingredients are the necessary requirement of generating the log-loop effects a la~\cite{Antoniadis:2019rkh}. In this context it is important to mention that this type IIB model based on the concrete CY orientifold at hand does not receive non-perturbative effects due to the absence of (diagonal) del-Pezzo divisors and the Wilson divisors. In addition, the chosen brane-setting does not generically allow the KK-type string-loop effects, nevertheless one can fix all the K\"ahler moduli using the BBHL and log-loop effects, purely perturbative in nature. Subsequently, it turns out that the resulting AdS in perturbative LVS can be uplifted to dS solution by using $D$-terms effects or $T$-brane uplifting prescriptions which are well-known in the literature.

\end{enumerate}

\noindent
For each of these four steps we have devoted separate sections to present the current state-of-the-art in a concise manner. This includes presenting a very simple and compact formulation of the F-term scalar potential taking into account the well-known BBHL's $\alpha'^{3}$-corrections and the one-loop logarithmic corrections, and subsequently incorporating a volume-dependent uplifting term in the effective potential. Using the full package of various possible corrections, it has been shown that metastable de-Sitter vacua can be achieved, with all the (K\"ahler) moduli being stabilised, except for some axions arising from the four-form potential $C_4$. The adopted analytic description in this work is instrumental towards a direct comparison among the various suggested schemes of moduli stabilisation such as KKLT, Racetrack, LVS, perturbative LVS or any other combinations of these where multiple ingredients across these schemes are adopted, e.g. see \cite{Basiouris:2020jgp,Basiouris:2021sdf}. The possible extensions of the log-loop models towards inflationary purpose have been an interesting direction to explore, and in this regard some initiatives are already being taken \cite{Antoniadis:2018ngr,Antoniadis:2019doc,Antoniadis:2020stf,Antoniadis:2021lhi,Antoniadis:2022owm,Ahmed:2022vlc,Ahmed:2022dhc}. However, for the current work, it is extremely difficult to extend the detailed discussion to a subject as diverse and vast as inflation, and for the time being we leave the interested readers with a couple of nice reviews \cite{Baumann:2014nda, Cicoli:2023opf}.

%%%%%%%%%%SSS%%%%%%%%%%%%%SSS%%%%%%%%%%%%%%SSS%%%%%%%%%%%%%%
%%%%%%%%%%SSS%%%%%%%%%%%%%SSS%%%%%%%%%%%%%%SSS%%%%%%%%%%%%%%
\vskip0.3cm
\noindent 
{\bf Acknowledgements:} The work of GKL is supported by the {\it Hellenic Foundation for
	Research and Innovation (H.F.R.I.)} under the {\it ``First Call for
	H.F.R.I. Research Projects to support Faculty members and
	Researchers and the procurement of high-cost research equipment
	grant'' (Project Number: 2251)''. 
	PS would like to thank the Department of Science and Technology (DST), India for the kind support.\\
The authors would like to thank their collaborators for a series of interesting works, and the useful discussions on the topics covered in this review. In this regard, it is pleasure to thank Shehu AbdusSalam, Steven Abel, Waqas Ahmed, Ignatios Antoniadis, Vasileios Basiouris, Ralph Blumenhagen, Federico Carta, Yifan Chen, Michele Cicoli, David Ciupke, Chiara Crinò, Victor A. Diaz, Inaki Garc{\'i}a Etxebarria, Xin Gao, Veronica Guidetti, Daniela Herschmann, Athanasios Karozas, Osmin Lacombe, Tianjun Li, Fernando Marchesano, Christoph Mayrhofer, Alessandro Mininno, Francesco Muia, David Prieto, Fernando Quevedo, Joan Quirant, Thorsten Rahn, Andreas Schachner, Rui Sun, Roberto Valandro, and Umer Zubair.
	}

%\newpage 

%\printbibliography
%\newpage
\bibliographystyle{JHEP}
\bibliography{reference}

\providecommand{\href}[2]{#2}\begingroup\raggedright\begin{thebibliography}{100}

\bibitem{Maldacena:2000mw}
J.~M. Maldacena and C.~Nunez, \emph{{Supergravity description of field theories
  on curved manifolds and a no go theorem}},
  \href{https://doi.org/10.1142/S0217751X01003937}{\emph{Int. J. Mod. Phys. A}
  {\bfseries 16} (2001) 822--855},
  [\href{https://arxiv.org/abs/hep-th/0007018}{{\ttfamily hep-th/0007018}}].

\bibitem{Hertzberg:2007wc}
M.~P. Hertzberg, S.~Kachru, W.~Taylor and M.~Tegmark, \emph{{Inflationary
  Constraints on Type IIA String Theory}},
  \href{https://doi.org/10.1088/1126-6708/2007/12/095}{\emph{JHEP} {\bfseries
  12} (2007) 095}, [\href{https://arxiv.org/abs/0711.2512}{{\ttfamily
  0711.2512}}].

\bibitem{Hertzberg:2007ke}
M.~P. Hertzberg, M.~Tegmark, S.~Kachru, J.~Shelton and O.~Ozcan,
  \emph{{Searching for Inflation in Simple String Theory Models: An
  Astrophysical Perspective}},
  \href{https://doi.org/10.1103/PhysRevD.76.103521}{\emph{Phys. Rev.}
  {\bfseries D76} (2007) 103521},
  [\href{https://arxiv.org/abs/0709.0002}{{\ttfamily 0709.0002}}].

\bibitem{Haque:2008jz}
S.~S. Haque, G.~Shiu, B.~Underwood and T.~Van~Riet, \emph{{Minimal simple de
  Sitter solutions}},
  \href{https://doi.org/10.1103/PhysRevD.79.086005}{\emph{Phys. Rev.}
  {\bfseries D79} (2009) 086005},
  [\href{https://arxiv.org/abs/0810.5328}{{\ttfamily 0810.5328}}].

\bibitem{Flauger:2008ad}
R.~Flauger, S.~Paban, D.~Robbins and T.~Wrase, \emph{{Searching for slow-roll
  moduli inflation in massive type IIA supergravity with metric fluxes}},
  \href{https://doi.org/10.1103/PhysRevD.79.086011}{\emph{Phys. Rev.}
  {\bfseries D79} (2009) 086011},
  [\href{https://arxiv.org/abs/0812.3886}{{\ttfamily 0812.3886}}].

\bibitem{Caviezel:2008tf}
C.~Caviezel, P.~Koerber, S.~Kors, D.~Lust, T.~Wrase and M.~Zagermann, \emph{{On
  the Cosmology of Type IIA Compactifications on SU(3)-structure Manifolds}},
  \href{https://doi.org/10.1088/1126-6708/2009/04/010}{\emph{JHEP} {\bfseries
  04} (2009) 010}, [\href{https://arxiv.org/abs/0812.3551}{{\ttfamily
  0812.3551}}].

\bibitem{Covi:2008ea}
L.~Covi, M.~Gomez-Reino, C.~Gross, J.~Louis, G.~A. Palma and C.~A. Scrucca,
  \emph{{de Sitter vacua in no-scale supergravities and Calabi-Yau string
  models}}, \href{https://doi.org/10.1088/1126-6708/2008/06/057}{\emph{JHEP}
  {\bfseries 06} (2008) 057},
  [\href{https://arxiv.org/abs/0804.1073}{{\ttfamily 0804.1073}}].

\bibitem{deCarlos:2009fq}
B.~de~Carlos, A.~Guarino and J.~M. Moreno, \emph{{Flux moduli stabilisation,
  Supergravity algebras and no-go theorems}},
  \href{https://doi.org/10.1007/JHEP01(2010)012}{\emph{JHEP} {\bfseries 01}
  (2010) 012}, [\href{https://arxiv.org/abs/0907.5580}{{\ttfamily 0907.5580}}].

\bibitem{Caviezel:2009tu}
C.~Caviezel, T.~Wrase and M.~Zagermann, \emph{{Moduli Stabilization and
  Cosmology of Type IIB on SU(2)-Structure Orientifolds}},
  \href{https://doi.org/10.1007/JHEP04(2010)011}{\emph{JHEP} {\bfseries 04}
  (2010) 011}, [\href{https://arxiv.org/abs/0912.3287}{{\ttfamily 0912.3287}}].

\bibitem{Danielsson:2009ff}
U.~H. Danielsson, S.~S. Haque, G.~Shiu and T.~Van~Riet, \emph{{Towards
  Classical de Sitter Solutions in String Theory}},
  \href{https://doi.org/10.1088/1126-6708/2009/09/114}{\emph{JHEP} {\bfseries
  09} (2009) 114}, [\href{https://arxiv.org/abs/0907.2041}{{\ttfamily
  0907.2041}}].

\bibitem{Danielsson:2010bc}
U.~H. Danielsson, P.~Koerber and T.~Van~Riet, \emph{{Universal de Sitter
  solutions at tree-level}},
  \href{https://doi.org/10.1007/JHEP05(2010)090}{\emph{JHEP} {\bfseries 05}
  (2010) 090}, [\href{https://arxiv.org/abs/1003.3590}{{\ttfamily 1003.3590}}].

\bibitem{Wrase:2010ew}
T.~Wrase and M.~Zagermann, \emph{{On Classical de Sitter Vacua in String
  Theory}}, \href{https://doi.org/10.1002/prop.201000053}{\emph{Fortsch. Phys.}
  {\bfseries 58} (2010) 906--910},
  [\href{https://arxiv.org/abs/1003.0029}{{\ttfamily 1003.0029}}].

\bibitem{Shiu:2011zt}
G.~Shiu and Y.~Sumitomo, \emph{{Stability Constraints on Classical de Sitter
  Vacua}}, \href{https://doi.org/10.1007/JHEP09(2011)052}{\emph{JHEP}
  {\bfseries 09} (2011) 052},
  [\href{https://arxiv.org/abs/1107.2925}{{\ttfamily 1107.2925}}].

\bibitem{McOrist:2012yc}
J.~McOrist and S.~Sethi, \emph{{M-theory and Type IIA Flux Compactifications}},
  \href{https://doi.org/10.1007/JHEP12(2012)122}{\emph{JHEP} {\bfseries 12}
  (2012) 122}, [\href{https://arxiv.org/abs/1208.0261}{{\ttfamily 1208.0261}}].

\bibitem{Dasgupta:2014pma}
K.~Dasgupta, R.~Gwyn, E.~McDonough, M.~Mia and R.~Tatar, \emph{{de Sitter Vacua
  in Type IIB String Theory: Classical Solutions and Quantum Corrections}},
  \href{https://doi.org/10.1007/JHEP07(2014)054}{\emph{JHEP} {\bfseries 07}
  (2014) 054}, [\href{https://arxiv.org/abs/1402.5112}{{\ttfamily 1402.5112}}].

\bibitem{Gautason:2015tig}
F.~F. Gautason, M.~Schillo, T.~Van~Riet and M.~Williams, \emph{{Remarks on
  scale separation in flux vacua}},
  \href{https://doi.org/10.1007/JHEP03(2016)061}{\emph{JHEP} {\bfseries 03}
  (2016) 061}, [\href{https://arxiv.org/abs/1512.00457}{{\ttfamily
  1512.00457}}].

\bibitem{Junghans:2016uvg}
D.~Junghans, \emph{{Tachyons in Classical de Sitter Vacua}},
  \href{https://doi.org/10.1007/JHEP06(2016)132}{\emph{JHEP} {\bfseries 06}
  (2016) 132}, [\href{https://arxiv.org/abs/1603.08939}{{\ttfamily
  1603.08939}}].

\bibitem{Andriot:2016xvq}
D.~Andriot and J.~Blåbäck, \emph{{Refining the boundaries of the classical de
  Sitter landscape}}, \href{https://doi.org/10.1007/JHEP03(2017)102,
  10.1007/JHEP03(2018)083}{\emph{JHEP} {\bfseries 03} (2017) 102},
  [\href{https://arxiv.org/abs/1609.00385}{{\ttfamily 1609.00385}}].

\bibitem{Andriot:2017jhf}
D.~Andriot, \emph{{On classical de Sitter and Minkowski solutions with
  intersecting branes}},
  \href{https://doi.org/10.1007/JHEP03(2018)054}{\emph{JHEP} {\bfseries 03}
  (2018) 054}, [\href{https://arxiv.org/abs/1710.08886}{{\ttfamily
  1710.08886}}].

\bibitem{Danielsson:2018ztv}
U.~H. Danielsson and T.~Van~Riet, \emph{{What if string theory has no de Sitter
  vacua?}}, \href{https://doi.org/10.1142/S0218271818300070}{\emph{Int. J. Mod.
  Phys. D} {\bfseries 27} (2018) 1830007},
  [\href{https://arxiv.org/abs/1804.01120}{{\ttfamily 1804.01120}}].

\bibitem{Shukla:2019dqd}
P.~Shukla, \emph{{$T$-dualizing de Sitter no-go scenarios}},
  \href{https://doi.org/10.1103/PhysRevD.102.026014}{\emph{Phys. Rev. D}
  {\bfseries 102} (2020) 026014},
  [\href{https://arxiv.org/abs/1909.08630}{{\ttfamily 1909.08630}}].

\bibitem{Shukla:2019akv}
P.~Shukla, \emph{{Rigid nongeometric orientifolds and the swampland}},
  \href{https://doi.org/10.1103/PhysRevD.103.086010}{\emph{Phys. Rev. D}
  {\bfseries 103} (2021) 086010},
  [\href{https://arxiv.org/abs/1909.10993}{{\ttfamily 1909.10993}}].

\bibitem{Marchesano:2020uqz}
F.~Marchesano, D.~Prieto, J.~Quirant and P.~Shukla, \emph{{Systematics of Type
  IIA moduli stabilisation}},
  \href{https://doi.org/10.1007/JHEP11(2020)113}{\emph{JHEP} {\bfseries 11}
  (2020) 113}, [\href{https://arxiv.org/abs/2007.00672}{{\ttfamily
  2007.00672}}].

\bibitem{Shukla:2022srx}
P.~Shukla, \emph{{On stable type IIA de-Sitter vacua with geometric flux}},
  \href{https://doi.org/10.1140/epjc/s10052-023-11361-w}{\emph{Eur. Phys. J. C}
  {\bfseries 83} (2023) 196},
  [\href{https://arxiv.org/abs/2202.12840}{{\ttfamily 2202.12840}}].

\bibitem{Kachru:2003aw}
S.~Kachru, R.~Kallosh, A.~D. Linde and S.~P. Trivedi, \emph{{De Sitter vacua in
  string theory}},
  \href{https://doi.org/10.1103/PhysRevD.68.046005}{\emph{Phys. Rev. D}
  {\bfseries 68} (2003) 046005},
  [\href{https://arxiv.org/abs/hep-th/0301240}{{\ttfamily hep-th/0301240}}].

\bibitem{Burgess:2003ic}
C.~P. Burgess, R.~Kallosh and F.~Quevedo, \emph{{De Sitter string vacua from
  supersymmetric D terms}},
  \href{https://doi.org/10.1088/1126-6708/2003/10/056}{\emph{JHEP} {\bfseries
  10} (2003) 056}, [\href{https://arxiv.org/abs/hep-th/0309187}{{\ttfamily
  hep-th/0309187}}].

\bibitem{Achucarro:2006zf}
A.~Achucarro, B.~de~Carlos, J.~A. Casas and L.~Doplicher, \emph{{De Sitter
  vacua from uplifting D-terms in effective supergravities from realistic
  strings}}, \href{https://doi.org/10.1088/1126-6708/2006/06/014}{\emph{JHEP}
  {\bfseries 06} (2006) 014},
  [\href{https://arxiv.org/abs/hep-th/0601190}{{\ttfamily hep-th/0601190}}].

\bibitem{Westphal:2006tn}
A.~Westphal, \emph{{de Sitter string vacua from Kahler uplifting}},
  \href{https://doi.org/10.1088/1126-6708/2007/03/102}{\emph{JHEP} {\bfseries
  03} (2007) 102}, [\href{https://arxiv.org/abs/hep-th/0611332}{{\ttfamily
  hep-th/0611332}}].

\bibitem{Silverstein:2007ac}
E.~Silverstein, \emph{{Simple de Sitter Solutions}},
  \href{https://doi.org/10.1103/PhysRevD.77.106006}{\emph{Phys. Rev. D}
  {\bfseries 77} (2008) 106006},
  [\href{https://arxiv.org/abs/0712.1196}{{\ttfamily 0712.1196}}].

\bibitem{Rummel:2011cd}
M.~Rummel and A.~Westphal, \emph{{A sufficient condition for de Sitter vacua in
  type IIB string theory}},
  \href{https://doi.org/10.1007/JHEP01(2012)020}{\emph{JHEP} {\bfseries 01}
  (2012) 020}, [\href{https://arxiv.org/abs/1107.2115}{{\ttfamily 1107.2115}}].

\bibitem{Cicoli:2012fh}
M.~Cicoli, A.~Maharana, F.~Quevedo and C.~P. Burgess, \emph{{De Sitter String
  Vacua from Dilaton-dependent Non-perturbative Effects}},
  \href{https://doi.org/10.1007/JHEP06(2012)011}{\emph{JHEP} {\bfseries 06}
  (2012) 011}, [\href{https://arxiv.org/abs/1203.1750}{{\ttfamily 1203.1750}}].

\bibitem{Louis:2012nb}
J.~Louis, M.~Rummel, R.~Valandro and A.~Westphal, \emph{{Building an explicit
  de Sitter}}, \href{https://doi.org/10.1007/JHEP10(2012)163}{\emph{JHEP}
  {\bfseries 10} (2012) 163},
  [\href{https://arxiv.org/abs/1208.3208}{{\ttfamily 1208.3208}}].

\bibitem{Cicoli:2013cha}
M.~Cicoli, D.~Klevers, S.~Krippendorf, C.~Mayrhofer, F.~Quevedo and
  R.~Valandro, \emph{{Explicit de Sitter Flux Vacua for Global String Models
  with Chiral Matter}},
  \href{https://doi.org/10.1007/JHEP05(2014)001}{\emph{JHEP} {\bfseries 05}
  (2014) 001}, [\href{https://arxiv.org/abs/1312.0014}{{\ttfamily 1312.0014}}].

\bibitem{Cicoli:2015ylx}
M.~Cicoli, F.~Quevedo and R.~Valandro, \emph{{De Sitter from T-branes}},
  \href{https://doi.org/10.1007/JHEP03(2016)141}{\emph{JHEP} {\bfseries 03}
  (2016) 141}, [\href{https://arxiv.org/abs/1512.04558}{{\ttfamily
  1512.04558}}].

\bibitem{Cicoli:2017shd}
M.~Cicoli, I.~n. Garc\`\i{}a-Etxebarria, C.~Mayrhofer, F.~Quevedo, P.~Shukla
  and R.~Valandro, \emph{{Global Orientifolded Quivers with Inflation}},
  \href{https://doi.org/10.1007/JHEP11(2017)134}{\emph{JHEP} {\bfseries 11}
  (2017) 134}, [\href{https://arxiv.org/abs/1706.06128}{{\ttfamily
  1706.06128}}].

\bibitem{Akrami:2018ylq}
Y.~Akrami, R.~Kallosh, A.~Linde and V.~Vardanyan, \emph{{The Landscape, the
  Swampland and the Era of Precision Cosmology}},
  \href{https://doi.org/10.1002/prop.201800075}{\emph{Fortsch. Phys.}
  {\bfseries 67} (2019) 1800075},
  [\href{https://arxiv.org/abs/1808.09440}{{\ttfamily 1808.09440}}].

\bibitem{Antoniadis:2018hqy}
I.~Antoniadis, Y.~Chen and G.~K. Leontaris, \emph{{Perturbative moduli
  stabilisation in type IIB/F-theory framework}},
  \href{https://doi.org/10.1140/epjc/s10052-018-6248-4}{\emph{Eur. Phys. J. C}
  {\bfseries 78} (2018) 766},
  [\href{https://arxiv.org/abs/1803.08941}{{\ttfamily 1803.08941}}].

\bibitem{Antoniadis:2018ngr}
I.~Antoniadis, Y.~Chen and G.~K. Leontaris, \emph{{Inflation from the internal
  volume in type IIB/F-theory compactification}},
  \href{https://doi.org/10.1142/S0217751X19500428}{\emph{Int. J. Mod. Phys.}
  {\bfseries A34} (2019) 1950042},
  [\href{https://arxiv.org/abs/1810.05060}{{\ttfamily 1810.05060}}].

\bibitem{Antoniadis:2019rkh}
I.~Antoniadis, Y.~Chen and G.~K. Leontaris, \emph{{Logarithmic loop
  corrections, moduli stabilisation and de Sitter vacua in string theory}},
  \href{https://doi.org/10.1007/JHEP01(2020)149}{\emph{JHEP} {\bfseries 01}
  (2020) 149}, [\href{https://arxiv.org/abs/1909.10525}{{\ttfamily
  1909.10525}}].

\bibitem{Basiouris:2020jgp}
V.~Basiouris and G.~K. Leontaris, \emph{{Note on de Sitter vacua from
  perturbative and non-perturbative dynamics in type IIB/F-theory
  compactifications}},
  \href{https://doi.org/10.1016/j.physletb.2020.135809}{\emph{Phys. Lett. B}
  {\bfseries 810} (2020) 135809},
  [\href{https://arxiv.org/abs/2007.15423}{{\ttfamily 2007.15423}}].

\bibitem{Antoniadis:2020stf}
I.~Antoniadis, O.~Lacombe and G.~K. Leontaris, \emph{{Inflation near a
  metastable de Sitter vacuum from moduli stabilisation}},
  \href{https://doi.org/10.1140/epjc/s10052-020-08581-9}{\emph{Eur. Phys. J. C}
  {\bfseries 80} (2020) 1014},
  [\href{https://arxiv.org/abs/2007.10362}{{\ttfamily 2007.10362}}].

\bibitem{Cicoli:2018kdo}
M.~Cicoli, S.~De~Alwis, A.~Maharana, F.~Muia and F.~Quevedo, \emph{{De Sitter
  vs Quintessence in String Theory}},
  \href{https://doi.org/10.1002/prop.201800079}{\emph{Fortsch. Phys.}
  {\bfseries 67} (2019) 1800079},
  [\href{https://arxiv.org/abs/1808.08967}{{\ttfamily 1808.08967}}].

\bibitem{Crino:2020qwk}
C.~Crin\`o, F.~Quevedo and R.~Valandro, \emph{{On de Sitter String Vacua from
  Anti-D3-Branes in the Large Volume Scenario}},
  \href{https://doi.org/10.1007/JHEP03(2021)258}{\emph{JHEP} {\bfseries 03}
  (2021) 258}, [\href{https://arxiv.org/abs/2010.15903}{{\ttfamily
  2010.15903}}].

\bibitem{Cicoli:2021dhg}
M.~Cicoli, I.~n.~G. Etxebarria, F.~Quevedo, A.~Schachner, P.~Shukla and
  R.~Valandro, \emph{{The Standard Model quiver in de Sitter string
  compactifications}},
  \href{https://doi.org/10.1007/JHEP08(2021)109}{\emph{JHEP} {\bfseries 08}
  (2021) 109}, [\href{https://arxiv.org/abs/2106.11964}{{\ttfamily
  2106.11964}}].

\bibitem{Heckman:2019dsj}
J.~J. Heckman, C.~Lawrie, L.~Lin, J.~Sakstein and G.~Zoccarato,
  \emph{{Pixelated Dark Energy}},
  \href{https://doi.org/10.1002/prop.201900071}{\emph{Fortsch. Phys.}
  {\bfseries 67} (2019) 1900071},
  [\href{https://arxiv.org/abs/1901.10489}{{\ttfamily 1901.10489}}].

\bibitem{Heckman:2018mxl}
J.~J. Heckman, C.~Lawrie, L.~Lin and G.~Zoccarato, \emph{{F-theory and Dark
  Energy}}, \href{https://doi.org/10.1002/prop.201900057}{\emph{Fortsch. Phys.}
  {\bfseries 67} (2019) 1900057},
  [\href{https://arxiv.org/abs/1811.01959}{{\ttfamily 1811.01959}}].

\bibitem{Andriot:2022way}
D.~Andriot, L.~Horer and P.~Marconnet, \emph{{Charting the landscape of (anti-)
  de Sitter and Minkowski solutions of 10d supergravities}},
  \href{https://arxiv.org/abs/2201.04152}{{\ttfamily 2201.04152}}.

\bibitem{Ooguri:2006in}
H.~Ooguri and C.~Vafa, \emph{{On the Geometry of the String Landscape and the
  Swampland}},
  \href{https://doi.org/10.1016/j.nuclphysb.2006.10.033}{\emph{Nucl. Phys. B}
  {\bfseries 766} (2007) 21--33},
  [\href{https://arxiv.org/abs/hep-th/0605264}{{\ttfamily hep-th/0605264}}].

\bibitem{Obied:2018sgi}
G.~Obied, H.~Ooguri, L.~Spodyneiko and C.~Vafa, \emph{{De Sitter Space and the
  Swampland}},  \href{https://arxiv.org/abs/1806.08362}{{\ttfamily
  1806.08362}}.

\bibitem{Cicoli:2023opf}
M.~Cicoli, J.~P. Conlon, A.~Maharana, S.~Parameswaran, F.~Quevedo and
  I.~Zavala, \emph{{String Cosmology: from the Early Universe to Today}},
  \href{https://arxiv.org/abs/2303.04819}{{\ttfamily 2303.04819}}.

\bibitem{Garg:2018reu}
S.~K. Garg and C.~Krishnan, \emph{{Bounds on Slow Roll and the de Sitter
  Swampland}}, \href{https://doi.org/10.1007/JHEP11(2019)075}{\emph{JHEP}
  {\bfseries 11} (2019) 075},
  [\href{https://arxiv.org/abs/1807.05193}{{\ttfamily 1807.05193}}].

\bibitem{Agrawal:2018own}
P.~Agrawal, G.~Obied, P.~J. Steinhardt and C.~Vafa, \emph{{On the Cosmological
  Implications of the String Swampland}},
  \href{https://doi.org/10.1016/j.physletb.2018.07.040}{\emph{Phys. Lett. B}
  {\bfseries 784} (2018) 271--276},
  [\href{https://arxiv.org/abs/1806.09718}{{\ttfamily 1806.09718}}].

\bibitem{Andriot:2018wzk}
D.~Andriot, \emph{{On the de Sitter swampland criterion}},
  \href{https://doi.org/10.1016/j.physletb.2018.09.022}{\emph{Phys. Lett. B}
  {\bfseries 785} (2018) 570--573},
  [\href{https://arxiv.org/abs/1806.10999}{{\ttfamily 1806.10999}}].

\bibitem{Andriot:2018ept}
D.~Andriot, \emph{{New constraints on classical de Sitter: flirting with the
  swampland}}, \href{https://doi.org/10.1002/prop.201800103}{\emph{Fortsch.
  Phys.} {\bfseries 67} (2019) 1800103},
  [\href{https://arxiv.org/abs/1807.09698}{{\ttfamily 1807.09698}}].

\bibitem{Denef:2018etk}
F.~Denef, A.~Hebecker and T.~Wrase, \emph{{de Sitter swampland conjecture and
  the Higgs potential}},
  \href{https://doi.org/10.1103/PhysRevD.98.086004}{\emph{Phys. Rev. D}
  {\bfseries 98} (2018) 086004},
  [\href{https://arxiv.org/abs/1807.06581}{{\ttfamily 1807.06581}}].

\bibitem{Conlon:2018eyr}
J.~P. Conlon, \emph{{The de Sitter swampland conjecture and supersymmetric AdS
  vacua}}, \href{https://doi.org/10.1142/S0217751X18501786}{\emph{Int. J. Mod.
  Phys. A} {\bfseries 33} (2018) 1850178},
  [\href{https://arxiv.org/abs/1808.05040}{{\ttfamily 1808.05040}}].

\bibitem{Roupec:2018mbn}
C.~Roupec and T.~Wrase, \emph{{de Sitter Extrema and the Swampland}},
  \href{https://doi.org/10.1002/prop.201800082}{\emph{Fortsch. Phys.}
  {\bfseries 67} (2019) 1800082},
  [\href{https://arxiv.org/abs/1807.09538}{{\ttfamily 1807.09538}}].

\bibitem{Murayama:2018lie}
H.~Murayama, M.~Yamazaki and T.~T. Yanagida, \emph{{Do We Live in the
  Swampland?}}, \href{https://doi.org/10.1007/JHEP12(2018)032}{\emph{JHEP}
  {\bfseries 12} (2018) 032},
  [\href{https://arxiv.org/abs/1809.00478}{{\ttfamily 1809.00478}}].

\bibitem{Choi:2018rze}
K.~Choi, D.~Chway and C.~S. Shin, \emph{{The dS swampland conjecture with the
  electroweak symmetry and QCD chiral symmetry breaking}},
  \href{https://doi.org/10.1007/JHEP11(2018)142}{\emph{JHEP} {\bfseries 11}
  (2018) 142}, [\href{https://arxiv.org/abs/1809.01475}{{\ttfamily
  1809.01475}}].

\bibitem{Hamaguchi:2018vtv}
K.~Hamaguchi, M.~Ibe and T.~Moroi, \emph{{The swampland conjecture and the
  Higgs expectation value}},
  \href{https://doi.org/10.1007/JHEP12(2018)023}{\emph{JHEP} {\bfseries 12}
  (2018) 023}, [\href{https://arxiv.org/abs/1810.02095}{{\ttfamily
  1810.02095}}].

\bibitem{Olguin-Tejo:2018pfq}
Y.~Olguin-Trejo, S.~L. Parameswaran, G.~Tasinato and I.~Zavala, \emph{{Runaway
  Quintessence, Out of the Swampland}},
  \href{https://doi.org/10.1088/1475-7516/2019/01/031}{\emph{JCAP} {\bfseries
  1901} (2019) 031}, [\href{https://arxiv.org/abs/1810.08634}{{\ttfamily
  1810.08634}}].

\bibitem{Blanco-Pillado:2018xyn}
J.~J. Blanco-Pillado, M.~A. Urkiola and J.~M. Wachter, \emph{{Racetrack
  Potentials and the de Sitter Swampland Conjectures}},
  \href{https://doi.org/10.1007/JHEP01(2019)187}{\emph{JHEP} {\bfseries 01}
  (2019) 187}, [\href{https://arxiv.org/abs/1811.05463}{{\ttfamily
  1811.05463}}].

\bibitem{Lin:2018kjm}
C.-M. Lin, K.-W. Ng and K.~Cheung, \emph{{Chaotic inflation on the brane and
  the Swampland Criteria}},
  \href{https://doi.org/10.1103/PhysRevD.100.023545}{\emph{Phys. Rev.}
  {\bfseries D100} (2019) 023545},
  [\href{https://arxiv.org/abs/1810.01644}{{\ttfamily 1810.01644}}].

\bibitem{Han:2018yrk}
C.~Han, S.~Pi and M.~Sasaki, \emph{{Quintessence Saves Higgs Instability}},
  \href{https://doi.org/10.1016/j.physletb.2019.02.037}{\emph{Phys. Lett.}
  {\bfseries B791} (2019) 314--318},
  [\href{https://arxiv.org/abs/1809.05507}{{\ttfamily 1809.05507}}].

\bibitem{Raveri:2018ddi}
M.~Raveri, W.~Hu and S.~Sethi, \emph{{Swampland Conjectures and Late-Time
  Cosmology}}, \href{https://doi.org/10.1103/PhysRevD.99.083518}{\emph{Phys.
  Rev.} {\bfseries D99} (2019) 083518},
  [\href{https://arxiv.org/abs/1812.10448}{{\ttfamily 1812.10448}}].

\bibitem{Dasgupta:2018rtp}
K.~Dasgupta, M.~Emelin, E.~McDonough and R.~Tatar, \emph{{Quantum Corrections
  and the de Sitter Swampland Conjecture}},
  \href{https://doi.org/10.1007/JHEP01(2019)145}{\emph{JHEP} {\bfseries 01}
  (2019) 145}, [\href{https://arxiv.org/abs/1808.07498}{{\ttfamily
  1808.07498}}].

\bibitem{Danielsson:2018qpa}
U.~Danielsson, \emph{{The quantum swampland}},
  \href{https://doi.org/10.1007/JHEP04(2019)095}{\emph{JHEP} {\bfseries 04}
  (2019) 095}, [\href{https://arxiv.org/abs/1809.04512}{{\ttfamily
  1809.04512}}].

\bibitem{Andriolo:2018yrz}
S.~Andriolo, G.~Shiu, H.~Triendl, T.~Van~Riet, G.~Venken and G.~Zoccarato,
  \emph{{Compact G2 holonomy spaces from SU(3) structures}},
  \href{https://doi.org/10.1007/JHEP03(2019)059}{\emph{JHEP} {\bfseries 03}
  (2019) 059}, [\href{https://arxiv.org/abs/1811.00063}{{\ttfamily
  1811.00063}}].

\bibitem{Dasgupta:2019gcd}
K.~Dasgupta, M.~Emelin, M.~M. Faruk and R.~Tatar, \emph{{de Sitter Vacua in the
  String Landscape}},  \href{https://arxiv.org/abs/1908.05288}{{\ttfamily
  1908.05288}}.

\bibitem{Andriot:2019wrs}
D.~Andriot, \emph{{Open problems on classical de Sitter solutions}},
  \href{https://doi.org/10.1002/prop.201900026}{\emph{Fortsch. Phys.}
  {\bfseries 67} (2019) 1900026},
  [\href{https://arxiv.org/abs/1902.10093}{{\ttfamily 1902.10093}}].

\bibitem{Palti:2019pca}
E.~Palti, \emph{{The Swampland: Introduction and Review}},
  \href{https://doi.org/10.1002/prop.201900037}{\emph{Fortsch. Phys.}
  {\bfseries 67} (2019) 1900037},
  [\href{https://arxiv.org/abs/1903.06239}{{\ttfamily 1903.06239}}].

\bibitem{Cicoli:2012tz}
M.~Cicoli, F.~G. Pedro and G.~Tasinato, \emph{{Natural Quintessence in String
  Theory}}, \href{https://doi.org/10.1088/1475-7516/2012/07/044}{\emph{JCAP}
  {\bfseries 07} (2012) 044},
  [\href{https://arxiv.org/abs/1203.6655}{{\ttfamily 1203.6655}}].

\bibitem{Junghans:2022exo}
D.~Junghans, \emph{{LVS de Sitter Vacua are probably in the Swampland}},
  \href{https://arxiv.org/abs/2201.03572}{{\ttfamily 2201.03572}}.

\bibitem{SupernovaSearchTeam:1998fmf}
{\scshape Supernova Search Team} collaboration, A.~G. Riess et~al.,
  \emph{{Observational evidence from supernovae for an accelerating universe
  and a cosmological constant}},
  \href{https://doi.org/10.1086/300499}{\emph{Astron. J.} {\bfseries 116}
  (1998) 1009--1038}, [\href{https://arxiv.org/abs/astro-ph/9805201}{{\ttfamily
  astro-ph/9805201}}].

\bibitem{Dimopoulos:2022wzo}
K.~Dimopoulos, \emph{{Introduction to Cosmic Inflation and Dark Energy}}.
\newblock CRC Press, 5, 2022.

\bibitem{Cunillera:2022hdb}
F.~Cunillera, \emph{{Dark energy: EFTs and supergravity}}, Ph.D. thesis,
  Nottingham U., University of Nottingham, Nottingham U., 2022.
\newblock \href{https://arxiv.org/abs/2208.00177}{{\ttfamily 2208.00177}}.

\bibitem{Gonzalo:2021zsp}
E.~Gonzalo, L.~E. Ib\'a\~nez and I.~Valenzuela, \emph{{Swampland constraints on
  neutrino masses}}, \href{https://doi.org/10.1007/JHEP02(2022)088}{\emph{JHEP}
  {\bfseries 02} (2022) 088},
  [\href{https://arxiv.org/abs/2109.10961}{{\ttfamily 2109.10961}}].

\bibitem{Baumann:2014nda}
D.~Baumann and L.~McAllister, \emph{{Inflation and String Theory}}.
\newblock Cambridge Monographs on Mathematical Physics. Cambridge University
  Press, 5, 2015,
  \href{https://doi.org/10.1017/CBO9781316105733}{10.1017/CBO9781316105733}.

\bibitem{Shukla:2022dhz}
P.~Shukla, \emph{{Classifying divisor topologies for string phenomenology}},
  \href{https://doi.org/10.1007/JHEP12(2022)055}{\emph{JHEP} {\bfseries 12}
  (2022) 055}, [\href{https://arxiv.org/abs/2205.05215}{{\ttfamily
  2205.05215}}].

\bibitem{Vafa:2005ui}
C.~Vafa, \emph{{The String landscape and the swampland}},
  \href{https://arxiv.org/abs/hep-th/0509212}{{\ttfamily hep-th/0509212}}.

\bibitem{Vafa:2019evj}
C.~Vafa, \emph{{Cosmic Predictions from the String Swampland}},
  \href{https://doi.org/10.1103/Physics.12.115}{\emph{APS Physics} {\bfseries
  12} (2019) 115}.

\bibitem{vanBeest:2021lhn}
M.~van Beest, J.~Calder\'on-Infante, D.~Mirfendereski and I.~Valenzuela,
  \emph{{Lectures on the Swampland Program in String Compactifications}},
  \href{https://doi.org/10.1016/j.physrep.2022.09.002}{\emph{Phys. Rept.}
  {\bfseries 989} (2022) 1--50},
  [\href{https://arxiv.org/abs/2102.01111}{{\ttfamily 2102.01111}}].

\bibitem{Blumenhagen:2017cxt}
R.~Blumenhagen, I.~Valenzuela and F.~Wolf, \emph{{The Swampland Conjecture and
  F-term Axion Monodromy Inflation}},
  \href{https://doi.org/10.1007/JHEP07(2017)145}{\emph{JHEP} {\bfseries 07}
  (2017) 145}, [\href{https://arxiv.org/abs/1703.05776}{{\ttfamily
  1703.05776}}].

\bibitem{Blumenhagen:2018nts}
R.~Blumenhagen, D.~Kläwer, L.~Schlechter and F.~Wolf, \emph{{The Refined
  Swampland Distance Conjecture in Calabi-Yau Moduli Spaces}},
  \href{https://doi.org/10.1007/JHEP06(2018)052}{\emph{JHEP} {\bfseries 06}
  (2018) 052}, [\href{https://arxiv.org/abs/1803.04989}{{\ttfamily
  1803.04989}}].

\bibitem{Blumenhagen:2018hsh}
R.~Blumenhagen, \emph{{Large Field Inflation/Quintessence and the Refined
  Swampland Distance Conjecture}},
  \href{https://doi.org/10.22323/1.318.0175}{\emph{PoS} {\bfseries CORFU2017}
  (2018) 175}, [\href{https://arxiv.org/abs/1804.10504}{{\ttfamily
  1804.10504}}].

\bibitem{Palti:2017elp}
E.~Palti, \emph{{The Weak Gravity Conjecture and Scalar Fields}},
  \href{https://doi.org/10.1007/JHEP08(2017)034}{\emph{JHEP} {\bfseries 08}
  (2017) 034}, [\href{https://arxiv.org/abs/1705.04328}{{\ttfamily
  1705.04328}}].

\bibitem{Conlon:2016aea}
J.~P. Conlon and S.~Krippendorf, \emph{{Axion decay constants away from the
  lamppost}}, \href{https://doi.org/10.1007/JHEP04(2016)085}{\emph{JHEP}
  {\bfseries 04} (2016) 085},
  [\href{https://arxiv.org/abs/1601.00647}{{\ttfamily 1601.00647}}].

\bibitem{Hebecker:2017lxm}
A.~Hebecker, P.~Henkenjohann and L.~T. Witkowski, \emph{{Flat Monodromies and a
  Moduli Space Size Conjecture}},
  \href{https://doi.org/10.1007/JHEP12(2017)033}{\emph{JHEP} {\bfseries 12}
  (2017) 033}, [\href{https://arxiv.org/abs/1708.06761}{{\ttfamily
  1708.06761}}].

\bibitem{Klaewer:2016kiy}
D.~Klaewer and E.~Palti, \emph{{Super-Planckian Spatial Field Variations and
  Quantum Gravity}}, \href{https://doi.org/10.1007/JHEP01(2017)088}{\emph{JHEP}
  {\bfseries 01} (2017) 088},
  [\href{https://arxiv.org/abs/1610.00010}{{\ttfamily 1610.00010}}].

\bibitem{Baume:2016psm}
F.~Baume and E.~Palti, \emph{{Backreacted Axion Field Ranges in String
  Theory}}, \href{https://doi.org/10.1007/JHEP08(2016)043}{\emph{JHEP}
  {\bfseries 08} (2016) 043},
  [\href{https://arxiv.org/abs/1602.06517}{{\ttfamily 1602.06517}}].

\bibitem{Landete:2018kqf}
A.~Landete and G.~Shiu, \emph{{Mass Hierarchies and Dynamical Field Range}},
  \href{https://doi.org/10.1103/PhysRevD.98.066012}{\emph{Phys. Rev.}
  {\bfseries D98} (2018) 066012},
  [\href{https://arxiv.org/abs/1806.01874}{{\ttfamily 1806.01874}}].

\bibitem{Cicoli:2018tcq}
M.~Cicoli, D.~Ciupke, C.~Mayrhofer and P.~Shukla, \emph{{A Geometrical Upper
  Bound on the Inflaton Range}},
  \href{https://doi.org/10.1007/JHEP05(2018)001}{\emph{JHEP} {\bfseries 05}
  (2018) 001}, [\href{https://arxiv.org/abs/1801.05434}{{\ttfamily
  1801.05434}}].

\bibitem{Font:2019cxq}
A.~Font, A.~Herráez and L.~E. Ibáñez, \emph{{The Swampland Distance
  Conjecture and Towers of Tensionless Branes}},
  \href{https://doi.org/10.1007/JHEP08(2019)044}{\emph{JHEP} {\bfseries 08}
  (2019) 044}, [\href{https://arxiv.org/abs/1904.05379}{{\ttfamily
  1904.05379}}].

\bibitem{Grimm:2018cpv}
T.~W. Grimm, C.~Li and E.~Palti, \emph{{Infinite Distance Networks in Field
  Space and Charge Orbits}},
  \href{https://doi.org/10.1007/JHEP03(2019)016}{\emph{JHEP} {\bfseries 03}
  (2019) 016}, [\href{https://arxiv.org/abs/1811.02571}{{\ttfamily
  1811.02571}}].

\bibitem{Hebecker:2018fln}
A.~Hebecker, D.~Junghans and A.~Schachner, \emph{{Large Field Ranges from
  Aligned and Misaligned Winding}},
  \href{https://doi.org/10.1007/JHEP03(2019)192}{\emph{JHEP} {\bfseries 03}
  (2019) 192}, [\href{https://arxiv.org/abs/1812.05626}{{\ttfamily
  1812.05626}}].

\bibitem{Banlaki:2018ayh}
A.~Banlaki, A.~Chowdhury, C.~Roupec and T.~Wrase, \emph{{Scaling limits of dS
  vacua and the swampland}},
  \href{https://doi.org/10.1007/JHEP03(2019)065}{\emph{JHEP} {\bfseries 03}
  (2019) 065}, [\href{https://arxiv.org/abs/1811.07880}{{\ttfamily
  1811.07880}}].

\bibitem{Junghans:2018gdb}
D.~Junghans, \emph{{Weakly Coupled de Sitter Vacua with Fluxes and the
  Swampland}}, \href{https://doi.org/10.1007/JHEP03(2019)150}{\emph{JHEP}
  {\bfseries 03} (2019) 150},
  [\href{https://arxiv.org/abs/1811.06990}{{\ttfamily 1811.06990}}].

\bibitem{Junghans:2020acz}
D.~Junghans, \emph{{O-Plane Backreaction and Scale Separation in Type IIA Flux
  Vacua}}, \href{https://doi.org/10.1002/prop.202000040}{\emph{Fortsch. Phys.}
  {\bfseries 68} (2020) 2000040},
  [\href{https://arxiv.org/abs/2003.06274}{{\ttfamily 2003.06274}}].

\bibitem{Apers:2022zjx}
F.~Apers, M.~Montero, T.~Van~Riet and T.~Wrase, \emph{{Comments on classical
  AdS flux vacua with scale separation}},
  \href{https://arxiv.org/abs/2202.00682}{{\ttfamily 2202.00682}}.

\bibitem{Bena:2020xrh}
I.~Bena, J.~Bl\r{a}b\"ack, M.~Gra\~na and S.~L\"ust, \emph{{The Tadpole
  Problem}},  \href{https://arxiv.org/abs/2010.10519}{{\ttfamily 2010.10519}}.

\bibitem{Plauschinn:2021hkp}
E.~Plauschinn, \emph{{The tadpole conjecture at large complex-structure}},
  \href{https://arxiv.org/abs/2109.00029}{{\ttfamily 2109.00029}}.

\bibitem{Plauschinn:2020ram}
E.~Plauschinn, \emph{{Moduli Stabilization with Non-Geometric Fluxes
  \textemdash{} Comments on Tadpole Contributions and de-Sitter Vacua}},
  \href{https://doi.org/10.1002/prop.202100003}{\emph{Fortsch. Phys.}
  {\bfseries 69} (2021) 2100003},
  [\href{https://arxiv.org/abs/2011.08227}{{\ttfamily 2011.08227}}].

\bibitem{Marchesano:2021gyv}
F.~Marchesano, D.~Prieto and M.~Wiesner, \emph{{F-theory flux vacua at large
  complex structure}},
  \href{https://doi.org/10.1007/JHEP08(2021)077}{\emph{JHEP} {\bfseries 08}
  (2021) 077}, [\href{https://arxiv.org/abs/2105.09326}{{\ttfamily
  2105.09326}}].

\bibitem{Candelas:1990pi}
P.~Candelas and X.~de~la Ossa, \emph{{Moduli Space of {Calabi-Yau} Manifolds}},
  \href{https://doi.org/10.1016/0550-3213(91)90122-E}{\emph{Nucl. Phys. B}
  {\bfseries 355} (1991) 455--481}.

\bibitem{Greene:1996cy}
B.~R. Greene, \emph{{String theory on Calabi-Yau manifolds}},  in
  \emph{{Theoretical Advanced Study Institute in Elementary Particle Physics
  (TASI 96): Fields, Strings, and Duality}}, pp.~543--726, 6, 1996,
  \href{https://arxiv.org/abs/hep-th/9702155}{{\ttfamily hep-th/9702155}}.

\bibitem{Lichnerowicz1963}
A.~Lichnerowicz, \emph{{Spineurs harmoniques}}, {\emph{C. R. Acad. Sci. Paris
  Sér. A-B} {\bfseries 257} (1963) 7--9}.

\bibitem{Hitchin}
N.~Hitchin, \emph{{Harmonic spinors}},
  \href{https://doi.org/10.1016/0001-8708(74)90021-8}{\emph{Advances in
  Mathematics} {\bfseries 14} (1974) 1--55}.

\bibitem{Green:1986ck}
P.~Green and T.~Hubsch, \emph{{Calabi-yau Manifolds as Complete Intersections
  in Products of Complex Projective Spaces}},
  \href{https://doi.org/10.1007/BF01205673}{\emph{Commun. Math. Phys.}
  {\bfseries 109} (1987) 99}.

\bibitem{Candelas:1987kf}
P.~Candelas, A.~M. Dale, C.~A. Lutken and R.~Schimmrigk, \emph{{Complete
  Intersection Calabi-Yau Manifolds}},
  \href{https://doi.org/10.1016/0550-3213(88)90352-5}{\emph{Nucl. Phys. B}
  {\bfseries 298} (1988) 493}.

\bibitem{Green:1987cr}
P.~S. Green, T.~Hubsch and C.~A. Lutken, \emph{{All Hodge Numbers of All
  Complete Intersection Calabi-Yau Manifolds}},
  \href{https://doi.org/10.1088/0264-9381/6/2/006}{\emph{Class. Quant. Grav.}
  {\bfseries 6} (1989) 105--124}.

\bibitem{Candelas:1993dm}
P.~Candelas, X.~De~La~Ossa, A.~Font, S.~H. Katz and D.~R. Morrison,
  \emph{{Mirror symmetry for two parameter models. 1.}},
  \href{https://doi.org/10.1016/0550-3213(94)90322-0}{\emph{Nucl. Phys. B}
  {\bfseries 416} (1994) 481--538},
  [\href{https://arxiv.org/abs/hep-th/9308083}{{\ttfamily hep-th/9308083}}].

\bibitem{Batyrev:1993oya}
V.~V. Batyrev, \emph{{Dual polyhedra and mirror symmetry for Calabi-Yau
  hypersurfaces in toric varieties}}, {\emph{J. Alg. Geom.} {\bfseries 3}
  (1994) 493--545}, [\href{https://arxiv.org/abs/alg-geom/9310003}{{\ttfamily
  alg-geom/9310003}}].

\bibitem{Candelas:1994hw}
P.~Candelas, A.~Font, S.~H. Katz and D.~R. Morrison, \emph{{Mirror symmetry for
  two parameter models. 2.}},
  \href{https://doi.org/10.1016/0550-3213(94)90155-4}{\emph{Nucl. Phys. B}
  {\bfseries 429} (1994) 626--674},
  [\href{https://arxiv.org/abs/hep-th/9403187}{{\ttfamily hep-th/9403187}}].

\bibitem{Hosono:1994ax}
S.~Hosono, A.~Klemm, S.~Theisen and S.-T. Yau, \emph{{Mirror symmetry, mirror
  map and applications to complete intersection Calabi-Yau spaces}},
  \href{https://doi.org/10.1016/0550-3213(94)00440-P}{\emph{Nucl. Phys. B}
  {\bfseries 433} (1995) 501--554},
  [\href{https://arxiv.org/abs/hep-th/9406055}{{\ttfamily hep-th/9406055}}].

\bibitem{Kreuzer:2000xy}
M.~Kreuzer and H.~Skarke, \emph{{Complete classification of reflexive polyhedra
  in four-dimensions}},
  \href{https://doi.org/10.4310/ATMP.2000.v4.n6.a2}{\emph{Adv. Theor. Math.
  Phys.} {\bfseries 4} (2000) 1209--1230},
  [\href{https://arxiv.org/abs/hep-th/0002240}{{\ttfamily hep-th/0002240}}].

\bibitem{Gray:2013mja}
J.~Gray, A.~S. Haupt and A.~Lukas, \emph{{All Complete Intersection Calabi-Yau
  Four-Folds}}, \href{https://doi.org/10.1007/JHEP07(2013)070}{\emph{JHEP}
  {\bfseries 07} (2013) 070},
  [\href{https://arxiv.org/abs/1303.1832}{{\ttfamily 1303.1832}}].

\bibitem{Anderson:2011ns}
L.~B. Anderson, J.~Gray, A.~Lukas and E.~Palti, \emph{{Two Hundred Heterotic
  Standard Models on Smooth Calabi-Yau Threefolds}},
  \href{https://doi.org/10.1103/PhysRevD.84.106005}{\emph{Phys. Rev. D}
  {\bfseries 84} (2011) 106005},
  [\href{https://arxiv.org/abs/1106.4804}{{\ttfamily 1106.4804}}].

\bibitem{Anderson:2012yf}
L.~B. Anderson, J.~Gray, A.~Lukas and E.~Palti, \emph{{Heterotic Line Bundle
  Standard Models}}, \href{https://doi.org/10.1007/JHEP06(2012)113}{\emph{JHEP}
  {\bfseries 06} (2012) 113},
  [\href{https://arxiv.org/abs/1202.1757}{{\ttfamily 1202.1757}}].

\bibitem{Anderson:2013xka}
L.~B. Anderson, A.~Constantin, J.~Gray, A.~Lukas and E.~Palti, \emph{{A
  Comprehensive Scan for Heterotic SU(5) GUT models}},
  \href{https://doi.org/10.1007/JHEP01(2014)047}{\emph{JHEP} {\bfseries 01}
  (2014) 047}, [\href{https://arxiv.org/abs/1307.4787}{{\ttfamily 1307.4787}}].

\bibitem{Anderson:2010mh}
L.~B. Anderson, J.~Gray, A.~Lukas and B.~Ovrut, \emph{{Stabilizing the Complex
  Structure in Heterotic Calabi-Yau Vacua}},
  \href{https://doi.org/10.1007/JHEP02(2011)088}{\emph{JHEP} {\bfseries 02}
  (2011) 088}, [\href{https://arxiv.org/abs/1010.0255}{{\ttfamily 1010.0255}}].

\bibitem{Anderson:2011cza}
L.~B. Anderson, J.~Gray, A.~Lukas and B.~Ovrut, \emph{{Stabilizing All
  Geometric Moduli in Heterotic Calabi-Yau Vacua}},
  \href{https://doi.org/10.1103/PhysRevD.83.106011}{\emph{Phys. Rev. D}
  {\bfseries 83} (2011) 106011},
  [\href{https://arxiv.org/abs/1102.0011}{{\ttfamily 1102.0011}}].

\bibitem{Bobkov:2010rf}
K.~Bobkov, V.~Braun, P.~Kumar and S.~Raby, \emph{{Stabilizing All Kahler Moduli
  in Type IIB Orientifolds}},
  \href{https://doi.org/10.1007/JHEP12(2010)056}{\emph{JHEP} {\bfseries 12}
  (2010) 056}, [\href{https://arxiv.org/abs/1003.1982}{{\ttfamily 1003.1982}}].

\bibitem{Carta:2021sms}
F.~Carta, A.~Mininno, N.~Righi and A.~Westphal, \emph{{Gopakumar-Vafa
  hierarchies in winding inflation and uplifts}},
  \href{https://doi.org/10.1007/JHEP05(2021)271}{\emph{JHEP} {\bfseries 05}
  (2021) 271}, [\href{https://arxiv.org/abs/2101.07272}{{\ttfamily
  2101.07272}}].

\bibitem{Carta:2021uwv}
F.~Carta, A.~Mininno, N.~Righi and A.~Westphal, \emph{{Thraxions: Towards Full
  String Models}},  \href{https://arxiv.org/abs/2110.02963}{{\ttfamily
  2110.02963}}.

\bibitem{Carta:2022web}
F.~Carta, A.~Mininno and P.~Shukla, \emph{{Divisor topologies of CICY 3-folds
  and their applications to phenomenology}},
  \href{https://doi.org/10.1007/JHEP05(2022)101}{\emph{JHEP} {\bfseries 05}
  (2022) 101}, [\href{https://arxiv.org/abs/2201.02165}{{\ttfamily
  2201.02165}}].

\bibitem{Carta:2022oex}
F.~Carta, A.~Mininno and P.~Shukla, \emph{{Systematics of perturbatively flat
  flux vacua for CICYs}},  \href{https://arxiv.org/abs/2201.10581}{{\ttfamily
  2201.10581}}.

\bibitem{Carta:2020ohw}
F.~Carta, J.~Moritz and A.~Westphal, \emph{{A landscape of orientifold vacua}},
  \href{https://doi.org/10.1007/JHEP05(2020)107}{\emph{JHEP} {\bfseries 05}
  (2020) 107}, [\href{https://arxiv.org/abs/2003.04902}{{\ttfamily
  2003.04902}}].

\bibitem{Blumenhagen:2008zz}
R.~Blumenhagen, V.~Braun, T.~W. Grimm and T.~Weigand, \emph{{GUTs in Type IIB
  Orientifold Compactifications}},
  \href{https://doi.org/10.1016/j.nuclphysb.2009.02.011}{\emph{Nucl. Phys. B}
  {\bfseries 815} (2009) 1--94},
  [\href{https://arxiv.org/abs/0811.2936}{{\ttfamily 0811.2936}}].

\bibitem{Collinucci:2008sq}
A.~Collinucci, M.~Kreuzer, C.~Mayrhofer and N.-O. Walliser, \emph{{Four-modulus
  'Swiss Cheese' chiral models}},
  \href{https://doi.org/10.1088/1126-6708/2009/07/074}{\emph{JHEP} {\bfseries
  07} (2009) 074}, [\href{https://arxiv.org/abs/0811.4599}{{\ttfamily
  0811.4599}}].

\bibitem{Cicoli:2011qg}
M.~Cicoli, C.~Mayrhofer and R.~Valandro, \emph{{Moduli Stabilisation for Chiral
  Global Models}}, \href{https://doi.org/10.1007/JHEP02(2012)062}{\emph{JHEP}
  {\bfseries 02} (2012) 062},
  [\href{https://arxiv.org/abs/1110.3333}{{\ttfamily 1110.3333}}].

\bibitem{Cicoli:2012vw}
M.~Cicoli, S.~Krippendorf, C.~Mayrhofer, F.~Quevedo and R.~Valandro,
  \emph{{D-Branes at del Pezzo Singularities: Global Embedding and Moduli
  Stabilisation}}, \href{https://doi.org/10.1007/JHEP09(2012)019}{\emph{JHEP}
  {\bfseries 09} (2012) 019},
  [\href{https://arxiv.org/abs/1206.5237}{{\ttfamily 1206.5237}}].

\bibitem{Blumenhagen:2012kz}
R.~Blumenhagen, X.~Gao, T.~Rahn and P.~Shukla, \emph{{A Note on Poly-Instanton
  Effects in Type IIB Orientifolds on Calabi-Yau Threefolds}},
  \href{https://doi.org/10.1007/JHEP06(2012)162}{\emph{JHEP} {\bfseries 06}
  (2012) 162}, [\href{https://arxiv.org/abs/1205.2485}{{\ttfamily 1205.2485}}].

\bibitem{Blumenhagen:2012ue}
R.~Blumenhagen, X.~Gao, T.~Rahn and P.~Shukla, \emph{{Moduli Stabilization and
  Inflationary Cosmology with Poly-Instantons in Type IIB Orientifolds}},
  \href{https://doi.org/10.1007/JHEP11(2012)101}{\emph{JHEP} {\bfseries 11}
  (2012) 101}, [\href{https://arxiv.org/abs/1208.1160}{{\ttfamily 1208.1160}}].

\bibitem{Cicoli:2013mpa}
M.~Cicoli, S.~Krippendorf, C.~Mayrhofer, F.~Quevedo and R.~Valandro,
  \emph{{D3/D7 Branes at Singularities: Constraints from Global Embedding and
  Moduli Stabilisation}},
  \href{https://doi.org/10.1007/JHEP07(2013)150}{\emph{JHEP} {\bfseries 07}
  (2013) 150}, [\href{https://arxiv.org/abs/1304.0022}{{\ttfamily 1304.0022}}].

\bibitem{Gao:2013rra}
X.~Gao and P.~Shukla, \emph{{F-term Stabilization of Odd Axions in LARGE Volume
  Scenario}},
  \href{https://doi.org/10.1016/j.nuclphysb.2013.11.015}{\emph{Nucl.Phys.}
  {\bfseries B878} (2014) 269--294},
  [\href{https://arxiv.org/abs/1307.1141}{{\ttfamily 1307.1141}}].

\bibitem{Cicoli:2016xae}
M.~Cicoli, F.~Muia and P.~Shukla, \emph{{Global Embedding of Fibre Inflation
  Models}}, \href{https://doi.org/10.1007/JHEP11(2016)182}{\emph{JHEP}
  {\bfseries 11} (2016) 182},
  [\href{https://arxiv.org/abs/1611.04612}{{\ttfamily 1611.04612}}].

\bibitem{Cicoli:2017axo}
M.~Cicoli, D.~Ciupke, V.~A. Diaz, V.~Guidetti, F.~Muia and P.~Shukla,
  \emph{{Chiral Global Embedding of Fibre Inflation Models}},
  \href{https://doi.org/10.1007/JHEP11(2017)207}{\emph{JHEP} {\bfseries 11}
  (2017) 207}, [\href{https://arxiv.org/abs/1709.01518}{{\ttfamily
  1709.01518}}].

\bibitem{AbdusSalam:2020ywo}
S.~AbdusSalam, S.~Abel, M.~Cicoli, F.~Quevedo and P.~Shukla, \emph{{A
  systematic approach to K\"ahler moduli stabilisation}},
  \href{https://doi.org/10.1007/JHEP08(2020)047}{\emph{JHEP} {\bfseries 08}
  (2020) 047}, [\href{https://arxiv.org/abs/2005.11329}{{\ttfamily
  2005.11329}}].

\bibitem{Leontaris:2022rzj}
G.~K. Leontaris and P.~Shukla, \emph{{Stabilising all K\"ahler moduli in
  perturbative LVS}},
  \href{https://doi.org/10.1007/JHEP07(2022)047}{\emph{JHEP} {\bfseries 07}
  (2022) 047}, [\href{https://arxiv.org/abs/2203.03362}{{\ttfamily
  2203.03362}}].

\bibitem{Braun:2017nhi}
A.~P. Braun, C.~Long, L.~McAllister, M.~Stillman and B.~Sung, \emph{{The Hodge
  Numbers of Divisors of Calabi-Yau Threefold Hypersurfaces}},
  \href{https://arxiv.org/abs/1712.04946}{{\ttfamily 1712.04946}}.

\bibitem{Demirtas:2018akl}
M.~Demirtas, C.~Long, L.~McAllister and M.~Stillman, \emph{{The Kreuzer-Skarke
  Axiverse}}, \href{https://doi.org/10.1007/JHEP04(2020)138}{\emph{JHEP}
  {\bfseries 04} (2020) 138},
  [\href{https://arxiv.org/abs/1808.01282}{{\ttfamily 1808.01282}}].

\bibitem{Demirtas:2020dbm}
M.~Demirtas, L.~McAllister and A.~Rios-Tascon, \emph{{Bounding the
  Kreuzer-Skarke Landscape}},
  \href{https://doi.org/10.1002/prop.202000086}{\emph{Fortsch. Phys.}
  {\bfseries 68} (2020) 2000086},
  [\href{https://arxiv.org/abs/2008.01730}{{\ttfamily 2008.01730}}].

\bibitem{Kreuzer:2002uu}
M.~Kreuzer and H.~Skarke, \emph{{PALP: A Package for analyzing lattice
  polytopes with applications to toric geometry}},
  \href{https://doi.org/10.1016/S0010-4655(03)00491-0}{\emph{Comput. Phys.
  Commun.} {\bfseries 157} (2004) 87--106},
  [\href{https://arxiv.org/abs/math/0204356}{{\ttfamily math/0204356}}].

\bibitem{Braun:2011ik}
A.~P. Braun and N.-O. Walliser, \emph{{A New offspring of PALP}},
  \href{https://arxiv.org/abs/1106.4529}{{\ttfamily 1106.4529}}.

\bibitem{Blumenhagen:2010pv}
R.~Blumenhagen, B.~Jurke, T.~Rahn and H.~Roschy, \emph{{Cohomology of Line
  Bundles: A Computational Algorithm}},
  \href{https://doi.org/10.1063/1.3501132}{\emph{J. Math. Phys.} {\bfseries 51}
  (2010) 103525}, [\href{https://arxiv.org/abs/1003.5217}{{\ttfamily
  1003.5217}}].

\bibitem{Blumenhagen:2011xn}
R.~Blumenhagen, B.~Jurke and T.~Rahn, \emph{{Computational Tools for Cohomology
  of Toric Varieties}}, \href{https://doi.org/10.1155/2011/152749}{\emph{Adv.
  High Energy Phys.} {\bfseries 2011} (2011) 152749},
  [\href{https://arxiv.org/abs/1104.1187}{{\ttfamily 1104.1187}}].

\bibitem{Lust:2006zg}
D.~Lust, S.~Reffert, E.~Scheidegger, W.~Schulgin and S.~Stieberger,
  \emph{{Moduli Stabilization in Type IIB Orientifolds (II)}},
  \href{https://doi.org/10.1016/j.nuclphysb.2006.12.017}{\emph{Nucl. Phys.}
  {\bfseries B766} (2007) 178--231},
  [\href{https://arxiv.org/abs/hep-th/0609013}{{\ttfamily hep-th/0609013}}].

\bibitem{Lust:2006zh}
D.~Lust, S.~Reffert, E.~Scheidegger and S.~Stieberger, \emph{{Resolved Toroidal
  Orbifolds and their Orientifolds}},
  \href{https://doi.org/10.4310/ATMP.2008.v12.n1.a2}{\emph{Adv. Theor. Math.
  Phys.} {\bfseries 12} (2008) 67--183},
  [\href{https://arxiv.org/abs/hep-th/0609014}{{\ttfamily hep-th/0609014}}].

\bibitem{Gao:2013pra}
X.~Gao and P.~Shukla, \emph{{On Classifying the Divisor Involutions in
  Calabi-Yau Threefolds}},
  \href{https://doi.org/10.1007/JHEP11(2013)170}{\emph{JHEP} {\bfseries 11}
  (2013) 170}, [\href{https://arxiv.org/abs/1307.1139}{{\ttfamily 1307.1139}}].

\bibitem{Altman:2021pyc}
R.~Altman, J.~Carifio, X.~Gao and B.~D. Nelson, \emph{{Orientifold Calabi-Yau
  threefolds with divisor involutions and string landscape}},
  \href{https://doi.org/10.1007/JHEP03(2022)087}{\emph{JHEP} {\bfseries 03}
  (2022) 087}, [\href{https://arxiv.org/abs/2111.03078}{{\ttfamily
  2111.03078}}].

\bibitem{Crino:2022zjk}
C.~Crin\`o, F.~Quevedo, A.~Schachner and R.~Valandro, \emph{{A Database of
  Calabi-Yau Orientifolds and the Size of D3-Tadpoles}},
  \href{https://arxiv.org/abs/2204.13115}{{\ttfamily 2204.13115}}.

\bibitem{Hosono:1994av}
S.~Hosono, A.~Klemm and S.~Theisen, \emph{{Lectures on mirror symmetry}},
  \href{https://arxiv.org/abs/hep-th/9403096}{{\ttfamily hep-th/9403096}}.

\bibitem{Benmachiche:2006df}
I.~Benmachiche and T.~W. Grimm, \emph{{Generalized N=1 orientifold
  compactifications and the Hitchin functionals}},
  \href{https://doi.org/10.1016/j.nuclphysb.2006.05.003}{\emph{Nucl. Phys. B}
  {\bfseries 748} (2006) 200--252},
  [\href{https://arxiv.org/abs/hep-th/0602241}{{\ttfamily hep-th/0602241}}].

\bibitem{Gukov:1999ya}
S.~Gukov, C.~Vafa and E.~Witten, \emph{{CFT's from Calabi-Yau four folds}},
  \href{https://doi.org/10.1016/S0550-3213(00)00373-4}{\emph{Nucl. Phys. B}
  {\bfseries 584} (2000) 69--108},
  [\href{https://arxiv.org/abs/hep-th/9906070}{{\ttfamily hep-th/9906070}}].

\bibitem{Shelton:2005cf}
J.~Shelton, W.~Taylor and B.~Wecht, \emph{{Nongeometric flux
  compactifications}},
  \href{https://doi.org/10.1088/1126-6708/2005/10/085}{\emph{JHEP} {\bfseries
  0510} (2005) 085}, [\href{https://arxiv.org/abs/hep-th/0508133}{{\ttfamily
  hep-th/0508133}}].

\bibitem{Ihl:2007ah}
M.~Ihl, D.~Robbins and T.~Wrase, \emph{{Toroidal orientifolds in IIA with
  general NS-NS fluxes}},
  \href{https://doi.org/10.1088/1126-6708/2007/08/043}{\emph{JHEP} {\bfseries
  0708} (2007) 043}, [\href{https://arxiv.org/abs/0705.3410}{{\ttfamily
  0705.3410}}].

\bibitem{Robbins:2007yv}
D.~Robbins and T.~Wrase, \emph{{D-terms from generalized NS-NS fluxes in type
  II}}, \href{https://doi.org/10.1088/1126-6708/2007/12/058}{\emph{JHEP}
  {\bfseries 0712} (2007) 058},
  [\href{https://arxiv.org/abs/0709.2186}{{\ttfamily 0709.2186}}].

\bibitem{Aldazabal:2008zza}
G.~Aldazabal, P.~G. Camara and J.~Rosabal, \emph{{Flux algebra, Bianchi
  identities and Freed-Witten anomalies in F-theory compactifications}},
  \href{https://doi.org/10.1016/j.nuclphysb.2009.01.006}{\emph{Nucl.Phys.}
  {\bfseries B814} (2009) 21--52},
  [\href{https://arxiv.org/abs/0811.2900}{{\ttfamily 0811.2900}}].

\bibitem{Grana:2006hr}
M.~Grana, J.~Louis and D.~Waldram, \emph{{SU(3) x SU(3) compactification and
  mirror duals of magnetic fluxes}},
  \href{https://doi.org/10.1088/1126-6708/2007/04/101}{\emph{JHEP} {\bfseries
  04} (2007) 101}, [\href{https://arxiv.org/abs/hep-th/0612237}{{\ttfamily
  hep-th/0612237}}].

\bibitem{Aldazabal:2006up}
G.~Aldazabal, P.~G. Camara, A.~Font and L.~Ibanez, \emph{{More dual fluxes and
  moduli fixing}},
  \href{https://doi.org/10.1088/1126-6708/2006/05/070}{\emph{JHEP} {\bfseries
  0605} (2006) 070}, [\href{https://arxiv.org/abs/hep-th/0602089}{{\ttfamily
  hep-th/0602089}}].

\bibitem{deCarlos:2009qm}
B.~de~Carlos, A.~Guarino and J.~M. Moreno, \emph{{Complete classification of
  Minkowski vacua in generalised flux models}},
  \href{https://doi.org/10.1007/JHEP02(2010)076}{\emph{JHEP} {\bfseries 1002}
  (2010) 076}, [\href{https://arxiv.org/abs/0911.2876}{{\ttfamily 0911.2876}}].

\bibitem{Aldazabal:2010ef}
G.~Aldazabal, E.~Andres, P.~G. Camara and M.~Grana, \emph{{U-dual fluxes and
  Generalized Geometry}},
  \href{https://doi.org/10.1007/JHEP11(2010)083}{\emph{JHEP} {\bfseries 11}
  (2010) 083}, [\href{https://arxiv.org/abs/1007.5509}{{\ttfamily 1007.5509}}].

\bibitem{Lombardo:2016swq}
D.~M. Lombardo, F.~Riccioni and S.~Risoli, \emph{{$P$ fluxes and exotic
  branes}}, \href{https://doi.org/10.1007/JHEP12(2016)114}{\emph{JHEP}
  {\bfseries 12} (2016) 114},
  [\href{https://arxiv.org/abs/1610.07975}{{\ttfamily 1610.07975}}].

\bibitem{Lombardo:2017yme}
D.~M. Lombardo, F.~Riccioni and S.~Risoli, \emph{{Non-geometric fluxes \&
  tadpole conditions for exotic branes}},
  \href{https://doi.org/10.1007/JHEP10(2017)134}{\emph{JHEP} {\bfseries 10}
  (2017) 134}, [\href{https://arxiv.org/abs/1704.08566}{{\ttfamily
  1704.08566}}].

\bibitem{Shukla:2019wfo}
P.~Shukla, \emph{{Dictionary for the type II nongeometric flux
  compactifications}},
  \href{https://doi.org/10.1103/PhysRevD.103.086009}{\emph{Phys. Rev. D}
  {\bfseries 103} (2021) 086009},
  [\href{https://arxiv.org/abs/1909.07391}{{\ttfamily 1909.07391}}].

\bibitem{Shukla:2016xdy}
P.~Shukla, \emph{{Revisiting the two formulations of Bianchi identities and
  their implications on moduli stabilization}},
  \href{https://doi.org/10.1007/JHEP08(2016)146}{\emph{JHEP} {\bfseries 08}
  (2016) 146}, [\href{https://arxiv.org/abs/1603.08545}{{\ttfamily
  1603.08545}}].

\bibitem{Gao:2018ayp}
X.~Gao, P.~Shukla and R.~Sun, \emph{{On Missing Bianchi Identities in
  Cohomology Formulation}},
  \href{https://doi.org/10.1140/epjc/s10052-019-7291-5}{\emph{Eur. Phys. J.}
  {\bfseries C79} (2019) 781},
  [\href{https://arxiv.org/abs/1805.05748}{{\ttfamily 1805.05748}}].

\bibitem{Witten:1996bn}
E.~Witten, \emph{{Nonperturbative superpotentials in string theory}},
  \href{https://doi.org/10.1016/0550-3213(96)00283-0}{\emph{Nucl. Phys. B}
  {\bfseries 474} (1996) 343--360},
  [\href{https://arxiv.org/abs/hep-th/9604030}{{\ttfamily hep-th/9604030}}].

\bibitem{Blumenhagen:2010ja}
R.~Blumenhagen, A.~Collinucci and B.~Jurke, \emph{{On Instanton Effects in
  F-theory}}, \href{https://doi.org/10.1007/JHEP08(2010)079}{\emph{JHEP}
  {\bfseries 08} (2010) 079},
  [\href{https://arxiv.org/abs/1002.1894}{{\ttfamily 1002.1894}}].

\bibitem{Denef:2004dm}
F.~Denef, M.~R. Douglas and B.~Florea, \emph{{Building a better racetrack}},
  \href{https://doi.org/10.1088/1126-6708/2004/06/034}{\emph{JHEP} {\bfseries
  06} (2004) 034}, [\href{https://arxiv.org/abs/hep-th/0404257}{{\ttfamily
  hep-th/0404257}}].

\bibitem{BlancoPillado:2006he}
J.~J. Blanco-Pillado, C.~P. Burgess, J.~M. Cline, C.~Escoda, M.~Gomez-Reino,
  R.~Kallosh et~al., \emph{{Inflating in a better racetrack}},
  \href{https://doi.org/10.1088/1126-6708/2006/09/002}{\emph{JHEP} {\bfseries
  09} (2006) 002}, [\href{https://arxiv.org/abs/hep-th/0603129}{{\ttfamily
  hep-th/0603129}}].

\bibitem{Balasubramanian:2005zx}
V.~Balasubramanian, P.~Berglund, J.~P. Conlon and F.~Quevedo,
  \emph{{Systematics of moduli stabilisation in Calabi-Yau flux
  compactifications}},
  \href{https://doi.org/10.1088/1126-6708/2005/03/007}{\emph{JHEP} {\bfseries
  03} (2005) 007}, [\href{https://arxiv.org/abs/hep-th/0502058}{{\ttfamily
  hep-th/0502058}}].

\bibitem{Gao:2013hn}
X.~Gao and P.~Shukla, \emph{{On Non-Gaussianities in Two-Field Poly-Instanton
  Inflation}}, \href{https://doi.org/10.1007/JHEP03(2013)061}{\emph{JHEP}
  {\bfseries 03} (2013) 061},
  [\href{https://arxiv.org/abs/1301.6076}{{\ttfamily 1301.6076}}].

\bibitem{Cicoli:2011ct}
M.~Cicoli, F.~G. Pedro and G.~Tasinato, \emph{{Poly-instanton Inflation}},
  \href{https://doi.org/10.1088/1475-7516/2011/12/022}{\emph{JCAP} {\bfseries
  1112} (2011) 022}, [\href{https://arxiv.org/abs/1110.6182}{{\ttfamily
  1110.6182}}].

\bibitem{Lust:2013kt}
D.~Lüst and X.~Zhang, \emph{{Four Kahler Moduli Stabilisation in type IIB
  Orientifolds with K3-fibred Calabi-Yau threefold compactification}},
  \href{https://doi.org/10.1007/JHEP05(2013)051}{\emph{JHEP} {\bfseries 05}
  (2013) 051}, [\href{https://arxiv.org/abs/1301.7280}{{\ttfamily 1301.7280}}].

\bibitem{Gao:2014fva}
X.~Gao, T.~Li and P.~Shukla, \emph{{Cosmological observables in multi-field
  inflation with a non-flat field space}},
  \href{https://doi.org/10.1088/1475-7516/2014/10/008}{\emph{JCAP} {\bfseries
  10} (2014) 008}, [\href{https://arxiv.org/abs/1403.0654}{{\ttfamily
  1403.0654}}].

\bibitem{RoblesLlana:2006is}
D.~Robles-Llana, M.~Rocek, F.~Saueressig, U.~Theis and S.~Vandoren,
  \emph{{Nonperturbative corrections to 4D string theory effective actions from
  SL(2,Z) duality and supersymmetry}},
  \href{https://doi.org/10.1103/PhysRevLett.98.211602}{\emph{Phys. Rev. Lett.}
  {\bfseries 98} (2007) 211602},
  [\href{https://arxiv.org/abs/hep-th/0612027}{{\ttfamily hep-th/0612027}}].

\bibitem{Grimm:2007xm}
T.~W. Grimm, \emph{{Non-Perturbative Corrections and Modularity in N=1 Type IIB
  Compactifications}},
  \href{https://doi.org/10.1088/1126-6708/2007/10/004}{\emph{JHEP} {\bfseries
  0710} (2007) 004}, [\href{https://arxiv.org/abs/0705.3253}{{\ttfamily
  0705.3253}}].

\bibitem{Demirtas:2019sip}
M.~Demirtas, M.~Kim, L.~Mcallister and J.~Moritz, \emph{{Vacua with Small Flux
  Superpotential}},
  \href{https://doi.org/10.1103/PhysRevLett.124.211603}{\emph{Phys. Rev. Lett.}
  {\bfseries 124} (2020) 211603},
  [\href{https://arxiv.org/abs/1912.10047}{{\ttfamily 1912.10047}}].

\bibitem{Demirtas:2021nlu}
M.~Demirtas, M.~Kim, L.~McAllister, J.~Moritz and A.~Rios-Tascon, \emph{{Small
  cosmological constants in string theory}},
  \href{https://doi.org/10.1007/JHEP12(2021)136}{\emph{JHEP} {\bfseries 12}
  (2021) 136}, [\href{https://arxiv.org/abs/2107.09064}{{\ttfamily
  2107.09064}}].

\bibitem{Broeckel:2021uty}
I.~Broeckel, M.~Cicoli, A.~Maharana, K.~Singh and K.~Sinha, \emph{{On the
  Search for Low $W_0$}},  \href{https://arxiv.org/abs/2108.04266}{{\ttfamily
  2108.04266}}.

\bibitem{Carta:2021kpk}
F.~Carta, A.~Mininno and P.~Shukla, \emph{{Systematics of perturbatively flat
  flux vacua}}, \href{https://doi.org/10.1007/JHEP02(2022)205}{\emph{JHEP}
  {\bfseries 02} (2022) 205},
  [\href{https://arxiv.org/abs/2112.13863}{{\ttfamily 2112.13863}}].

\bibitem{Candelas:1990rm}
P.~Candelas, X.~C. De~La~Ossa, P.~S. Green and L.~Parkes, \emph{{A Pair of
  Calabi-Yau manifolds as an exactly soluble superconformal theory}},
  \href{https://doi.org/10.1016/0550-3213(91)90292-6}{\emph{Nucl. Phys. B}
  {\bfseries 359} (1991) 21--74}.

\bibitem{Becker:2002nn}
K.~Becker, M.~Becker, M.~Haack and J.~Louis, \emph{{Supersymmetry breaking and
  alpha-prime corrections to flux induced potentials}},
  \href{https://doi.org/10.1088/1126-6708/2002/06/060}{\emph{JHEP} {\bfseries
  06} (2002) 060}, [\href{https://arxiv.org/abs/hep-th/0204254}{{\ttfamily
  hep-th/0204254}}].

\bibitem{Conlon:2005ki}
J.~P. Conlon, F.~Quevedo and K.~Suruliz, \emph{{Large-volume flux
  compactifications: Moduli spectrum and D3/D7 soft supersymmetry breaking}},
  \href{https://doi.org/10.1088/1126-6708/2005/08/007}{\emph{JHEP} {\bfseries
  08} (2005) 007}, [\href{https://arxiv.org/abs/hep-th/0505076}{{\ttfamily
  hep-th/0505076}}].

\bibitem{Cicoli:2007xp}
M.~Cicoli, J.~P. Conlon and F.~Quevedo, \emph{{Systematics of String Loop
  Corrections in Type IIB Calabi-Yau Flux Compactifications}},
  \href{https://doi.org/10.1088/1126-6708/2008/01/052}{\emph{JHEP} {\bfseries
  01} (2008) 052}, [\href{https://arxiv.org/abs/0708.1873}{{\ttfamily
  0708.1873}}].

\bibitem{Burgess:2020qsc}
C.~P. Burgess, M.~Cicoli, D.~Ciupke, S.~Krippendorf and F.~Quevedo, \emph{{UV
  Shadows in EFTs: Accidental Symmetries, Robustness and No-Scale
  Supergravity}}, \href{https://doi.org/10.1002/prop.202000076}{\emph{Fortsch.
  Phys.} {\bfseries 68} (2020) 2000076},
  [\href{https://arxiv.org/abs/2006.06694}{{\ttfamily 2006.06694}}].

\bibitem{Ciupke:2015msa}
D.~Ciupke, J.~Louis and A.~Westphal, \emph{{Higher-Derivative Supergravity and
  Moduli Stabilization}},
  \href{https://doi.org/10.1007/JHEP10(2015)094}{\emph{JHEP} {\bfseries 10}
  (2015) 094}, [\href{https://arxiv.org/abs/1505.03092}{{\ttfamily
  1505.03092}}].

\bibitem{Cicoli:2023abc}
M.~Cicoli, M.~Licheri, P.~Piantadosi, F.~Quevedo and P.~Shukla, \emph{{To
  appear}},  \href{https://arxiv.org/abs/xxxx.xxxxx}{{\ttfamily xxxx.xxxxx}}.

\bibitem{Grimm:2017okk}
T.~W. Grimm, K.~Mayer and M.~Weissenbacher, \emph{{Higher derivatives in Type
  II and M-theory on Calabi-Yau threefolds}},
  \href{https://doi.org/10.1007/JHEP02(2018)127}{\emph{JHEP} {\bfseries 02}
  (2018) 127}, [\href{https://arxiv.org/abs/1702.08404}{{\ttfamily
  1702.08404}}].

\bibitem{AbdusSalam:2022krp}
S.~AbdusSalam, C.~Crin\`o and P.~Shukla, \emph{{On K3-fibred LARGE Volume
  Scenario with de Sitter vacua from anti-D3-branes}},
  \href{https://doi.org/10.1007/JHEP03(2023)132}{\emph{JHEP} {\bfseries 03}
  (2023) 132}, [\href{https://arxiv.org/abs/2206.12889}{{\ttfamily
  2206.12889}}].

\bibitem{Berg:2004ek}
M.~Berg, M.~Haack and B.~Kors, \emph{{Loop corrections to volume moduli and
  inflation in string theory}},
  \href{https://doi.org/10.1103/PhysRevD.71.026005}{\emph{Phys. Rev. D}
  {\bfseries 71} (2005) 026005},
  [\href{https://arxiv.org/abs/hep-th/0404087}{{\ttfamily hep-th/0404087}}].

\bibitem{Berg:2005ja}
M.~Berg, M.~Haack and B.~Kors, \emph{{String loop corrections to Kahler
  potentials in orientifolds}},
  \href{https://doi.org/10.1088/1126-6708/2005/11/030}{\emph{JHEP} {\bfseries
  11} (2005) 030}, [\href{https://arxiv.org/abs/hep-th/0508043}{{\ttfamily
  hep-th/0508043}}].

\bibitem{Berg:2005yu}
M.~Berg, M.~Haack and B.~Kors, \emph{{On volume stabilization by quantum
  corrections}},
  \href{https://doi.org/10.1103/PhysRevLett.96.021601}{\emph{Phys. Rev. Lett.}
  {\bfseries 96} (2006) 021601},
  [\href{https://arxiv.org/abs/hep-th/0508171}{{\ttfamily hep-th/0508171}}].

\bibitem{Berg:2007wt}
M.~Berg, M.~Haack and E.~Pajer, \emph{{Jumping Through Loops: On Soft Terms
  from Large Volume Compactifications}},
  \href{https://doi.org/10.1088/1126-6708/2007/09/031}{\emph{JHEP} {\bfseries
  09} (2007) 031}, [\href{https://arxiv.org/abs/0704.0737}{{\ttfamily
  0704.0737}}].

\bibitem{vonGersdorff:2005bf}
G.~von Gersdorff and A.~Hebecker, \emph{{Kahler corrections for the volume
  modulus of flux compactifications}},
  \href{https://doi.org/10.1016/j.physletb.2005.08.024}{\emph{Phys. Lett. B}
  {\bfseries 624} (2005) 270--274},
  [\href{https://arxiv.org/abs/hep-th/0507131}{{\ttfamily hep-th/0507131}}].

\bibitem{Gao:2022uop}
X.~Gao, A.~Hebecker, S.~Schreyer and G.~Venken, \emph{{Loops, Local Corrections
  and Warping in the LVS and other Type IIB Models}},
  \href{https://arxiv.org/abs/2204.06009}{{\ttfamily 2204.06009}}.

\bibitem{Kiritsis:1997em}
E.~Kiritsis and B.~Pioline, \emph{{On R**4 threshold corrections in IIb string
  theory and (p, q) string instantons}},
  \href{https://doi.org/10.1016/S0550-3213(97)00645-7}{\emph{Nucl. Phys. B}
  {\bfseries 508} (1997) 509--534},
  [\href{https://arxiv.org/abs/hep-th/9707018}{{\ttfamily hep-th/9707018}}].

\bibitem{Antoniadis:2002tr}
I.~Antoniadis, R.~Minasian and P.~Vanhove, \emph{{Noncompact Calabi-Yau
  manifolds and localized gravity}},
  \href{https://doi.org/10.1016/S0550-3213(02)00974-4}{\emph{Nucl. Phys. B}
  {\bfseries 648} (2003) 69--93},
  [\href{https://arxiv.org/abs/hep-th/0209030}{{\ttfamily hep-th/0209030}}].

\bibitem{Blumenhagen:2009qh}
R.~Blumenhagen, M.~Cvetic, S.~Kachru and T.~Weigand, \emph{{D-Brane Instantons
  in Type II Orientifolds}},
  \href{https://doi.org/10.1146/annurev.nucl.010909.083113}{\emph{Ann. Rev.
  Nucl. Part. Sci.} {\bfseries 59} (2009) 269--296},
  [\href{https://arxiv.org/abs/0902.3251}{{\ttfamily 0902.3251}}].

\bibitem{Ben-Dayan:2013fva}
I.~Ben-Dayan, S.~Jing, A.~Westphal and C.~Wieck, \emph{{Accidental inflation
  from K\"ahler uplifting}},
  \href{https://doi.org/10.1088/1475-7516/2014/03/054}{\emph{JCAP} {\bfseries
  03} (2014) 054}, [\href{https://arxiv.org/abs/1309.0529}{{\ttfamily
  1309.0529}}].

\bibitem{Conlon:2005jm}
J.~P. Conlon and F.~Quevedo, \emph{{Kahler moduli inflation}},
  \href{https://doi.org/10.1088/1126-6708/2006/01/146}{\emph{JHEP} {\bfseries
  01} (2006) 146}, [\href{https://arxiv.org/abs/hep-th/0509012}{{\ttfamily
  hep-th/0509012}}].

\bibitem{Cicoli:2008va}
M.~Cicoli, J.~P. Conlon and F.~Quevedo, \emph{{General Analysis of LARGE Volume
  Scenarios with String Loop Moduli Stabilisation}},
  \href{https://doi.org/10.1088/1126-6708/2008/10/105}{\emph{JHEP} {\bfseries
  10} (2008) 105}, [\href{https://arxiv.org/abs/0805.1029}{{\ttfamily
  0805.1029}}].

\bibitem{Derendinger:2004jn}
J.-P. Derendinger, C.~Kounnas, P.~M. Petropoulos and F.~Zwirner,
  \emph{{Superpotentials in IIA compactifications with general fluxes}},
  \href{https://doi.org/10.1016/j.nuclphysb.2005.02.038}{\emph{Nucl.Phys.}
  {\bfseries B715} (2005) 211--233},
  [\href{https://arxiv.org/abs/hep-th/0411276}{{\ttfamily hep-th/0411276}}].

\bibitem{Grana:2012rr}
M.~Gra\~{n}a and D.~Marques, \emph{{Gauged Double Field Theory}},
  \href{https://doi.org/10.1007/JHEP04(2012)020}{\emph{JHEP} {\bfseries 1204}
  (2012) 020}, [\href{https://arxiv.org/abs/1201.2924}{{\ttfamily 1201.2924}}].

\bibitem{Dibitetto:2012rk}
G.~Dibitetto, J.~Fernandez-Melgarejo, D.~Marques and D.~Roest, \emph{{Duality
  orbits of non-geometric fluxes}},
  \href{https://doi.org/10.1002/prop.201200078}{\emph{Fortsch.Phys.} {\bfseries
  60} (2012) 1123--1149}, [\href{https://arxiv.org/abs/1203.6562}{{\ttfamily
  1203.6562}}].

\bibitem{Danielsson:2012by}
U.~Danielsson and G.~Dibitetto, \emph{{On the distribution of stable de Sitter
  vacua}}, \href{https://doi.org/10.1007/JHEP03(2013)018}{\emph{JHEP}
  {\bfseries 1303} (2013) 018},
  [\href{https://arxiv.org/abs/1212.4984}{{\ttfamily 1212.4984}}].

\bibitem{Blaback:2013ht}
J.~Blåbäck, U.~Danielsson and G.~Dibitetto, \emph{{Fully stable dS vacua from
  generalised fluxes}},
  \href{https://doi.org/10.1007/JHEP08(2013)054}{\emph{JHEP} {\bfseries 1308}
  (2013) 054}, [\href{https://arxiv.org/abs/1301.7073}{{\ttfamily 1301.7073}}].

\bibitem{Damian:2013dq}
C.~Damian, L.~R. Diaz-Barron, O.~Loaiza-Brito and M.~Sabido, \emph{{Slow-Roll
  Inflation in Non-geometric Flux Compactification}},
  \href{https://doi.org/10.1007/JHEP06(2013)109}{\emph{JHEP} {\bfseries 1306}
  (2013) 109}, [\href{https://arxiv.org/abs/1302.0529}{{\ttfamily 1302.0529}}].

\bibitem{Damian:2013dwa}
C.~Damian and O.~Loaiza-Brito, \emph{{More stable de Sitter vacua from S-dual
  nongeometric fluxes}},
  \href{https://doi.org/10.1103/PhysRevD.88.046008}{\emph{Phys.Rev.} {\bfseries
  D88} (2013) 046008}, [\href{https://arxiv.org/abs/1304.0792}{{\ttfamily
  1304.0792}}].

\bibitem{Hassler:2014mla}
F.~Hassler, D.~Lust and S.~Massai, \emph{{On Inflation and de Sitter in
  Non‐Geometric String Backgrounds}},
  \href{https://doi.org/10.1002/prop.201700062}{\emph{Fortsch. Phys.}
  {\bfseries 65} (2017) 1700062},
  [\href{https://arxiv.org/abs/1405.2325}{{\ttfamily 1405.2325}}].

\bibitem{Blaback:2015zra}
J.~Blåbäck, U.~H. Danielsson, G.~Dibitetto and S.~C. Vargas, \emph{{Universal
  dS vacua in STU-models}},
  \href{https://doi.org/10.1007/JHEP10(2015)069}{\emph{JHEP} {\bfseries 10}
  (2015) 069}, [\href{https://arxiv.org/abs/1505.04283}{{\ttfamily
  1505.04283}}].

\bibitem{Dibitetto:2011qs}
G.~Dibitetto, A.~Guarino and D.~Roest, \emph{{Vacua Analysis in Extended
  Supersymmetry Compactifications}},
  \href{https://doi.org/10.1002/prop.201200004}{\emph{Fortsch. Phys.}
  {\bfseries 60} (2012) 987--990},
  [\href{https://arxiv.org/abs/1112.1306}{{\ttfamily 1112.1306}}].

\bibitem{Guarino:2008ik}
A.~Guarino and G.~J. Weatherill, \emph{{Non-geometric flux vacua, S-duality and
  algebraic geometry}},
  \href{https://doi.org/10.1088/1126-6708/2009/02/042}{\emph{JHEP} {\bfseries
  0902} (2009) 042}, [\href{https://arxiv.org/abs/0811.2190}{{\ttfamily
  0811.2190}}].

\bibitem{Blumenhagen:2015kja}
R.~Blumenhagen, A.~Font, M.~Fuchs, D.~Herschmann, E.~Plauschinn, Y.~Sekiguchi
  et~al., \emph{{A Flux-Scaling Scenario for High-Scale Moduli Stabilization in
  String Theory}},
  \href{https://doi.org/10.1016/j.nuclphysb.2015.06.003}{\emph{Nucl. Phys. B}
  {\bfseries 897} (2015) 500--554},
  [\href{https://arxiv.org/abs/1503.07634}{{\ttfamily 1503.07634}}].

\bibitem{Shukla:2015hpa}
P.~Shukla, \emph{{A symplectic rearrangement of the four dimensional
  non-geometric scalar potential}},
  \href{https://doi.org/10.1007/JHEP11(2015)162}{\emph{JHEP} {\bfseries 11}
  (2015) 162}, [\href{https://arxiv.org/abs/1508.01197}{{\ttfamily
  1508.01197}}].

\bibitem{Shukla:2016hyy}
P.~Shukla, \emph{{Reading off the nongeometric scalar potentials via the
  topological data of the compactifying Calabi-Yau manifolds}},
  \href{https://doi.org/10.1103/PhysRevD.94.086003}{\emph{Phys. Rev.}
  {\bfseries D94} (2016) 086003},
  [\href{https://arxiv.org/abs/1603.01290}{{\ttfamily 1603.01290}}].

\bibitem{Blumenhagen:2015xpa}
R.~Blumenhagen, C.~Damian, A.~Font, D.~Herschmann and R.~Sun, \emph{{The
  Flux-Scaling Scenario: De Sitter Uplift and Axion Inflation}},
  \href{https://doi.org/10.1002/prop.201600030}{\emph{Fortsch. Phys.}
  {\bfseries 64} (2016) 536--550},
  [\href{https://arxiv.org/abs/1510.01522}{{\ttfamily 1510.01522}}].

\bibitem{Font:2008vd}
A.~Font, A.~Guarino and J.~M. Moreno, \emph{{Algebras and non-geometric flux
  vacua}}, \href{https://doi.org/10.1088/1126-6708/2008/12/050}{\emph{JHEP}
  {\bfseries 0812} (2008) 050},
  [\href{https://arxiv.org/abs/0809.3748}{{\ttfamily 0809.3748}}].

\bibitem{Hull:2004in}
C.~Hull, \emph{{A Geometry for non-geometric string backgrounds}},
  \href{https://doi.org/10.1088/1126-6708/2005/10/065}{\emph{JHEP} {\bfseries
  0510} (2005) 065}, [\href{https://arxiv.org/abs/hep-th/0406102}{{\ttfamily
  hep-th/0406102}}].

\bibitem{Kumar:1996zx}
A.~Kumar and C.~Vafa, \emph{{U manifolds}},
  \href{https://doi.org/10.1016/S0370-2693(97)00108-1}{\emph{Phys.Lett.}
  {\bfseries B396} (1997) 85--90},
  [\href{https://arxiv.org/abs/hep-th/9611007}{{\ttfamily hep-th/9611007}}].

\bibitem{Hull:2003kr}
C.~M. Hull and A.~Catal-Ozer, \emph{{Compactifications with S duality twists}},
  \href{https://doi.org/10.1088/1126-6708/2003/10/034}{\emph{JHEP} {\bfseries
  0310} (2003) 034}, [\href{https://arxiv.org/abs/hep-th/0308133}{{\ttfamily
  hep-th/0308133}}].

\bibitem{Shukla:2015rua}
P.~Shukla, \emph{{On modular completion of generalized flux orbits}},
  \href{https://doi.org/10.1007/JHEP11(2015)075}{\emph{JHEP} {\bfseries 11}
  (2015) 075}, [\href{https://arxiv.org/abs/1505.00544}{{\ttfamily
  1505.00544}}].

\bibitem{Blumenhagen:2013hva}
R.~Blumenhagen, X.~Gao, D.~Herschmann and P.~Shukla, \emph{{Dimensional
  Oxidation of Non-geometric Fluxes in Type II Orientifolds}},
  \href{https://doi.org/10.1007/JHEP10(2013)201}{\emph{JHEP} {\bfseries 1310}
  (2013) 201}, [\href{https://arxiv.org/abs/1306.2761}{{\ttfamily 1306.2761}}].

\bibitem{Gao:2015nra}
X.~Gao and P.~Shukla, \emph{{Dimensional oxidation and modular completion of
  non-geometric type IIB action}},
  \href{https://doi.org/10.1007/JHEP05(2015)018}{\emph{JHEP} {\bfseries 1505}
  (2015) 018}, [\href{https://arxiv.org/abs/1501.07248}{{\ttfamily
  1501.07248}}].

\bibitem{Shukla:2015bca}
P.~Shukla, \emph{{Implementing odd-axions in dimensional oxidation of 4D
  non-geometric type IIB scalar potential}},
  \href{https://doi.org/10.1016/j.nuclphysb.2015.11.020}{\emph{Nucl. Phys.}
  {\bfseries B902} (2016) 458--482},
  [\href{https://arxiv.org/abs/1507.01612}{{\ttfamily 1507.01612}}].

\bibitem{Andriot:2013xca}
D.~Andriot and A.~Betz, \emph{{$\beta$-supergravity: a ten-dimensional theory
  with non-geometric fluxes, and its geometric framework}},
  \href{https://doi.org/10.1007/JHEP12(2013)083}{\emph{JHEP} {\bfseries 1312}
  (2013) 083}, [\href{https://arxiv.org/abs/1306.4381}{{\ttfamily 1306.4381}}].

\bibitem{Andriot:2011uh}
D.~Andriot, M.~Larfors, D.~Lust and P.~Patalong, \emph{{A ten-dimensional
  action for non-geometric fluxes}},
  \href{https://doi.org/10.1007/JHEP09(2011)134}{\emph{JHEP} {\bfseries 1109}
  (2011) 134}, [\href{https://arxiv.org/abs/1106.4015}{{\ttfamily 1106.4015}}].

\bibitem{Blumenhagen:2015lta}
R.~Blumenhagen, A.~Font and E.~Plauschinn, \emph{{Relating double field theory
  to the scalar potential of N = 2 gauged supergravity}},
  \href{https://doi.org/10.1007/JHEP12(2015)122}{\emph{JHEP} {\bfseries 12}
  (2015) 122}, [\href{https://arxiv.org/abs/1507.08059}{{\ttfamily
  1507.08059}}].

\bibitem{Villadoro:2005cu}
G.~Villadoro and F.~Zwirner, \emph{{N=1 effective potential from dual type-IIA
  D6/O6 orientifolds with general fluxes}},
  \href{https://doi.org/10.1088/1126-6708/2005/06/047}{\emph{JHEP} {\bfseries
  0506} (2005) 047}, [\href{https://arxiv.org/abs/hep-th/0503169}{{\ttfamily
  hep-th/0503169}}].

\bibitem{Blumenhagen:2015qda}
R.~Blumenhagen, A.~Font, M.~Fuchs, D.~Herschmann and E.~Plauschinn,
  \emph{{Towards Axionic Starobinsky-like Inflation in String Theory}},
  \href{https://doi.org/10.1016/j.physletb.2015.05.001}{\emph{Phys. Lett.}
  {\bfseries B746} (2015) 217--222},
  [\href{https://arxiv.org/abs/1503.01607}{{\ttfamily 1503.01607}}].

\bibitem{Blumenhagen:2015jva}
R.~Blumenhagen, A.~Font, M.~Fuchs, D.~Herschmann and E.~Plauschinn,
  \emph{{Large field inflation and string moduli stabilization}}, {\emph{PoS}
  {\bfseries PLANCK2015} (2015) 021},
  [\href{https://arxiv.org/abs/1510.04059}{{\ttfamily 1510.04059}}].

\bibitem{Li:2015taa}
T.~Li, Z.~Li and D.~V. Nanopoulos, \emph{{Helical Phase Inflation via
  Non-Geometric Flux Compactifications: from Natural to Starobinsky-like
  Inflation}}, \href{https://doi.org/10.1007/JHEP10(2015)138}{\emph{JHEP}
  {\bfseries 10} (2015) 138},
  [\href{https://arxiv.org/abs/1507.04687}{{\ttfamily 1507.04687}}].

\bibitem{Bianchi:2011qh}
M.~Bianchi, A.~Collinucci and L.~Martucci, \emph{{Magnetized E3-brane
  instantons in F-theory}},
  \href{https://doi.org/10.1007/JHEP12(2011)045}{\emph{JHEP} {\bfseries 12}
  (2011) 045}, [\href{https://arxiv.org/abs/1107.3732}{{\ttfamily 1107.3732}}].

\bibitem{Bianchi:2012pn}
M.~Bianchi, A.~Collinucci and L.~Martucci, \emph{{Freezing E3-brane instantons
  with fluxes}}, \href{https://doi.org/10.1002/prop.201200030}{\emph{Fortsch.
  Phys.} {\bfseries 60} (2012) 914--920},
  [\href{https://arxiv.org/abs/1202.5045}{{\ttfamily 1202.5045}}].

\bibitem{Cicoli:2011it}
M.~Cicoli, M.~Kreuzer and C.~Mayrhofer, \emph{{Toric K3-Fibred Calabi-Yau
  Manifolds with del Pezzo Divisors for String Compactifications}},
  \href{https://doi.org/10.1007/JHEP02(2012)002}{\emph{JHEP} {\bfseries 02}
  (2012) 002}, [\href{https://arxiv.org/abs/1107.0383}{{\ttfamily 1107.0383}}].

\bibitem{Altman:2014bfa}
R.~Altman, J.~Gray, Y.-H. He, V.~Jejjala and B.~D. Nelson, \emph{{A Calabi-Yau
  Database: Threefolds Constructed from the Kreuzer-Skarke List}},
  \href{https://doi.org/10.1007/JHEP02(2015)158}{\emph{JHEP} {\bfseries 02}
  (2015) 158}, [\href{https://arxiv.org/abs/1411.1418}{{\ttfamily 1411.1418}}].

\bibitem{Cicoli:2008gp}
M.~Cicoli, C.~P. Burgess and F.~Quevedo, \emph{{Fibre Inflation: Observable
  Gravity Waves from IIB String Compactifications}},
  \href{https://doi.org/10.1088/1475-7516/2009/03/013}{\emph{JCAP} {\bfseries
  03} (2009) 013}, [\href{https://arxiv.org/abs/0808.0691}{{\ttfamily
  0808.0691}}].

\bibitem{OGUISO:1993}
K.~Oguiso, \emph{{On Algebraic Fiber Space Structures on a Calabi-Yau 3-Fold}},
  \href{https://doi.org/10.1142/S0129167X93000248}{\emph{International Journal
  of Mathematics} {\bfseries 04} (1993) 439--465}.

\bibitem{Schulz:2004tt}
M.~B. Schulz, \emph{{Calabi-Yau duals of torus orientifolds}},
  \href{https://doi.org/10.1088/1126-6708/2006/05/023}{\emph{JHEP} {\bfseries
  05} (2006) 023}, [\href{https://arxiv.org/abs/hep-th/0412270}{{\ttfamily
  hep-th/0412270}}].

\bibitem{Gao:2022fdi}
X.~Gao, A.~Hebecker, S.~Schreyer and G.~Venken, \emph{{The LVS Parametric
  Tadpole Constraint}},  \href{https://arxiv.org/abs/2202.04087}{{\ttfamily
  2202.04087}}.

\bibitem{Antoniadis:2019doc}
I.~Antoniadis, Y.~Chen and G.~K. Leontaris, \emph{{Moduli stabilisation and
  inflation in type IIB/F-theory}},
  \href{https://doi.org/10.22323/1.347.0068}{\emph{PoS} {\bfseries CORFU2018}
  (2019) 068}, [\href{https://arxiv.org/abs/1901.05075}{{\ttfamily
  1901.05075}}].

\bibitem{Antoniadis:2020ryh}
I.~Antoniadis, Y.~Chen and G.~K. Leontaris, \emph{{String loop corrections and
  de Sitter vacua}}, \href{https://doi.org/10.22323/1.376.0099}{\emph{PoS}
  {\bfseries CORFU2019} (2020) 099}.

\bibitem{Antoniadis:2021lhi}
I.~Antoniadis, O.~Lacombe and G.~K. Leontaris, \emph{{Hybrid inflation and
  waterfall field in string theory from D7-branes}},
  \href{https://doi.org/10.1007/JHEP01(2022)011}{\emph{JHEP} {\bfseries 01}
  (2022) 011}, [\href{https://arxiv.org/abs/2109.03243}{{\ttfamily
  2109.03243}}].

\bibitem{Basiouris:2021sdf}
V.~Basiouris and G.~K. Leontaris, \emph{{Remarks on the Effects of Quantum
  Corrections on Moduli Stabilization and de Sitter Vacua in Type IIB String
  Theory}}, \href{https://doi.org/10.1002/prop.202100181}{\emph{Fortsch. Phys.}
  {\bfseries 70} (2022) 2100181},
  [\href{https://arxiv.org/abs/2109.08421}{{\ttfamily 2109.08421}}].

\bibitem{Antoniadis:2022owm}
I.~Antoniadis, O.~Lacombe and G.~K. Leontaris, \emph{{Type IIB moduli
  stabilization, inflation and waterfall fields}},
  \href{https://doi.org/10.1142/S0217751X22440018}{\emph{Int. J. Mod. Phys. A}
  {\bfseries 37} (2022) 2244001}.

\bibitem{Ahmed:2022vlc}
W.~Ahmed, A.~Karozas, G.~K. Leontaris and U.~Zubair, \emph{{Smooth hybrid
  inflation with low reheat temperature and observable gravity waves in
  SU(5)xU(1)super-GUT}},
  \href{https://doi.org/10.1088/1475-7516/2022/06/027}{\emph{JCAP} {\bfseries
  06} (2022) 027}, [\href{https://arxiv.org/abs/2201.12789}{{\ttfamily
  2201.12789}}].

\bibitem{Ahmed:2022dhc}
W.~Ahmed, A.~Karozas, G.~K. Leontaris and I.~Tavellaris, \emph{{Moduli
  Stabilisation, de Sitter Vacua and Hybrid Inflation in Large Volume
  Compactifications}},  \href{https://arxiv.org/abs/2301.00329}{{\ttfamily
  2301.00329}}.

\end{thebibliography}\endgroup

%\newpage
%\bibliographystyle{utphys}
%\bibliography{reference}

\end{document}